\documentclass[twocolumn,amsmath,amssymb,prd]{revtex4}

\usepackage{epsfig}

\def\al{\alpha}
\def\be{\beta}
\def\ga{\gamma}
\def\de{\delta}
\def\ep{\epsilon}

\def\et{\eta}
\def\th{\theta}

\def\ka{\kappa}
\def\la{\lambda}

\def\rh{\rho}

\def\si{\sigma}

\def\ta{\tau}

\def\ph{\phi}

\def\ch{\chi}
\def\ps{\psi}
\def\om{\omega}
\def\Ga{\Gamma}
\def\De{\Delta}

\def\Ph{\Phi}
\def\Ps{\Psi}

\def\cl{{\cal L}}

\def\mn{{\mu\nu}}

\def\fr#1#2{{{#1}\over{#2}}}
\def\frac#1#2{{\textstyle{{#1}\over{#2}}}}
\def\half{{\textstyle{1\over 2}}}
\def\ol{\overline}
\def\prt{\partial}

\def\re{{\rm Re}}
\def\im{{\rm Im}}

\def\lsim{\mathrel{\rlap{\lower4pt\hbox{\hskip1pt$\sim$}}
    \raise1pt\hbox{$<$}}}
\def\gsim{\mathrel{\rlap{\lower4pt\hbox{\hskip1pt$\sim$}}
    \raise1pt\hbox{$>$}}}

\def\vev#1{\langle {#1}\rangle}

\def\etal{{\it et al.}}

\newcommand{\beq}{\begin{equation}}
\newcommand{\eeq}{\end{equation}}
\newcommand{\bea}{\begin{eqnarray}}
\newcommand{\eea}{\end{eqnarray}}
\newcommand{\rf}[1]{(\ref{#1})}
\newcommand{\bM}{\begin{pmatrix}}
\newcommand{\eM}{\end{pmatrix}}

\def\nn{\nonumber}

\def\Psb{\ol\Ps{}}
\def\psb{\ol\ps{}}

\def\ring#1{{\mathaccent'27 #1}}
\def\ari{{\ring{a}}}
\def\cri{{\ring{c}}}

\def\hri{{\ring{h}}}
\def\Eri{{\ring{E}}}

\def\C#1{\cri^{(#1)}}

\def\heff{h_{\rm eff}}
\def\deh{\de h}
\def\hosc{h_{\rm osc}}

\def\ceafm{$c_8a_5m$}

\def\nub{\ol\nu}

\def\Q{\mathcal Q}
\def\S{\mathcal S}
\def\P{\mathcal P}
\def\V{\mathcal V}
\def\A{\mathcal A}
\def\T{\mathcal T}
\def\C{\mathcal C}

\def\B{\mathcal B}
\def\Qhat{\widehat\Q}
\def\Shat{\widehat\S}
\def\Phat{\widehat\P}
\def\Vhat{\widehat\V}
\def\Ahat{\widehat\A}
\def\That{\widehat\T}
\def\Tdual{\widetilde{\widehat\T}\phantom{}}

\def\mhat{\widehat m}
\def\mfivehat{\widehat m_5{}}
\def\ahat{\widehat a}
\def\bhat{\widehat b}
\def\chat{\widehat c}
\def\dhat{\widehat d}
\def\ehat{\widehat e}
\def\fhat{\widehat f}
\def\ghat{\widehat g}
\def\Hhat{\widehat H}
\def\gdual{\widetilde{\widehat g}\phantom{}}
\def\Hdual{\widetilde{\widehat H}\phantom{}}

\def\mbf#1{\boldsymbol #1}
\def\syjm#1#2{{}_{#1}Y_{#2}}

\def\bsi{\overline\si}

\def\eom{E}
\def\mbfp{|\mbf p|}
\def\phat{\hat {\mbf p}}

\def\xhat{\hat x}

\def\pvec{\mbf p}
\def\pmag{\mbfp}
\def\kvec{\boldsymbol k}
\def\kmag{|\boldsymbol k|}

\def\nuTemplate#1#2#3#4{\big(#1^{(#2)}\big)_{#3}^{#4}}
\def\cLcoef#1#2#3{\nuTemplate{c_L}{#1}{#2}{#3}}
\def\aLcoef#1#2#3{\nuTemplate{a_L}{#1}{#2}{#3}}
\def\mlcoef#1#2#3{\nuTemplate{m_l}{#1}{#2}{#3}}
\def\elcoef#1#2#3{\nuTemplate{e_l}{#1}{#2}{#3}}
\def\Hlcoef#1#2#3{\nuTemplate{H_l}{#1}{#2}{#3}}
\def\glcoef#1#2#3{\nuTemplate{g_l}{#1}{#2}{#3}}
\def\HMcoef#1#2#3{\nuTemplate{H_{M+}}{#1}{#2}{#3}}
\def\gMcoef#1#2#3{\nuTemplate{g_{M+}}{#1}{#2}{#3}}
\def\clcoef#1#2#3{\nuTemplate{c_l}{#1}{#2}{#3}}
\def\alcoef#1#2#3{\nuTemplate{a_l}{#1}{#2}{#3}}

\def\ae{\widehat a_{\rm eff}}
\def\ce{\widehat c_{\rm eff}}
\def\He{\widehat H_{\rm eff}}
\def\ge{\widehat g_{\rm eff}}
\def\ceff#1#2#3{\nuTemplate{c_\text{eff}}{#1}{#2}{#3}}
\def\aeff#1#2#3{\nuTemplate{a_\text{eff}}{#1}{#2}{#3}}
\def\Heff#1#2#3{\nuTemplate{H_\text{eff}}{#1}{#2}{#3}}
\def\geff#1#2#3{\nuTemplate{g_\text{eff}}{#1}{#2}{#3}}
\def\cfb#1#2{\nuTemplate{c_\text{fb}}{#1}{#2}{}}
\def\afb#1#2{\nuTemplate{a_\text{fb}}{#1}{#2}{}}
\def\gfb#1#2{\nuTemplate{g_\text{fb}}{#1}{#2}{}}
\def\cdia#1#2#3{\nuTemplate{c_\text{d}}{#1}{#2}{#3}}
\def\adia#1#2#3{\nuTemplate{a_\text{d}}{#1}{#2}{#3}}
\def\gdia#1#2#3{\nuTemplate{g_\text{d}}{#1}{#2}{#3}}
\def\cof#1#2{\nuTemplate{c_\text{of}}{#1}{#2}{}}
\def\aof#1#2{\nuTemplate{a_\text{of}}{#1}{#2}{}}
\def\cfc#1#2{\cri^{(#1)}_{#2}}
\def\afc#1#2{\ari^{(#1)}_{#2}}
\def\ccfc#1#2{\cri^{(#1)}_{#2}}
\def\acfc#1#2{\ari^{(#1)}_{#2}}

\def\SBamp#1#2#3#4{\big({\mathcal A}_{#1}^{(#2)}\big)_{#3}^{#4}}

\def\M#1#2#3{({\mathcal M}^{(#1)}_{#2})_{#3}}
\def\N#1#2{{}_{#1}{\mathcal N}_{#2}}
\def\H#1#2{{\mathcal H}^{(#1)}_{#2}}

\def\LBamp#1#2#3#4{\big({\mathcal A}_{#1}^{(#2)}\big)_{#3}^{#4}}
\def\LBAMP#1#2#3#4{\big({\mathcal B}_{#1}^{(#2)}\big)_{#3}^{#4}}

\def\ACPT#1{{\mathcal A}^{CPT}_{#1}}

\def\TOFamp#1#2#3#4{\big({\mathcal A}_{#1}^{(#2)}\big)_{#3}^{#4}}

\def\vg{v}
\def\vgof{\vg^{\rm of}}
\def\vgri{\ring\vg{}}

\def\msm{3$\nu$SM} 
\def\cv{\v Cerenkov}

\begin{document}
\title{Neutrinos with Lorentz-violating operators 
of arbitrary dimension}

\author{V.\ Alan Kosteleck\'y$^1$ and Matthew Mewes$^2$}

\affiliation{$^1$Physics Department, Indiana University, 
Bloomington, Indiana 47405, USA\\
$^2$Physics Department, Swarthmore College,
Swarthmore, Pennsylvania 19081, USA}

\date{IUHET 567, December 2011;
published in 
Phys.\ Rev.\ D {\bf 85}, 096005 (2012) 
} 

\begin{abstract}

The behavior of fermions in the presence
of Lorentz and CPT violation is studied.
Allowing for operators of any mass dimension,
we classify all Lorentz-violating terms
in the quadratic Lagrange density for free fermions.
The result is adapted to obtain the effective hamiltonian 
describing the propagation and mixing 
of three flavors of left-handed neutrinos
in the presence of Lorentz violation 
involving operators of arbitrary mass dimension.
A characterization of the neutrino 
coefficients for Lorentz violation is provided 
via a decomposition using spin-weighted spherical harmonics.
The restriction of the general theory
to various special cases is discussed,
including among others
the renormalizable limit,
the massless scenario,
flavor-blind and oscillation-free models,
the diagonalizable case,
and several isotropic limits.
The formalism is combined with existing data 
on neutrino oscillations and kinematics 
to extract a variety of measures 
of coefficients for Lorentz and CPT violation.
For oscillations,
we use results from the short-baseline experiments
LSND and MiniBooNE to obtain explicit sensitivities 
to effects from flavor-mixing Lorentz-violating operators 
up to mass dimension 10,
and we present methods to analyze data 
from long-baseline experiments.
For propagation,
we use time-of-flight measurements
from the supernova SN1987A 
and from a variety of experiments including MINOS and OPERA
to constrain oscillation-free Lorentz-violating operators 
up to mass dimension 10,
and we discuss constraints from
threshold effects in meson decays and \cv\ emission. 

\end{abstract}

\maketitle

\section{Introduction}

Nongravitational phenomena are well described by
the minimal Standard Model (SM) of particle physics
\cite{pdg}.
However,
the existence of physics beyond the SM is established 
by the confirmed observation of neutrino oscillations,
while the SM itself is believed to be 
a low-energy effective theory 
emerging from an underlying unified description 
of gravity and quantum physics at the Planck scale.
Potential experimental signals
exposing foundational Planck-scale physics
are therefore of great interest
but are challenging to identify.
One proposed class of signals 
involves the breaking of Lorentz symmetry
associated with tiny deviations from relativity
\cite{ksp}.
In recent years,
searches for Lorentz violation
and related CPT violation
have been performed with a wide range of systems
and at impressive sensitivities
\cite{tables}.

Effective field theory can be used to describe Lorentz violation
in realistic models at attainable energies
\cite{kp}.
In this approach,
CPT violation is associated with Lorentz violation
\cite{owg}.
The comprehensive effective field theory
incorporating the SM and General Relativity
and characterizing general Lorentz violation 
is the Standard-Model Extension (SME)
\cite{ck,akgrav}.
Its Lagrange density behaves as a scalar density 
under observer transformations, 
so the SME action is coordinate independent.
Each Lorentz-violating term is formed by contracting 
a Lorentz-violating operator of a given mass dimension $d$ 
with a controlling coefficient.
The mass dimension $d$ provides
a partial classification of the operators,
and it offers a rough sense 
of the size of associated experimental effects.

In this work,
we focus on Lorentz violation in neutrinos.
The interferometric nature of neutrinos
and their tiny apparent mass scale
makes them natural probes for Planck-scale effects
such as Lorentz and CPT violation.
Various searches for Lorentz violation with neutrinos
have achieved sensitivities to SME coefficients
at levels comparable to Planck-suppressed effects,
including ones by 
the LSND
\cite{lsndlv}, 
Super-Kamiokande (SK)
\cite{sklv},
MINOS
\cite{minoslv1,minoslv2},
IceCube 
\cite{IceCube},
and MiniBooNE
\cite{MiniBooNElv}
collaborations,
while evidence for superluminal neutrinos
has recently been presented by the OPERA collaboration
\cite{opera}.

The introduction of 
neutrino coefficients for Lorentz and CPT violation
\cite{ck}
has led to numerous efforts to understand 
their implications for neutrino behavior.
The general effective hamiltonian 
describing neutrino propagation and mixing
in the presence of Lorentz-violating operators 
of renormalizable dimension
contains four types of coefficients
\cite{km},
leading to many novel effects
that can be revealed in suitable experiments
\cite{km-sb,dkm}.
One interesting theoretical challenge is the construction 
of a global Lorentz-violating model 
describing all established neutrino behavior 
and perhaps also one or more of the known anomalies,
without the two usual neutrino masses 
\cite{bicycle,tandem,bmw,puma,blmw}.
Many of the SME-based phenomenological studies 
of neutrino behavior focus on the special case 
of isotropic Lorentz violation in a preferred frame 
\cite{cg99,bpww,bbm,dg,gl,dgms,dgpg,gghm,fk,%
gktw,dges,jk,rnknkm,xm,bgpg,ba,ym,ag,as,bcgw,ltz,cmh},
sometimes in the two-flavor limit.
Several works treat anisotropic effects 
without lepton-number violation 
\cite{mp,bef,gl2,hmw,bb,akm,bapion,qm}.
Models also exist 
that incorporate nonconservation of lepton number,
which leads to neutrino-antineutrino mixing
\cite{scsk,hmp,hmp2}.
The propagation of free neutrinos in the SME
follows geodesics in a pseudo-Riemann-Finsler geometry 
\cite{finsler}.

The primary goal of this paper
is to extend available techniques for handling 
Lorentz violation in the neutrino sector 
to include operators of arbitrary mass dimension $d$.
We are motivated partly by the notion that the usual SM
represents the dominant component 
of a low-energy effective theory,
with subdominant terms involving Lorentz violation
given by the minimal SME with $d=3$ or $d=4$.
In this picture,
higher-order corrections involve 
Lorentz-violating operators of larger $d$,
which are expected to grow in significance 
as energies increase
and may thereby provide a link to the underlying theory
\cite{kle}.
In some cases,
such as supersymmetric Lorentz-violating theories
\cite{susy},
operators with $d>4$ can represent 
the dominant corrections to the SM.
Similarly,
in noncommutative quantum electrodynamics
\cite{hayakawa},
the action written in terms 
of the conventional fermion and photon fields
contains only SME operators of dimension $d\geq 6$
\cite{chklo}.

The analysis performed in this work 
has many parallels to the treatment
for Lorentz-violating operators at arbitrary $d$
performed for the photon sector
\cite{kmnonmin},
and much of the methodology established for that work
is applicable here.
The formalism contains as a limiting case
the earlier systematic investigation
of neutrino operators of renormalizable dimension
\cite{km}
and the corresponding SME-based studies 
of neutrino Lorentz violation mentioned above.
However,
the present work generalizes the existing treatment
to include effects at leading order in both mass
and Lorentz violation,
uncovering novel Lorentz-violating effects
involving neutrino helicity flip.
It also incorporates as limiting cases
studies of nonrenormalizable neutrino operators
\cite{km,amu,jlm,jc,jp,pbmp,sehmr,mmgls},
including discussions of modified dispersion relations
in the context of superluminal neutrinos
\cite{ftlnu1,ftlnu2,ftlnu3,ftlnu4,jr,re,mrad,dmpt,jw,kz,%
gss,aem,cg,lgm,byyy,cns,ndhd,jcjc,afks,mlm,ssx,%
frk,aabp,mr1,llmwz,histy,msch,mmvv,hllnq,fbhl,mr2}.

To achieve a reasonable scope
while covering neutrino propagation and mixing 
in the presence of general Lorentz and CPT violation, 
we take the Lagrange density of interest 
to be quadratic in free fermion fields.
Our methods and results are applicable to any type of fermion
and hence are also relevant for other sectors of the SME.
Implications of this and of possible interaction terms 
are considered elsewhere
\cite{akmmnrint}.
In this work,
our primary focus is the application 
to multiple generations of left-handed neutrinos.
We obtain the effective hamiltonian
describing neutrino propagation and mixing
in the presence of Lorentz- and CPT-violating operators
of arbitrary mass dimension.
Since rotations form a subgroup of the Lorentz group,
careful use of rotational properties 
often benefits experimental analyses,
and so we develop a decomposition in spherical harmonics
to characterize effects.
We find that all Lorentz-violating features  
of neutrino propagation and mixing
are determined by four sets of 
effective spherical coefficients for Lorentz violation,
which can physically be distinguished
by their Dirac or Majorana nature
and by their CPT properties. 
Using this classification scheme,
various limiting models can readily be identified and studied,
and explicit measurements of coefficients for Lorentz violation
can be extracted from observational and experimental data.
Here,
we use existing results 
from experiments on oscillations, times of flight, thresholds,
and \cv\ emission
to tabulate measurements and maximal attained sensitivities 
for Lorentz-violating operators of mass dimensions $d \leq 10$.
Many of our results represent the first available constraints
on the corresponding Lorentz-violating effects.
 
The structure of this paper is as follows.
The general quadratic action for a set of fermion fields 
in the presence of arbitrary Lorentz and CPT violation
is presented in Sec.\ \ref{Fermions}.
The specialization of this analysis 
to the neutrino sector,
which allows for multiple flavors of left-handed neutrinos,
is derived in Sec.\ \ref{Neutrinos}.
The spherical decomposition of the effective hamiltonian
for neutrino propagation 
is considered in Sec.\ \ref{Spherical decomposition}.
We discuss a variety of special cases 
in Sec.\ \ref{Special models},
including the renormalizable limit,
the massless scenario,
flavor-blind and single-flavor models,
the diagonalizable case,
and isotropic models.
The results are applied 
in Secs.\ \ref{Applications to oscillations}
and \ref{Applications to kinematics}
to extract numerous limits on coefficients for Lorentz violation
from existing data on neutrino oscillations and propagation.

\section{Fermions}
\label{Fermions}

The construction 
of a realistic low-energy effective theory for fermions 
that is coordinate independent
and describes general Lorentz violation 
can be achieved by adding appropriate terms 
to the conventional fermion Lagrange density.
Each additional term is formed by contracting
a coefficient for Lorentz violation
with a tensor operator,
and all possible terms are included
\cite{kp,ck,akgrav}.
In this work we focus on noninteracting fermions,
which corresponds to restricting Lorentz-violating terms
in the action to fermion bilinears.
The formalism presented in this section
holds for arbitrary fermions,
but our primary interest in subsequent sections 
lies in applications to neutrino physics
for which chiral components are physically relevant.

Consider the case of $N$ spinor fields $\ps_a$,
where $a$ ranges over $n$ spinor flavors 
$a = 1,2,\ldots, N$.
To allow for Majorana couplings in the construction,
it is convenient to combine the $N$ spinors $\ps_a$
together with their charge conjugates $\ps^C_a = C\psb^T_a$
into a $2N$-dimensional multiplet of spinors,
\beq
\Ps_A = 
\begin{pmatrix}
\ps_a \\
\ps^C_a
\end{pmatrix} ,
\label{Psi}
\eeq
where $A$ ranges over $2N$ values. 
The redundancy in $\Ps$
implies that it obeys the relationship
\beq
\Ps^C = \C \Ps ,
\quad
\C = \begin{pmatrix}
0 & 1 \\ 1 & 0
\end{pmatrix},
\label{cccond}
\eeq
where the $2N\times 2N$ matrix $\C$ is defined 
in terms of $N\times N$ blocks in flavor space.

In terms of the spinor $\Ps_A$,
we can write the general Lagrange density
incorporating Lorentz and CPT violation in the form
\bea
S &=& \int \cl d^4x ,
\nn\\
\cl &=& 
\half \Psb_A (
\ga^\mu i\prt_\mu \de_{AB}
- M_{AB} + \Qhat_{AB}
) \Ps_B 
+ {\rm h.c.}
\quad
\label{multilag}
\eea
The first part of this expression 
generates the usual kinetic term,
while the second part involves an arbitrary mass matrix $M_{AB}$. 
The third part contains
the Lorentz-violating operator $\Qhat_{AB}$,
which is a general $4\times 4$ matrix in spinor space
and a $2N\times 2N$ matrix in flavor space
that involves derivatives $i\prt_\mu$.

Since the effects from Lorentz violation
are generically expected to be small,
possibly arising as Planck-suppressed effects, 
it is reasonable to treat $\Qhat_{AB}$ 
as a perturbative contribution when necessary.
This approach is adopted in the present work.
In principle,
the unperturbed theory could be taken as unconventional
if desired.
For example,
some models of superluminal neutrinos
adopt a kinetic term involving 
the replacement $i \ga^\mu \prt_\mu \to i\ga_5 \ga^\mu \prt_\mu$
\cite{ftlnu1,ftlnu2,ftlnu3,ftlnu4,jr,re,mrad,dmpt,jw,kz}.
Models of this kind can be incorporated in
the present formalism
with a nonperturbative choice of $\Qhat_{AB}$.

The hermiticity of $\cl$
implies that the general form of $M_{AB}$ can be written 
\bea
M_{AB}&=&m_{AB}+im_{5AB}\ga_5 ,
\eea
where $m$ and $m_5$ are hermitian $2N\times 2N$ matrices.
The relationship \rf{cccond} implies the conditions 
\beq
m = \C m^T \C,
\quad
m_5 = \C m_5^T \C,
\eeq
where the transpose acts in flavor space.

The operator $\Qhat_{AB}$ can in general 
depend on spacetime position,
either in a prescribed way or through dynamical fields.
For instance,
explicit Lorentz violation occurs when $\Qhat_{AB}$ 
contains a fixed background with nontrival Lorentz properties,
while spontaneous Lorentz violation can arise
if $\Qhat_{AB}$ involves field variables 
with dynamics generating tensor vacuum values.
An analysis incorporating spacetime dependence
would be of interest but would be burdened 
by theoretical and experimental complexities 
beyond Lorentz violation,
so it is advantageous to focus on operators
that conserve energy and momentum. 
This can be assured
by requiring the invariance of the action $S$
under spacetime translations,
which is achieved when $\Qhat_{AB}$ is spacetime independent.
For spontaneous Lorentz breaking,
requiring invariance under spacetime translations
implies neglecting soliton solutions,
along with any massive or Nambu-Goldstone (NG) modes
\cite{ng}. 
The latter can be interpreted 
as the photon in Einstein-Maxwell theory 
\cite{akgrav,bumblebee},
as the graviton
\cite{cardinal},
or as a variety of other forces
\cite{newforces}.

We remark in passing that spacetime independence 
may be a natural feature of a model,
or it may be a useful approximation 
describing dominant contributions or averaged effects
from known or hypothesized forces
in the vicinity of the Earth.
Even Lorentz-invariant interactions
typically generate effective Lorentz violation in this way.
For example,
couplings to a tiny and previously unknown 
Lorentz-invariant inverse-square force
in the vicinity of the Earth
would generate effective Lorentz-violating behavior
described by the SME.
This idea is the basis for some of the sharpest sensitivities 
obtained on torsion to date
\cite{torsion}
and on some Lorentz-invariant effects from quantum gravity
\cite{bs}.
Models for new Lorentz-invariant gravitational interactions
are viable only if they are compatible 
with the numerous existing constraints for Lorentz violation
\cite{tables}.
Experimental bounds on spacetime-independent $\Qhat_{AB}$
obtained in this work
therefore also constrain Lorentz-invariant models 
involving neutrinos. 

A decomposition of $\Qhat_{AB}$
permits the Lorentz-violating operators in $\cl$
to be classified and enumerated.
The spin part of $\Qhat_{AB}$ can be characterized 
by expanding in the basis of 16 Dirac matrices $\ga_I$,
\bea
\Qhat_{AB} &=& 
\sum_I \Qhat_{AB}^I \ga_I
\nn\\
&&
\hskip -25pt
=\Shat_{AB}
+i\Phat_{AB} \ga_5
+\Vhat_{AB}^\mu \ga_\mu
+\Ahat_{AB}^\mu \ga_5\ga_\mu
+\half \That_{AB}^\mn \si_\mn ,
\nn\\
\label{multiqhatsplit}
\eea
where the $2N\times 2N$ derivative-dependent 
matrix operators $\Qhat_{AB}^I$
are hermitian in flavor space.
The derivative dependence 
can be revealed by expressing each $\Qhat_{AB}^I$
as a sum of operators of definite mass dimension $d$,
\beq
\Qhat_{AB}^I= \sum_{d=3}^\infty
\Q_{AB}^{(d)I\al_1\al_2\ldots\al_{d-3}}
p_{\al_1}p_{\al_2}\ldots p_{\al_{d-3}} ,
\label{multiQI_expansion}
\eeq
where $p_\mu = i\prt_\mu$.
Since each $\Qhat_{AB}^I$ has mass dimension one,
the coefficients
$\Q_{AB}^{(d)I\al_1\al_2\ldots\al_{d-3}}$
have mass dimension $4-d$.
Following the discussion above,
these coefficients can be taken as spacetime constants.

A useful refinement of the above decomposition
involves first splitting $\Qhat_{AB}$ as 
\beq
\ga^\nu p_\nu \de_{AB}
- M_{AB} + \Qhat_{AB}
= \widehat\Ga^\nu_{AB} p_\nu - \widehat M_{AB} ,
\label{qhat}
\eeq
in analogy to the usual split 
in the single-fermion limit of the minimal SME
\cite{ck}.
The combination $\widehat\Ga^\nu_{AB} p_\nu$
contains all operators of even mass dimension,
while $\widehat M_{AB}$ contains all those of odd mass dimension. 
Expanding these combinations using Dirac matrices gives
\bea
\widehat\Ga^\nu_{AB} &=& 
\ga^\nu \de_{AB}
+\chat^{\mn}_{AB} \ga_\mu
+\dhat^{\mn}_{AB} \ga_5\ga_\mu
\nn\\
&&
+\ehat^\nu_{AB}
+i\fhat^\nu_{AB} \ga_5
+\half \ghat^{\ka\la\nu}_{AB} \si_{\ka\la} ,
\nn\\
\widehat M_{AB} &=& 
m_{AB} 
+ i m_{5AB} \ga_5
+ \mhat_{AB}
+i\mfivehat_{AB} \ga_5
\nn\\
&&
+\ahat^\mu_{AB} \ga_\mu
+\bhat^\mu_{AB} \ga_5\ga_\mu
+\half \Hhat^\mn_{AB} \si_\mn .
\label{GaM}
\eea
The dimensionless operators
$\chat^{\mn}_{AB}$, $\dhat^{\mn}_{AB}$
are CPT even,
while the dimensionless operators 
$\ehat^\mu_{AB}$, $\fhat^\mu_{AB}$, $\ghat^{\mu\rh\nu}_{AB}$
are CPT odd.
The remaining operators have mass dimension one,
with $\mhat_{AB}$, $\mfivehat_{AB}$, $\Hhat^\mn_{AB}$ 
being CPT even and 
$\ahat^\mu_{AB}$, $\bhat^\mu_{AB}$
being CPT odd.
Note that all the operators in Eq.\ \rf{GaM}
have counterparts in the minimal SME
except for $\mhat_{AB}$ and $\mfivehat_{AB}$,
which contain only terms
of nonrenormalizable dimension.

Since $\widehat\Ga^\nu_{AB}$ appears contracted with $p_\nu$
in Eq.\ \rf{qhat},
the operators 
$\chat^{\mn}_{AB}$, $\dhat^{\mn}_{AB}$,
$\ehat^\mu_{AB}$, $\fhat^\mu_{AB}$, $\ghat^{\mu\rh\nu}_{AB}$
are also automatically contracted with $p_\nu$. 
It is therefore natural and convenient to define
the contracted operators
\bea
\chat^\mu_{AB} &=& \chat^{\mn}_{AB} p_\nu  , 
\quad
\dhat^\mu_{AB} = \dhat^{\mn}_{AB} p_\nu ,
\nn\\
\ehat_{AB} &=& \ehat^\nu_{AB} p_\nu ,
\quad
\fhat_{AB} = \fhat^\nu_{AB} p_\nu , 
\quad
\ghat^{\ka\la}_{AB} = \ghat^{\ka\la\nu}_{AB} p_\nu .
\qquad
\eea
The CPT properties of these contracted operators
matches those of their counterparts in the minimal SME.
Using this definition reveals the relationships 
\bea
\Shat_{AB} &=& \ehat_{AB} - \mhat_{AB} ,
\quad
\Phat_{AB} = \fhat_{AB} - \mfivehat_{AB},
\nn\\ 
\Vhat^\mu_{AB} &=& \chat^\mu_{AB} - \ahat^\mu_{AB} , 
\quad
\Ahat^\mu_{AB} = \dhat^\mu_{AB} - \bhat^\mu_{AB} , 
\nn\\ 
\That^\mn_{AB} &=& \ghat^\mn_{AB} - \Hhat^\mn_{AB} 
\eea
between the expansions \rf{multiqhatsplit} and \rf{qhat}.

We can also take advantage of the property \rf{cccond} 
to separate operators into Dirac and Majorana pieces.
For each operator of mass dimension $d$ in $\Qhat_{AB}$,
the property \rf{cccond} yields the constraint
\beq
\Qhat = (-1)^{d-3}\C C \Qhat^T C^{-1} \C ,
\eeq
where the transpose acts in both spinor and flavor spaces.
This implies the conditions 
\bea
\Shat &=& (-1)^{d+1}\C \Shat^T \C , 
\quad
\Phat = (-1)^{d+1}\C \Phat^T \C , 
\nn\\
\Vhat^\mu &=& (-1)^d\C (\Vhat^\mu)^T \C , 
\quad
\Ahat^\mu = (-1)^{d+1}\C (\Ahat^\mu)^T \C , 
\nn\\
\That^\mn &=& (-1)^d\C (\That^\mn)^T \C ,
\eea
where now the transpose acts only in flavor space.
Using these results,
we can write the component operators $\Qhat^I_{AB}$
in terms of four $N\times N$ block matrices
that can be designated 
as being of Dirac or Majorana type,
\bea
\Shat &=& \begin{pmatrix}
\Shat_D & \Shat_M \\
\Shat_M^\dag & (-1)^{d+1}\Shat_D^T
\end{pmatrix},
\nn\\
\Phat &=& \begin{pmatrix}
\Phat_D & \Phat_M \\
\Phat_M^\dag & (-1)^{d+1}\Phat_D^T
\end{pmatrix},
\nn\\
\Vhat^\mu &=& \begin{pmatrix}
\Vhat^\mu_D & \Vhat^\mu_M \\
(\Vhat^\mu_M)^\dag & (-1)^d(\Vhat^\mu_D)^T
\end{pmatrix},
\nn\\
\Ahat^\mu &=& \begin{pmatrix}
\Ahat^\mu_D & \Ahat^\mu_M \\
(\Ahat^\mu_M)^\dag & (-1)^{d+1}(\Ahat^\mu_D)^T
\end{pmatrix},
\nn\\
\That^\mn &=& \begin{pmatrix}
\That^\mn_D & \That^\mn_M \\
(\That^\mn_M)^\dag & (-1)^d(\That^\mn_D)^T
\end{pmatrix}.
\label{diracmajorana}
\eea
In these expressions,
all the Dirac-like matrices are hermitian in flavor space.
Depending on the mass dimension,
each Majorana matrix operator
is either symmetric or antisymmetric in flavor space, 
\bea
\Shat_M &=& (-1)^{d+1} \Shat_M^T ,
\quad
\Phat_M = (-1)^{d+1} \Phat_M^T,
\nn\\
\Vhat_M^\mu &=& (-1)^d (\Vhat_M^\mu)^T,
\quad
\Ahat_M^\mu = (-1)^{d+1} (\Ahat_M^\mu)^T,
\nn\\
\That_M^\mn &=& (-1)^d (\That_M^\mn)^T.
\label{majoranaconds}
\eea
Using the designations \rf{diracmajorana},
each component operator
in the expansions \rf{GaM}
can also be split into four $N\times N$ block matrices 
of Dirac or Majorana type
obeying the conditions \rf{majoranaconds}. 

Many physical features of fermions 
are most conveniently understood
in terms of a hamiltonian formulation 
rather than an approach based on the Lagrange density \rf{multilag}.
The presence of arbitrary Lorentz violation
and the concomitant higher-order time derivatives 
complicates the construction of the hamiltonian.
However, 
we can find an effective $2N\times 2N$ hamiltonian $H_{AB}$ 
that correctly describes the physics 
at leading order in Lorentz violation.
Starting with the modified Dirac equation
\beq
(p\cdot\ga \de_{AB} - M_{AB} + \Qhat_{AB}) \Ps_B = 0,
 \eeq
we can multiply on the left by $\ga_0$
and then define $H_{AB}$ by the condition
\beq
(\eom \de_{AB} - H_{AB}) \Ps_B 
= \ga_0 (p\cdot\ga \de_{AB} - M_{AB} + \Qhat_{AB}) \Ps_B
= 0 ,
\eeq
where $\eom = p_0$.
We can thereby identify 
\bea
H_{AB} &=& \ga_0 ( \mbf p \cdot \mbf \ga \de_{AB} + M_{AB} 
- \Qhat_{AB}) 
\nn\\
&=& (H_0)_{AB} + \de H_{AB},
\label{genham}
\eea
where $(H_0)_{AB}=\ga_0 ( \mbf p \cdot \mbf \ga \de_{AB} + M_{AB})$ 
is the usual hamiltonian
with conventional energy $\eom_0$
and $\de H_{AB} = - \ga_0 \Qhat_{AB}$
is the Lorentz-violating perturbation.
Note that the latter term
typically depends on $\eom$. 
However, 
the changes to the energy $\eom_0$ induced by $\de H_{AB}$
are perturbative by construction,
so at leading order $\de H_{AB}$ can be evaluated 
at the conventional energy $\eom_0$.
The leading-order effective hamiltonian
can therefore be written as
\bea
H_{AB} &=& (H_0)_{AB} 
\nn\\
&&
\hskip-25pt
- \ga_0\big(
\Shat_{AB}
+i\Phat_{AB} \ga_5
\nn\\
&&
+\Vhat_{AB}^\mu \ga_\mu
+\Ahat_{AB}^\mu \ga_5\ga_\mu
+\half \That_{AB}^\mn \si_\mn 
\big)\big\vert_{\eom \to \eom_0}. 
\qquad
\label{leadingham}
\eea

\section{Neutrinos}
\label{Neutrinos}

Our primary interest in this work
lies in the neutrino sector of the SME.
Since the observed neutrinos are chiral fermions,
describing their properties in the presence 
of Lorentz and CPT violation 
requires projecting the general formalism presented above
onto left-handed fields.
In what follows,
we retain all leading-order terms from Lorentz violation
arising from operators of arbitrary mass dimension,
including terms linear in neutrino mass.
This incorporates and extends our earlier analysis 
for operators of renormalizable dimension
\cite{km},
which treated as negligible all terms involving the product 
of a coefficient for Lorentz violation with a neutrino mass.

To proceed with the analysis,
it is convenient to introduce 
left- and right-handed mass matrices $m_L$ and $m_R$
satisfying $m_R=(m_L)^\dag=m+im_5$,
which combine to form $M$ according to 
\bea
M &=& m_L P_L + m_R P_R,
\eea
where
$P_L = (1-\ga_5)/2$ and $P_R = (1+\ga_5)/2$
are the usual chiral projection operators.
The components of the matrix $m_R=m_L^\dag$
can be identified with Dirac- or Majorana-type masses 
by separating $m_R$ into four $N \times N$ submatrices
according to
\beq
m_R \C =
\left(\begin{array}{cc}
L & D \\
D^T & R
\end{array}\right) .
\label{usumass}
\eeq
Here,
$R$ and $L$ are the
right- and left-handed Majorana-mass matrices,
while $D$ is the Dirac-mass matrix.
The complex matrices $R$, $L$, $D$ are 
restricted only by the requirement that $R$ and $L$ are symmetric.

In the absence of Lorentz violation,
the general equation describing massive left-handed fermions is
\beq
p\cdot\ga \ps_L - L \ps_L^C - D \ps_R = 0 ,
\eeq
and mixing between left- and right-handed neutrinos 
vanishes if $D=0$.
For nonzero $D$, 
a seesaw mechanism is usually invoked
to suppress left-right mixing
\cite{seesaw},
based on the assumption that $R$ is large.
Since the right-handed neutrinos obey
\beq
p\cdot\ga \ps_R^C - R \ps_R - D^T \ps_L^C = 0 ,
\eeq
a large $R$ implies $\ps_R \approx -R^{-1} D^T \ps_L^C$.
The behavior of left-handed neutrinos is therefore well approximated
by the equation 
\bea
p\cdot\ga \ps_L - m_l \ps_L^C = 0 ,
\label{leftdirac}
\eea
where the effective left-handed mass matrix $m_l$ is given by 
\bea
m_l = L - DR^{-1}D^T.
\eea
Note that $m_l$ is symmetric, 
$m_l = m_l^T$.

Since experiment shows that propagating neutrinos are left-handed
and that any right-handed components play a negligible role,
applying a left-handed projection
produces an excellent approximation
to the physical neutrino behavior.
In $N\times N$ block form,
the relevant projection of the hamiltonian $H$ is 
\bea
H_L &=& 
\begin{pmatrix} P_L & 0 \\ 0 & P_R \end{pmatrix}
H
\begin{pmatrix} P_L & 0 \\ 0 & P_R \end{pmatrix}.
\eea
As usual,
the projectors imply that 
this expression can be reduced to an operator
acting on two-dimensional Weyl spinors.
We introduce $\si^\mu = (\si^0, \si^j)$,
where $\si^0$ is the $2\times 2$ identity matrix
and $\si^j$ are the usual three Pauli matrices
with adjoint matrices 
$\bsi^\mu = (\si^0, -\si^j)$.
Denoting by $\ph$ the two-component Weyl spinor associated
with a four-component Dirac spinor $P_L \ps$,
the $2N$-dimensional multiplet $\Ps$ in Eq.\ \rf{Psi}
can be replaced with a $2N$-dimensional multiplet $\Ph_{\rm W}$
of the form
\beq
\Ph_{\rm W} = \begin{pmatrix}
\ph \\ \ph^C \end{pmatrix} ,
\eeq
where $\ph^C = i\si^2\ph^*$.
Flavor indices are suppressed in these expressions.  
Similarly,
the hamiltonian $H_L$ can be replaced with
its Weyl counterpart $H_{\rm W}$.

In the absence of Lorentz violation, 
Eq.\ \rf{leftdirac} becomes
\beq
p\cdot\bsi \ph - m_l \ph^C = 0 ,
\eeq
and the hamiltonian takes the form 
\beq
(H_{\rm W})_0 =
\begin{pmatrix}
-\mbf p\cdot\mbf\si & m_l \\
m_l^\dag & \mbf p\cdot\mbf\si
\end{pmatrix} .
\eeq
The Lorentz-violating piece $\de H$ in Eq.\ \rf{genham} becomes
\bea
\de H_{\rm W} &=&
\begin{pmatrix}
-\Vhat^\mu_L \bsi_\mu & 
-\Shat_L - \tfrac{i}{2}\That^\mn_M \bsi_\mu\si_\nu \\
\noalign{\medskip}
- \Shat_L^\dag - \tfrac{i}{2} (\That^\mn_M)^\dag \si_\mu\bsi_\nu&
(-1)^{(d+1)}\Vhat^\mu_L{}^T \si_\mu 
\end{pmatrix} ,
\nn\\
\label{dehw}
\eea
where
\bea
\Shat_L &=& \Shat_M + i \Phat_M,
\quad
\Vhat^\mu_L = \Vhat^\mu_D + \Ahat^\mu_D.
\eea
Note that the preservation of chirality
ensures that the orthogonal combinations
$\Vhat^\mu_R$
and $\Shat_R$
are absent from $H_{\rm W}$.

The full hamiltonian 
\beq
H_{\rm W}= (H_{\rm W})_0 +\de H_{\rm W}
\label{rellvham}
\eeq
can be block diagonalized 
within a suitable approximation.
We proceed here treating neutrinos as relativistic particles,
but performing a nonrelativistic diagonalization 
of $H_{\rm W}$ could also be of interest in certain contexts
beyond our present scope.
These could include,
for example,
experiments with neutrinos of ultra-low energy
such as measurements of the beta-decay endpoint,
or studies of the cosmic neutrino background.

For the relativistic case,
we can block diagonalize $H_{\rm W}$
to order $m_l^2$
using the transformation
\bea
U = \begin{pmatrix}
1 - \fr{m_l m_l^\dag}{8\mbf p^2} &
-\fr{m_l\mbf p\cdot\mbf\si}{2\mbf p^2} \\
\fr{m_l^\dag\mbf p\cdot\mbf\si}{2\mbf p^2} &
1 - \fr{m_l^\dag m_l}{8\mbf p^2}
\end{pmatrix} .
\eea
Consider first the Lorentz-invariant piece $(H_{\rm W})_0$.
This becomes
\bea
(H'_{\rm W})_0 &=& U (H_{\rm W})_0 U^\dag 
\nn\\
&=& 
\begin{pmatrix}
-\mbf p\cdot\mbf\si\big(1+\fr{m_l m_l^\dag}{2\mbf p^2}\big) & 0 \\
0 & \mbf p\cdot\mbf\si\big(1+\fr{m_l^\dag m_l}{2\mbf p^2}\big)
\end{pmatrix} ,
\qquad
\eea
which acts on the transformed Weyl doublet
\beq
\Ph'_{\rm W} = U \Ph_{\rm W} 
= \begin{pmatrix}
\ph' \\ (\ph')^C
\end{pmatrix} .
\eeq
To find the effective hamiltonian
governing the propagation of neutrinos,
we expand $\Ph'_W$ in helicity components as
\bea
\ph' &=& [A(t,\mbf p) e^{i\mbf x\cdot\mbf p} 
+ B^*(t,\mbf p) e^{-i\mbf x\cdot\mbf p}]
\xi_{\mbf p},
\nn\\
(\ph')^C &=& 
[A^*(t,\mbf p) e^{-i\mbf x\cdot\mbf p}
+ B(t,\mbf p) e^{i\mbf x\cdot\mbf p}]
\xi_{\mbf p}^C ,
\eea
where $\xi_{\mbf p}$ is a
normalized negative-helicity spinor satisfying 
$\mbf p\cdot\mbf\si \xi_{\mbf p}
= -|\mbf p| \xi_{\mbf p}$.
The amplitude $A(t,\mbf p)$
is associated with negative-helicity neutrinos,
while $B(t,\mbf p)$ 
is associated with positive-helicity antineutrinos.
In the Lorentz-invariant limit,
restricting to the positive-energy part of the
Schr\"odinger equation gives 
\beq
i\fr{\prt}{\prt t}
\begin{pmatrix}A\\B\end{pmatrix}
= (\heff)_0
\begin{pmatrix}A\\B\end{pmatrix},
\eeq
where
\beq
(\heff)_0
= 
|\mbf p| 
\begin{pmatrix}
1 & 0 \\ 
0 & 1
\end{pmatrix}
+ \fr 1 {2|\mbf p|}
\begin{pmatrix}
m_l m_l^\dag & 0 \\
0 & m_l^\dag m_l
\end{pmatrix}
\label{effliham}
\eeq
is the usual effective hamiltonian
for neutrino and antineutrino amplitudes
in the Lorentz-invariant limit.

To obtain the Lorentz-violating contribution $\deh$
to the effective hamiltonian,
we first diagonalize $\de H_{\rm W}$ as
\bea
\de H'_{\rm W} &=& U ~ \de H_{\rm W} ~ U^\dag
\nn\\
&=& \de H_{\rm W}  
+ [ \de U, \de H_{\rm W} ] + O(m_l^2) ,
\eea
where
\beq
\de U = 
\fr {\mbf p\cdot\mbf\si}{2\mbf p^2}
\begin{pmatrix}
0 & -m_l \\
m_l^\dag & 0 
\end{pmatrix} .
\eeq
We then obtain $\deh$ by 
projecting onto the positive-energy piece, 
\bea
\deh &=&
\begin{pmatrix}
\xi_{\mbf p}^\dag & 0 \\
0 & \xi_{\mbf p}^{C\dag}
\end{pmatrix}
\de H'_{\rm W}
\begin{pmatrix}
\xi_{\mbf p} & 0 \\
0 & \xi_{\mbf p}^C
\end{pmatrix}.
\eea
An explicit result can be obtained using the identities
\bea
\xi_{\mbf p}^\dag \bsi_\mu \xi_{\mbf p} 
&=&  \xi_{\mbf p}^{C\dag} \si_\mu \xi^C_{\mbf p} 
\approx \fr{p_\mu}{|\mbf p|} ,
\nn\\
\xi_{\mbf p}^\dag \si_\mu \xi_{\mbf p}^C 
&=& -\xi_{\mbf p}^\dag \bsi_\mu \xi_{\mbf p}^C
= \sqrt2 \ep_\mu ,
\nn\\
\xi_{\mbf p}^{C\dag} \si_\mu \xi_{\mbf p}
&=& -\xi_{\mbf p}^{C\dag} \bsi_\mu \xi_{\mbf p}
= \sqrt2 \ep^*_\mu ,
\eea
where the polarization vector $\ep^\mu$ can be taken as 
\beq
\ep^\mu = 
\frac{1}{\sqrt2}(0; \hat{\mbf e}_1 + i\hat{\mbf e}_2 ) ,
\quad
\ep_\mu(-\mbf p) = \ep_\mu^*(\mbf p) .
\eeq
Here,
$\hat{\mbf e}_1$ and $\hat{\mbf e}_2$
are arbitrary unit vectors chosen so that
$\{\phat,\hat{\mbf e}_1,\hat{\mbf e}_2\}$
form a right-handed orthonormal triad.
Adopting the correspondence
$\phat = \hat {\mbf r}$,
$\hat{\mbf e}_1 = \hat{\mbf\th}$,
$\hat{\mbf e}_2 = \hat{\mbf\ph}$
to the usual spherical-coordinate unit vectors
implies the spatial part of $\ep^\mu$ 
is the helicity unit vector $\hat {\mbf \ep}_+$
introduced in Appendix A 2 of Ref.\ \cite{kmnonmin}.

Some calculation along the above lines reveals that 
to order $O(m_l)$ 
the Lorentz-violating piece $\deh$
of the effective hamiltonian takes the form 
\beq
\deh = \fr{1}{|\mbf p|}
\begin{pmatrix}
\ae - \ce & -\ge + \He \\
-\ge^\dag +\He^\dag & -\ae^T -\ce^T
\end{pmatrix} ,
\label{efflvham}
\eeq
where conjugation and transposition are flavor-space operations.
For convenience and clarity,
this expression splits each $N\times N$ hamiltonian block 
into CPT-odd and CPT-even parts,
where the notation reflects the CPT properties
of the corresponding operators in the minimal SME.
The CPT-odd parts take the form 
\bea
\ae &=& 
p_\mu\ahat^\mu_L - \ehat_l + 2i\ep_\mu\ep_\nu^* \ghat_l^\mn ,
\nn\\
\ge &=& 
i\sqrt2\, p_\mu \ep_\nu \ghat^\mn_{M+} 
+ \sqrt2\, \ep_\mu \ahat_l^\mu ,
\label{cptoddcombos}
\eea
while the CPT-even terms are 
\bea
\ce &=& 
p_\mu \chat^\mu_L - \mhat_l 
+ 2i\ep_\mu\ep_\nu^* \Hhat_l^\mn ,
\nn\\
\He &=& i\sqrt2\, p_\mu \ep_\nu \Hhat^\mn_{M+} 
+ \sqrt2\, \ep_\mu \chat_l^\mu .
\label{cptevencombos}
\eea
In these expressions,
each quantity $\That_{M+}^\mn$ 
is defined as the combination
$\That_{M+}^\mn = \half(\That^\mn + i \Tdual^\mn )$,
and it obeys the identity
$\ep_\mu \ep_\nu^* \That_{M+}^\mn \approx
- p^j \That_{M+}^{0j}/|\mbf p|$.  
Here and below,
a tilde denotes the usual dual 
with $\ep^{\mu\nu\al\be}/2$.
The operators independent of the mass matrix $m_l$
are defined as 
\bea
\ahat_L^\mu &=&
\ahat_D^\mu +\bhat_D^\mu  ,
\quad
\ghat^\mn_{M+} = 
\half\big(\ghat^\mn_M +i \gdual^\mn_M\big) ,
\nn\\
\chat_L^\mu &=& 
\chat_D^\mu +\dhat_D^\mu ,
\quad
\Hhat^\mn_{M+} = 
\half\big(\Hhat^\mn_M +i \Hdual^\mn_M\big),
\label{mindepops}
\eea
and they obey the hermiticity and symmetry conditions
\bea
\ahat_L^\mu &=& (\ahat_L^\mu)^\dag , 
\quad
\ghat^\mn_{M+} = 
i\gdual^\mn_{M+} = \big(\ghat^\mn_{M+}\big)^T ,
\nn\\
\chat_L^\mu &=& \big(\chat_L^\mu\big)^\dag , 
\quad
\Hhat^\mn_{M+} = 
i\Hdual^\mn_{M+} = -\big(\Hhat^\mn_{M+}\big)^T .
\eea
The operators linear in $m_l$ are given by  
\bea
\mhat_l &=& 
\half \big(\mhat_M +i \mhat_{5M}\big)m_l^\dag
+\half m_l\big(\mhat_M +i \mhat_{5M}\big)^\dag ,
\nn\\
\ahat_l^\mu &=& 
\half \ahat_L^\mu m_l + \half m_l \big(\ahat_L^\mu\big)^T,
\nn\\
\chat_l^\mu &=& 
\half \chat_L^\mu m_l - \half m_l \big(\chat_L^\mu\big)^T,
\nn\\
\ehat_l &=&  
\half \big(\ehat_M +i \fhat_M\big)m_l^\dag
+\half m_l\big(\ehat_M +i \fhat_M\big)^\dag,
\nn\\
\ghat_l^\mn &=& 
\half \ghat_{M+}^\mn m_l^\dag 
+\half  m_l \big(\ghat_{M+}^\mn\big)^\dag ,
\nn\\
\Hhat_l^\mn &=& 
\half \Hhat_{M+}^\mn m_l^\dag 
+ \half m_l \big(\Hhat_{M+}^\mn\big)^\dag ,
\label{linmops}
\eea
and they satisfy
\bea
\mhat_l &=& \mhat_l^\dag ,
\quad
\ahat_l^\mu = \big(\ahat_l^\mu\big)^T ,
\quad
\chat_l^\mu = -\big(\chat_l^\mu\big)^T ,
\nn\\
\ehat_l &=& \ehat_l^\dag ,
\quad
\ghat_l^\mn = \big(\ghat_l^\mn\big)^\dag ,
\quad
\Hhat_l^\mn = \big(\Hhat_l^\mn\big)^\dag .
\eea
Generically,
all the above operators depend on the 4-momentum. 

The net effective hamiltonian $\heff$ 
is the $2N \times 2N$ matrix
given as the sum of Eqs.\ \rf{effliham} and \rf{efflvham},
\beq
\heff = (\heff)_0 + \deh.
\label{hresult}
\eeq 
Note that neutrinos and antineutrinos have identical mass spectra
despite the presence of CPT violation
\cite{owg}.
Note also
that the mass-induced operators \rf{linmops}
are in principle all determined 
by the mass-independent operators \rf{mindepops}
once the mass matrix $m_l$ is known.
However,
inspection of Eqs.\ \rf{cptoddcombos} and \rf{cptevencombos}
reveals that the two kinds of operators enter
$\heff$ through different projections
with $p_\mu$ and $\ep_\mu$,
and they therefore represent independent observable effects.

The terms in $\deh$ can be classified
according to their operator dimension $d$
and their properties under discrete transformations.
Neutrinos maximally break C and P symmetry
because these transformations reverse chirality,
so we consider here 
only the chirality-preserving operators CP, T, and CPT.
They transform a Weyl spinor $\ph$ according to
\bea
{\rm CP}: && \ph\to
\ph^{CP}(t,\mbf x) = 
\et_{CP}^{} \si^2 \ph^*(t,-\mbf x) ,
\nn\\
{\rm T}: && \ph\to
\ph^T(t,\mbf x) = 
\et_T^{} \si^2 \ph^*(-t,\mbf x),
\nn\\
{\rm CPT}: && \ph\to
\ph^{CPT}(t,\mbf x) = 
\et_{CPT}^{} \ph(-t,-\mbf x) ,
\eea
where the phases $\et_{CP}^{}$, $\et_T^{}$ are arbitrary
but combine to give 
$\et_{CPT}^{} = -\et_{CP}^{}\et_T^*$.
For definiteness,
we choose $\et_{CP} = 1, \et_T =\et_{CPT} = i$.
Table \ref{cpttable} summarizes the behavior 
of the coefficients for Lorentz violation 
under CP, T, and CPT.
Both CP and T act to complex-conjugate each coefficient
and multiply it by a factor of $(-1)^n$,
where $n$ is the number of spatial indices
on the coefficient. 
The first six coefficients listed in the table
enter $\deh$ in the on-diagonal blocks,
which control 
$\nu\leftrightarrow\nu$ and $\nub\leftrightarrow\nub$ mixing.
The remaining four appear in the off-diagonal blocks,
which are associated with $\nu\leftrightarrow\nub$ mixing.
Notice that each class of coefficients 
has a unique set of properties.

\begin{table}
\begin{tabular}{@{\quad}c@{\quad}|@{\quad}c@{\quad}c@{\quad}c@{\quad}c}															
coefficient	&	$d$	&		CP		&		T		&		CPT		\\
\hline															
\hline															
$a_L^{\mu\al_1\ldots\al_{d-3}}$	&	odd	&	$	-	$	&	$	+	$	&	$	-	$	\\
$c_L^{\mu\al_1\ldots\al_{d-3}}$	&	even	&	$	+	$	&	$	+	$	&	$	+	$	\\
$m_l^{\al_1\ldots\al_{d-3}}$ 	&	odd	&	$	+	$	&	$	+	$	&	$	+	$	\\
$e_l^{\al_1\ldots\al_{d-3}}$	&	even	&	$	-	$	&	$	+	$	&	$	-	$	\\
$g_l^{\mn\al_1\ldots\al_{d-3}}$	&	even	&	$	+	$	&	$	-	$	&	$	-	$	\\
$H_l^{\mn\al_1\ldots\al_{d-3}}$	&	odd	&	$	-	$	&	$	-	$	&	$	+	$	\\
\hline															
$g_{M+}^{\mn\al_1\ldots\al_{d-3}}$	&	even	&	$	+	$	&	$	-	$	&	$	-	$	\\
$H_{M+}^{\mn\al_1\ldots\al_{d-3}}$	&	odd	&	$	-	$	&	$	-	$	&	$	+	$	\\
$a_l^{\mu\al_1\ldots\al_{d-3}}$	&	odd	&	$	-	$	&	$	+	$	&	$	-	$	\\
$c_l^{\mu\al_1\ldots\al_{d-3}}$	&	even	&	$	+	$	&	$	+	$	&	$	+	$	\\
\hline															
\end{tabular}															
\caption{\label{cpttable}
Properties of neutrino coefficients under discrete transformations.
For CP and T,
each coefficient must also be complex conjugated
and multiplied by an additional factor of $(-1)^n$,
where $n$ is the number of spatial indices.}
\end{table}

\section{Spherical decomposition}
\label{Spherical decomposition}

\begin{table*}
\renewcommand{\arraystretch}{1.5}
\begin{tabular}{c||c|c|c||c|c|c}
coefficient & $d$ & $j$ & number & CP & T & CPT\\
\hline
$\aLcoef{d}{jm}{ab}$ & odd, $\geq 3$  & $d-2\geq j\geq 0$ & $9(d-1)^2$ &
$(-1)^{j+1}\aLcoef{d}{jm}{ba}$ & $(-1)^j\aLcoef{d}{jm}{ba}$ & $-\aLcoef{d}{jm}{ab}$ \\
$\cLcoef{d}{jm}{ab}$ & even, $\geq 4$ & $d-2\geq j\geq 0$ & $9(d-1)^2$ &
$(-1)^j\cLcoef{d}{jm}{ba}$ & $(-1)^j\cLcoef{d}{jm}{ba}$ & $\cLcoef{d}{jm}{ab}$ \\
$\mlcoef{d}{jm}{ab}$ & odd, $\geq 5$  & $d-3\geq j\geq 0$ & $9(d-2)^2$ &
$(-1)^j\mlcoef{d}{jm}{ba}$ & $(-1)^j\mlcoef{d}{jm}{ba}$ & $\mlcoef{d}{jm}{ab}$ \\
$\elcoef{d}{jm}{ab}$ & even, $\geq 4$ & $d-3\geq j\geq 0$ & $9(d-2)^2$ &
$(-1)^{j+1}\elcoef{d}{jm}{ba}$ & $(-1)^j\elcoef{d}{jm}{ba}$ & $-\elcoef{d}{jm}{ab}$ \\
$\glcoef{d}{jm}{ab}$ & even, $\geq 4$ & $d-2\geq j\geq 0$ & $9(d-1)^2$ &
$(-1)^{j+1}\glcoef{d}{jm}{ba}$ & $(-1)^j\glcoef{d}{jm}{ba}$ & $-\glcoef{d}{jm}{ab}$ \\
$\Hlcoef{d}{jm}{ab}$ & 
$\begin{cases} d=3 & \\ \mbox{odd, } \geq 5 \end{cases}$ & 
$\begin{array}{c} j=1\\[-2pt] d-2\geq j\geq 0\end{array}$ & 
$\begin{array}{c} 27\\[-2pt] 9(d-1)^2\end{array}$ &
$(-1)^j\Hlcoef{d}{jm}{ba}$ & $(-1)^j\Hlcoef{d}{jm}{ba}$ & $\Hlcoef{d}{jm}{ab}$ \\
[2pt]
\hline
&&&&&&\\[-14pt]
$\gMcoef{d}{jm}{ab}$ & even, $\geq 4$ & $d-2\geq j\geq 1$ & $12d(d-2)$ &
$(-1)^{j+m+1} \Big(\gMcoef{d}{j(-m)}{ab}\Big)^*$ &
$(-1)^{j+m} \Big(\gMcoef{d}{j(-m)}{ab}\Big)^*$ &
$-\gMcoef{d}{jm}{ab}$ \\
$\HMcoef{d}{jm}{ab}$ & odd, $\geq 3$  & $d-2\geq j\geq 1$ & $6d(d-2)$ &
$(-1)^{j+m} \Big(\HMcoef{d}{j(-m)}{ab}\Big)^*$ &
$(-1)^{j+m} \Big(\HMcoef{d}{j(-m)}{ab}\Big)^*$ &
$\HMcoef{d}{jm}{ab}$ \\
$\alcoef{d}{jm}{ab}$ & odd, $\geq 3$  & $d-2\geq j\geq 1$ & $12d(d-2)$ &
$(-1)^{j+m+1} \Big(\alcoef{d}{j(-m)}{ab}\Big)^*$ &
$(-1)^{j+m} \Big(\alcoef{d}{j(-m)}{ab}\Big)^*$ &
$-\alcoef{d}{jm}{ab}$ \\
$\clcoef{d}{jm}{ab}$ & even, $\geq 4$ & $d-2\geq j\geq 1$ & $6d(d-2)$ &
$(-1)^{j+m} \Big(\clcoef{d}{j(-m)}{ab}\Big)^*$ &
$(-1)^{j+m} \Big(\clcoef{d}{j(-m)}{ab}\Big)^*$ &
$\clcoef{d}{jm}{ab}$ \\
\hline\hline
$\aeff{d}{jm}{ab}$ & odd, $\geq 3$  & $d-1\geq j\geq 0$ & $9d^2$ &
$(-1)^{j+1}\aeff{d}{jm}{ba}$ & $(-1)^j\aeff{d}{jm}{ba}$ & $-\aeff{d}{jm}{ab}$ \\
$\ceff{d}{jm}{ab}$ & 
$\begin{cases} d=2 & \\ \mbox{even, } \geq 4 & \end{cases}$ & 
$\begin{array}{c} j=1\\[-2pt] d-1\geq j\geq 0\end{array}$ & 
$\begin{array}{c} 27\\[-2pt] 9d^2\end{array}$ &
$(-1)^j\ceff{d}{jm}{ba}$ & $(-1)^j\ceff{d}{jm}{ba}$ & $\ceff{d}{jm}{ab}$ \\
$\geff{d}{jm}{ab}$ & even, $\geq 2$  & $d-1\geq j\geq 1$ & $12(d^2-1)$ &
$(-1)^{j+m+1} \Big(\geff{d}{j(-m)}{ab}\Big)^*$ &
$(-1)^{j+m} \Big(\geff{d}{j(-m)}{ab}\Big)^*$ &
$-\geff{d}{jm}{ab}$ \\
$\Heff{d}{jm}{ab}$ & odd, $\geq 3$  & $d-1\geq j\geq 1$ & $6(d^2-1)$ &
$(-1)^{j+m} \Big(\Heff{d}{j(-m)}{ab}\Big)^*$ &
$(-1)^{j+m} \Big(\Heff{d}{j(-m)}{ab}\Big)^*$ &
$\Heff{d}{jm}{ab}$ 
\end{tabular}
\caption{\label{sphercoeff}
Spherical coefficients and their properties
under discrete transformations. }
\end{table*}

Many experimental tests of Lorentz invariance
rely on searching for anisotropies
associated with violations of rotation symmetry.
Searches of this type require knowledge 
of the transformation properties
of the coefficients for Lorentz violation under rotations.
In principle,
any given rotation can be performed 
on the cartesian coefficients for Lorentz violation
discussed in the previous subsection.
However, 
in practice this may require significant calculation,
while the results can be cumbersome
and can disguise basic aspects of rotation symmetry.

An alternative approach involves 
decomposing the coefficients for Lorentz violation
in spherical harmonics.
This emphasizes the importance of rotations,
and it ensures comparatively simple properties 
under rotation transformations.
It is useful both in searches for violations of isotropy
and in theoretical treatments of certain models,
such as those exhibiting isotropy in a preferred frame.
A decomposition of this type has already been used to
classify and enumerate photon-sector operators
of arbitrary mass dimension
\cite{kmnonmin}.

To perform this decomposition for neutrinos,
we expand in spherical harmonics
the $p_\mu$-dependent combinations
appearing in the Lorentz-violating piece \rf{efflvham}
of the hamiltonian 
$\deh$.
The terms appearing in the diagonal blocks of $\deh$
are rotational scalars, 
so they can be expanded 
using the standard spherical harmonics 
$Y_{jm} \equiv \syjm{0}{jm}$.
For example, 
the contribution involving the coefficients $\chat_L^\mu$ 
can be written as
\beq
p_\mu (\chat_L^\mu)_{ab} = 
\sum_{djmn} E^{d-2-n} |\mbf p|^n
\, Y_{jm}(\phat) \,
\cLcoef{d}{njm}{ab} ,
\eeq
where the indices $a,b$ range over neutrino flavors as before.
Note that $E$ can be approximated 
as $E\approx |\mbf p|$
to second order in the small mass $m_l$.
Since we are interested in Lorentz-violating effects 
to first order in $m_l$,
the second-order terms can be discarded 
and $E \approx |\mbf p|$ can be assumed 
for the expansion of the Lorentz-violating operators.
This approximation is analogous to that used
to obtain the vacuum coefficients in the photon sector
\cite{kmnonmin}.

Using this approximation,
the relevant spherical decompositions 
for the six types of coefficients 
appearing in the diagonal blocks of $\deh$
can be written as 
\bea
p_\mu (\ahat_L^\mu)_{ab} &=& \sum_{djm} 
|\mbf p|^{d-2}\, Y_{jm}(\phat)\, \aLcoef{d}{jm}{ab} ,
\nn\\
p_\mu (\chat_L^\mu)_{ab} &=& \sum_{djm} 
|\mbf p|^{d-2}\, Y_{jm}(\phat)\, \cLcoef{d}{jm}{ab} , 
\nn\\
(\mhat_l)_{ab} &=& \sum_{djm} 
|\mbf p|^{d-3}\, Y_{jm}(\phat)\, \mlcoef{d}{jm}{ab} ,
\nn\\
(\ehat_l)_{ab} &=& \sum_{djm} 
|\mbf p|^{d-3}\, Y_{jm}(\phat)\,  \elcoef{d}{jm}{ab} ,
\nn\\
2i\ep_\mu\ep_\nu^* (\ghat_l^\mn)_{ab} &=& \sum_{djm} 
|\mbf p|^{d-3}\, Y_{jm}(\phat)\,  \glcoef{d}{jm}{ab} ,
\nn\\
2i\ep_\mu\ep_\nu^* (\Hhat_l^\mn)_{ab} &=& \sum_{djm} 
|\mbf p|^{d-3}\, Y_{jm}(\phat)\,  \Hlcoef{d}{jm}{ab} .
\label{fundcoeff1}
\eea
The coefficients 
$\aLcoef{d}{jm}{ab}$ and $\cLcoef{d}{jm}{ab}$ 
have mass dimension $4-d$, 
while the derived coefficients
$\mlcoef{d}{jm}{ab}$,
$\elcoef{d}{jm}{ab}$,
$\glcoef{d}{jm}{ab}$,
and 
$\Hlcoef{d}{jm}{ab}$
have mass dimension $5-d$.
All these coefficients are hermitian in flavor space,
which implies they obey a relation of the form
\beq
\big({\mathcal K}_{jm}^{ab}\big)^* = 
(-1)^m {\mathcal K}_{j(-m)}^{ba} .  
\eeq

The off-diagonal blocks of $\deh$
induce mixing between neutrino and antineutrino states 
with opposite helicities,
with the operators in the upper right block
having helicity $-1$.
Functions with integral helicity can be expanded in
spin-weighted spherical harmonics $\syjm{s}{jm}$,
where the spin weight $s$ is the negative of the helicity. 
The definitions and properties of the spin-weighted functions
used here can be found in Appendix A of Ref.\ \cite{kmnonmin},
and the expansion procedure parallels that 
adopted for the photon sector.
The result of the decomposition can be written as
\bea
i \sqrt2 p_\mu\ep_\nu (\ghat_{M+}^\mn)_{ab} &=& \sum_{djm} 
|\mbf p|^{d-2}\, \syjm{+1}{jm}(\phat)\, \gMcoef{d}{jm}{ab} ,
\nn\\
i \sqrt2 p_\mu\ep_\nu (\Hhat_{M+}^\mn)_{ab} &=& \sum_{djm} 
|\mbf p|^{d-2}\, \syjm{+1}{jm}(\phat)\, \HMcoef{d}{jm}{ab} ,
\nn\\
\sqrt2 \ep_\mu (\ahat_l^\mu)_{ab} &=& \sum_{djm} 
|\mbf p|^{d-3}\, \syjm{+1}{jm}(\phat)\, \alcoef{d}{jm}{ab} ,
\nn\\
\sqrt2 \ep_\mu (\chat_l^\mu)_{ab} &=& \sum_{djm} 
|\mbf p|^{d-3}\, \syjm{+1}{jm}(\phat)\, \clcoef{d}{jm}{ab} .
\qquad
\label{fundcoeff2}
\eea
The CPT-even coefficients 
$\HMcoef{d}{jm}{ab}$ and $\clcoef{d}{jm}{ab}$
are antisymmetric in flavor space,
while the CPT-odd coefficients 
$\gMcoef{d}{jm}{ab}$ and $\alcoef{d}{jm}{ab}$
are symmetric.
The coefficients 
$\gMcoef{d}{jm}{ab}$ and $\HMcoef{d}{jm}{ab}$
have mass dimension $4-d$,
while the mass-induced coefficients 
$\alcoef{d}{jm}{ab}$ and $\clcoef{d}{jm}{ab}$ 
have dimension $5-d$.

Since the coefficients \rf{fundcoeff1} and \rf{fundcoeff2}
contribute to $\deh$
only through the combinations $\ae$, $\ce$, $\ge$, and $\He$, 
experiments are sensitive only to the latter.
This suggests that for practical applications
it is useful to consider instead the expansions
\bea
\ae^{ab} &=& \sum_{djm}  
|\mbf p|^{d-2}\, Y_{jm}(\phat) \aeff{d}{jm}{ab} ,
\nn\\
\ce^{ab} &=& \sum_{djm}  
|\mbf p|^{d-2}\, Y_{jm}(\phat) \ceff{d}{jm}{ab} ,
\nn\\
\ge^{ab} &=& \sum_{djm}  
|\mbf p|^{d-2}\, \syjm{+1}{jm}(\phat) \geff{d}{jm}{ab} ,
\nn\\
\He^{ab} &=& \sum_{djm}  
|\mbf p|^{d-2}\, \syjm{+1}{jm}(\phat) \Heff{d}{jm}{ab} .
\label{effcoeff}
\eea
The four sets of effective spherical coefficients 
appearing in these expansions
are related to the ten sets of fundamental ones 
\rf{fundcoeff1} and \rf{fundcoeff2}
by 
\bea
\aeff{d}{jm}{ab} &=& 
\aLcoef{d}{jm}{ab} -\elcoef{d+1}{jm}{ab} +\glcoef{d+1}{jm}{ab} ,
\nn\\
\ceff{d}{jm}{ab} &=& 
\cLcoef{d}{jm}{ab} -\mlcoef{d+1}{jm}{ab} +\Hlcoef{d+1}{jm}{ab} ,
\nn\\
\geff{d}{jm}{ab} &=& 
\gMcoef{d}{jm}{ab} + \alcoef{d+1}{jm}{ab} ,
\nn\\
\Heff{d}{jm}{ab} &=& 
\HMcoef{d}{jm}{ab} + \clcoef{d+1}{jm}{ab} .
\label{sphereffcoeff}
\eea

The reader is cautioned that 
the $d$ superscript on the effective coefficients 
for Lorentz violation
may differ from the dimension of the underlying operator.
Indeed,
the effective coefficients
are typically the sum of a fundamental coefficient labeled by $d$
and a mass-induced coefficient labeled by $d+1$.
Since the latter set is null in the case of massless neutrinos,
most of the mass effects at first order
can be absorbed into the zeroth-order terms.
The exceptions are the coefficients 
having the maximal $j$ value of $j = d-1$,
for which the effects arise purely from the mass interactions.
For example, 
the coefficients 
$\ceff{2}{1m}{ab}$ and $\geff{2}{1m}{ab}$
with superscripts $d=2$ 
arise entirely as mass-induced violations
from operators of mass dimension 3.

Table \ref{sphercoeff}
compiles some properties of the fundamental and effective
spherical coefficients for Lorentz violation.
The first six rows concern 
the operators \rf{fundcoeff1} 
on the diagonal blocks
of $\deh$,
while the next four concern 
the operators \rf{fundcoeff2}
on the off-diagonal blocks.
The final four rows present information about
the effective spherical coefficients \rf{effcoeff}.
For each coefficient,
the table provides the ranges of $d$ and $j$,
the number of independent real components
for the case of $N=3$ neutrino flavors,
and the CP, T, and CPT transformation properties.

\section{Special models}
\label{Special models}

Specific searches may have enhanced ability
to detect certain types of spherical coefficients.
Moreover,
the general analysis involves many independent components,
which may make some studies challenging to perform.
Special models representing limiting cases
may therefore be useful for certain experiments
and for theoretical purposes such as modeling signals
potentially associated with Lorentz and CPT violation.

This subsection considers some special limiting cases
of the general treatment above.
Five classes of limiting models are discussed.
We begin with renormalizable models,
in which all operators have mass dimension four or less. 
The second category is massless models,
in which all deviations of neutrino oscillation and propagation
from the usual lightlike behavior 
can be attributed solely to Lorentz violation.
Another class is the flavor-blind and oscillation-free models,
in which either mixing is absent 
or only single-flavor neutrino-antineutrino mixing occurs.
The fourth is diagonalizable models,
where a constant mixing matrix can simultaneously diagonalize
the mass matrix and all Lorentz- and CPT-violation contributions
to the effective hamiltonian.
Finally,
we consider several kinds of isotropic models,
for which a preferred frame
preserving rotation invariance exists.
For definiteness,
we take $N=3$ neutrino flavors throughout.

\subsection{Renormalizable models}
\label{Renormalizable models}

\begin{table}
\begin{tabular}{c|c|c|c}
spherical & cartesian & $j$ & number \\
\hline
$\aeff{3}{jm}{}$ & $a_L^{(3)\mu}$, $e_l^{(4)\mu}$, $g_l^{(4)\mu\nu\rh}$ & $0,1,2$ & $81$ \\
$\ceff{2}{jm}{}$ & $H_l^{(3)\mu\nu}$ & $1$ & $27$ \\
$\ceff{4}{jm}{}$ & $c_L^{(4)\mu\nu}$ & $0,1,2$ & $81$ \\
$\geff{2}{jm}{}$ & $a_l^{(3)\mu}$ & $1$ & $36$  \\
$\geff{4}{jm}{}$ & $g_{M+}^{(4)\mu\nu\rh}$ & $1,2$ & $96$ \\
$\Heff{3}{jm}{}$ & $H_{M+}^{(3)\mu\nu}$, $c_l^{(4)\mu\nu}$ & $1,2$ & $48$ \\
\end{tabular}
\caption{\label{rencoeff}
Spherical coefficients for renormalizable models.}
\end{table}

The limit of the SME
in which the only nonzero Lorentz-violating operators
are of mass dimension $d\leq 4$ 
is renormalizable to at least one loop
\cite{renorm}.
Theoretical aspects of renormalizable SME-based models 
for neutrinos have been extensively studied 
\cite{ck,akgrav,km,km-sb,dkm,bicycle,bmw,blmw,%
cg99,bpww,bbm,dg,gl,dgms,dgpg,gghm,fk,%
gktw,dges,jk,rnknkm,xm,bgpg,ba,ym,ag,as,bcgw,ltz,cmh,%
mp,bef,gl2,hmw,bb,akm,bapion,qm,scsk,hmp}
and several experimental collaborations
have measured numerous renormalizable coefficients 
for Lorentz violation
\cite{lsndlv,sklv,minoslv1,minoslv2,IceCube,MiniBooNElv,tables}.
Here,
adopting a fixed inertial frame
with cartesian coordinates $(t,x,y,z)$,
we establish the linear combination of cartesian coefficients
that corresponds to the spherical ones.
Throughout this subsection,
flavor indices are suppressed for simplicity.
 
To date,
discussions in the literature 
have been restricted to mass-independent operators
of renormalizable dimension.
Inspection of Eqs.\ \rf{fundcoeff1} and \rf{fundcoeff2}
reveals that the cartesian coefficients 
controlling these operators are
$a_L^{(3)\mu}$,
$c_L^{(4)\mn}$,
$g_{M+}^{(4)\mu\nu\rh}$,
and $H_{M+}^{(3)\mu\nu}$.
They enter the Lorentz-violating piece $\deh$
of the effective hamiltonian
in the combinations 
\rf{cptoddcombos} and \rf{cptevencombos}.
This structure permits a match to the established notation 
introduced in Eq.\ (14) of Ref.\ \cite{km},
for which the coefficients are conventionally denoted as
$(a_L)^\mu$, $(c_L)^\mn$, $g^{\mu\nu\rh}$, and $H^{\mn}$.
The correspondence is immediate,
with 
$a_L^{(3)\mu} \equiv (a_L)^\mu$,
$c_L^{(4)\mn} \equiv (c_L)^\mn$,
$g_{M+}^{(4)\mu\nu\rh} \equiv g^{\mu\nu\rh}$, 
and $H_{M+}^{(3)\mu\nu} \equiv H^{\mn}$.

The inclusion of mass-induced effects
introduces novel types of neutrino helicity flip
and thereby leads to additional cartesian coefficients 
for operators of renormalizable dimension.
Using Eqs.\ \rf{fundcoeff1} and \rf{fundcoeff2},
these are found to be
$a_l^{(3)\mu}$,
$c_l^{(4)\mu\nu}$,
$e_l^{(4)\mu}$, 
$g_l^{(4)\mu\nu\rh}$,
and $H_l^{(3)\mu\nu}$.
As can be seen from Eqs.\ \rf{cptoddcombos} and \rf{cptevencombos},
they contribute to the Lorentz-violating piece $\deh$
of the effective hamiltonian
through momentum and polarization projections
that differ from those for the mass-independent coefficients.
This implies the mass-induced coefficients
generate observationally distinct effects, 
so they offer additional possibilities for model building
along with new arenas for experimental searches. 

The cartesian coefficients for operators
of renormalizable dimension 
are related to spherical ones
through the expansions \rf{fundcoeff1}, \rf{fundcoeff2},
and \rf{sphereffcoeff}.
Table \ref{rencoeff}
lists the correspondence between these
two sets of coefficients,
along with the allowed $j$ values 
and the number of independent real components
for the spherical coefficients.

To express these relationships explicitly,
it is convenient to introduce the combinations
of cartesian unit vectors 
\beq
\hat{\mbf{x}}_\pm = \hat{\mbf{x}} \mp i \hat{\mbf{y}} .
\eeq
With this notation,
the connections between the spherical coefficients $\aeff{3}{jm}{}$ 
and the cartesian ones is 
\bea
\aeff{3}{00}{} &=& 
\sqrt{4\pi} \big(a_L^{(3)t} - e_l^{(4)t} -2 \tilde g_l^{(4)ttt}\big)
-\tfrac{4\sqrt\pi}{3} \tilde g_l^{(4)tjj} , 
\nn\\
\aeff{3}{1-1}{} &=& -\sqrt{\tfrac{2\pi}{3}}\, \xhat_-^j \,
\big(a_L^{(3)j} - e_l^{(4)j} -2 \tilde g_l^{(4)tjt}\big) ,
\nn\\
\aeff{3}{10}{} &=& -\sqrt{\tfrac{4\pi}{3}}
\big(a_L^{(3)z} - e_l^{(4)z} -2 \tilde g_l^{(4)tzt}\big) , 
\nn\\
\aeff{3}{11}{} &=& \sqrt{\tfrac{2\pi}{3}}\, \xhat_+^j \,
\big(a_L^{(3)j} - e_l^{(4)j} -2 \tilde g_l^{(4)tjt}\big) ,
\nn\\
\aeff{3}{2-2}{} &=&
-\sqrt{\tfrac{8\pi}{15}}\, \xhat_-^j\xhat_-^k\, \tilde g_l^{(4)tjk} , 
\nn\\
\aeff{3}{2-1}{} &=& -\sqrt{\tfrac{8\pi}{15}}\, \xhat_-^j\,
\big(\tilde g_l^{(4)tzj} + \tilde g_l^{(4)tjz} \big) , 
\nn\\
\aeff{3}{20}{} &=& -\sqrt{\tfrac{16\pi}{5}}
\big(\tilde g_l^{(4)tzz} -\tfrac13 \tilde g_l^{(4)tjj} \big) , 
\nn\\
\aeff{3}{21}{} &=& \sqrt{\tfrac{8\pi}{15}}\, \xhat_+^j\,
\big(\tilde g_l^{(4)tzj} + \tilde g_l^{(4)tjz} \big) , 
\nn\\
\aeff{3}{22}{} &=&
-\sqrt{\tfrac{8\pi}{15}}\, \xhat_+^j\xhat_+^k\, \tilde g_l^{(4)tjk} .
\eea
The purely mass-induced coefficients $\ceff{2}{jm}{}$
are given by 
\bea
\ceff{2}{1-1}{} &=& 
\sqrt{\tfrac{8\pi}{3}} \xhat_-^j \tilde H_l^{(3)tj} , 
\nn\\
\ceff{2}{10}{} &=& 
\sqrt{\tfrac{16\pi}{3}} \tilde H_l^{(3)tz} , 
\nn\\
\ceff{2}{11}{} &=& 
-\sqrt{\tfrac{8\pi}{3}} \xhat_+^j \tilde H_l^{(3)tj} ,
\eea
while the coefficients $\ceff{4}{jm}{}$ are 
\bea
\ceff{4}{00}{} &=& 
\sqrt{4\pi} \big( c_L^{(4)tt} + \tfrac13 c_L^{(4)jj} \big) , 
\nn\\
\ceff{4}{1-1}{} &=& -\sqrt{\tfrac{4\pi}{6}}\, \xhat_-^j\,
\big( c_L^{(4)tj} + c_L^{(4)jt} \big) , 
\nn\\
\ceff{4}{10}{} &=& -\sqrt{\tfrac{4\pi}{3}}
\big( c_L^{(4)tz} + c_L^{(4)zt} \big) , 
\nn\\
\ceff{4}{11}{} &=& \sqrt{\tfrac{4\pi}{6}}\, \xhat_+^j\,
\big( c_L^{(4)tj} + c_L^{(4)jt} \big) , 
\nn\\
\ceff{4}{2-2}{} &=& 
\sqrt{\tfrac{2\pi}{15}}\, \xhat_-^j \xhat_-^k \, c_L^{(4)jk} , 
\nn\\
\ceff{4}{2-1}{} &=& \sqrt{\tfrac{2\pi}{15}}\, \xhat_-^j \, 
\big( c_L^{(4)zj} + c_L^{(4)jz} \big) , 
\nn\\
\ceff{4}{20}{} &=& \sqrt{\tfrac{4\pi}{5}}
\big( c_L^{(4)zz} -\tfrac13 c_L^{(4)jj} \big) , 
\nn\\
\ceff{4}{21}{} &=& -\sqrt{\tfrac{2\pi}{15}}\, \xhat_+^j \, 
\big( c_L^{(4)zj} + c_L^{(4)jz} \big) , 
\nn\\
\ceff{4}{22}{} &=& 
\sqrt{\tfrac{2\pi}{15}}\, \xhat_+^j \xhat_+^k \, c_L^{(4)jk} .
\eea

The off-diagonal spherical coefficients
$\geff{2}{jm}{}$, $\geff{4}{jm}{}$, and $\Heff{3}{jm}{}$ 
control neutrino-antineutrino mixing.
The purely mass-induced coefficients $\geff{2}{jm}{}$
are given by  
\bea
\geff{2}{1-1}{} &=& 
\sqrt{\tfrac{4\pi}{3}} \, \xhat_-^j \, a_l^{(3)j} , 
\nn\\
\geff{2}{10}{} &=& \sqrt{\tfrac{8\pi}{3}} \, a_l^{(3)z} , 
\nn\\
\geff{2}{11}{} &=& 
-\sqrt{\tfrac{4\pi}{3}} \, \xhat_+^j \, a_l^{(3)j} ,
\eea
and the coefficients $\geff{4}{jm}{}$ are
\bea
\geff{4}{1-1}{} &=& 
i\sqrt{\tfrac{16\pi}{3}} \, \xhat_-^j \,
\big( g_{M+}^{(4)tjt} 
+ \tfrac12 g_{M+}^{(4)tzj} - \tfrac12 g_{M+}^{(4)tjz} \big) , 
\nn\\
\geff{4}{10}{} &=& i \sqrt{\tfrac{32\pi}{3}}
\big( g_{M+}^{(4)tzt} 
- \tfrac{i}{2} g_{M+}^{(4)txy} + \tfrac{i}{2} g_{M+}^{(4)tyx} \big) , 
\nn\\
\geff{4}{11}{} &=& -i\sqrt{\tfrac{16\pi}{3}} \, \xhat_+^j \,
\big( g_{M+}^{(4)tjt} 
- \tfrac12 g_{M+}^{(4)tzj} + \tfrac12 g_{M+}^{(4)tjz} \big) , 
\nn\\
\geff{4}{2-2}{} &=& 
-i\sqrt{\tfrac{4\pi}{5}} \, \xhat_-^j\xhat_-^k \, g_{M+}^{(4)tjk} , 
\nn\\
\geff{4}{2-1}{} &=& -i\sqrt{\tfrac{4\pi}{5}} \, \xhat_-^j \,  
\big(g_{M+}^{(4)tzj} + g_{M+}^{(4)tzj}\big) , 
\nn\\
\geff{4}{20}{} &=& -i\sqrt{\tfrac{24\pi}{5}}
\big( g_{M+}^{(4)tzz} -\tfrac13 g_{M+}^{(4)tjj}\big) , 
\nn\\
\geff{4}{21}{} &=& i\sqrt{\tfrac{4\pi}{5}} \, \xhat_+^j \,  
\big(g_{M+}^{(4)tzj} + g_{M+}^{(4)tzj}\big) , 
\nn\\
\geff{4}{22}{} &=& 
-i\sqrt{\tfrac{4\pi}{5}} \, \xhat_+^j\xhat_+^k \, g_{M+}^{(4)tjk} .
\eea
Lastly,
the coefficients $\Heff{3}{jm}{}$ are given by 
\bea
\Heff{3}{1-1}{} &=& 
\sqrt{\tfrac{4\pi}{3}} \, \xhat_-^j \,
\big( 2i H_{M+}^{(3)tj} 
\nn\\
&&
\hskip 40pt
+ c_l^{(4)jt} 
+ \tfrac12 c_l^{(4)zj} - \tfrac12 c_l^{(4)jz} \big) , 
\nn\\
\Heff{3}{10}{} &=& \sqrt{\tfrac{8\pi}{3}} 
\big( 2i H_{M+}^{(3)tz} 
\nn\\
&&
\hskip 40pt
+ c_l^{(4)zt} 
- \tfrac{i}{2} c_l^{(4)xy} + \tfrac{i}{2} c_l^{(4)yx} \big) , 
\nn\\
\Heff{3}{11}{} &=& -\sqrt{\tfrac{4\pi}{3}} \, \xhat_+^j \,
\big( 2i H_{M+}^{(3)tj} 
\nn\\
&&
\hskip 40pt
+ c_l^{(4)jt} 
- \tfrac12 c_l^{(4)zj} + \tfrac12 c_l^{(4)jz} \big) , 
\nn\\
\Heff{3}{2-2}{} &=& 
-\sqrt{\tfrac{\pi}{5}} \, \xhat_-^j\xhat_-^k \, c_l^{(4)jk} , 
\nn\\
\Heff{3}{2-1}{} &=& -\sqrt{\tfrac{\pi}{5}} \, \xhat_-^j \,
\big( c_l^{(4)zj} +c_l^{(4)jz} \big) , 
\nn\\
\Heff{3}{20}{} &=& -\sqrt{\tfrac{6\pi}{5}}
\big( c_l^{(4)zz} - \tfrac13 c_l^{(4)jj} \big) , 
\nn\\
\Heff{3}{21}{} &=& \sqrt{\tfrac{\pi}{5}} \, \xhat_+^j \,
\big( c_l^{(4)zj} +c_l^{(4)jz} \big) , 
\nn\\
\Heff{3}{22}{} &=& 
-\sqrt{\tfrac{\pi}{5}} \, \xhat_+^j\xhat_+^k \, c_l^{(4)jk} .
\eea

\subsection{Massless models}
\label{Massless models}

An interesting case of the general formalism
is the massless limit,
$m_l\to 0$.
Given the compelling experimental evidence 
that neutrinos oscillate amassed in recent years,
it is reasonable and conservative to adopt the perspective
that these oscillations 
arise from a small nonzero neutrino mass matrix.
However,
the frequently encountered claim 
that oscillations prove neutrinos have mass is false,
as oscillations can arise from Lorentz and CPT violation
even when all masses vanish.
Massless models may therefore be of interest for model building. 
They are also relevant in the ultrarelativistic limit,
where masses can be neglected
but Lorentz-violation operators of dimension four or greater 
remain important.

When $m_l\to 0$,
the mass-induced operators
$m_l$,
$\ehat_l$,
$\ahat_l^\mu$,
$\chat_l^\mu$,
$\ghat_l^\mn$,
$\Hhat_l^\mn$
and the associated coefficients 
in Eqs.\ \rf{fundcoeff1} and \rf{fundcoeff2}
all vanish.
The effective spherical coefficients
appearing in Eq.\ \rf{sphereffcoeff}
therefore reduce in massless models to
$\aeff{d}{jm}{ab} = \aLcoef{d}{jm}{ab}$, 
$\ceff{d}{jm}{ab} = \cLcoef{d}{jm}{ab}$, 
$\geff{d}{jm}{ab} = \gMcoef{d}{jm}{ab}$, 
and $\Heff{d}{jm}{ab} = \HMcoef{d}{jm}{ab}$.
This implies the angular-momentum index 
labeling the effective coefficients
is limited to $j\leq d-2$ rather than to $j\leq d-1$.
Table \ref{masslesscoeff}
lists the coefficients for massless models,
together with the allowed range of $d$ and $j$
and the number of independent real components.

\begin{table}
\renewcommand{\arraystretch}{1.5}
\begin{tabular}{c|c|c|c}
coefficient & $d$ & $j$ & number \\
\hline
$\aLcoef{d}{jm}{ab}$
& odd, $\geq 3$  & $d-2\geq j\geq 0$ & $9(d-1)^2$ \\
$\cLcoef{d}{jm}{ab}$
& even, $\geq 4$ & $d-2\geq j\geq 0$ & $9(d-1)^2$ \\
$\gMcoef{d}{jm}{ab}$ 
& even, $\geq 4$ & $d-2\geq j\geq 1$ & $12d(d-2)$ \\
$\HMcoef{d}{jm}{ab}$
& odd, $\geq 3$  & $d-2\geq j\geq 1$ & $6d(d-2)$ \\
\end{tabular}
\caption{\label{masslesscoeff}
Spherical coefficients for massless models.}
\end{table}

Lorentz-violating massless models studied in the literature
include the bicycle model
\cite{bicycle,km},
its generalization by Barger, Marfatia, and Whisnant
\cite{bmw},
and the isotropic subset of the minimal SME
\cite{blmw}.
All these massless models 
involve operators of renormalizable dimension,
and they can reproduce many observed features
of neutrino oscillations.
However,
to date no fully satisfactory massless model 
has been presented.
The primary issue faced by model builders
is simultaneously reproducing 
both the KamLAND data 
\cite{KamLAND(L/E)} 
and the observed shape of the solar neutrino spectrum
\cite{Borexino}
in the energy range 1-20 MeV.
The KamLAND results can be reproduced
using the massless Lorentz-violating seesaw mechanism
\cite{km},
but the $\simeq$1 MeV scale at which this must be triggered
is challenging to reconcile with the $\simeq$10 MeV scale
at which the solar-neutrino survival probability passes from
higher to lower values.
The isotropic minimal SME cannot accommodate
both these features
\cite{blmw},
whereas even a single mass parameter suffices 
\cite{puma}.

Given the above issues,
it is worth emphasizing that the existing literature 
concerns only a tiny portion 
of the available coefficient space for massless models.
The potential role of direction-dependent coefficients,
including the coefficients
$\gMcoef{4}{jm}{ab}$ and $\HMcoef{3}{jm}{ab}$
for operators of renormalizable dimension,
remains largely unexplored.
Also,
many new options exist for realistic model building 
using the nonminimal mass-independent operators
classified in the present work.
Constructing a phenomenologically viable massless model 
for neutrino oscillations 
remains an interesting and worthwhile open challenge.

\subsection{Flavor-blind and oscillation-free models}
\label{Flavor-blind and oscillation-free models}

A particularly simple limit of the general formalism
is the flavor-blind limit,
obtained by assuming that the mass-squared matrix 
and the Lorentz violation 
affect all flavors in the same way.
This limit is unrealistic as a global description of neutrinos
because no neutrino-neutrino oscillations appear.
However,
under suitable circumstances
it may represent a useful approximation
to the physics of neutrino propagation. 
For example,
it can be physically relevant 
when the dominant effects of Lorentz violation are flavor blind,
with oscillations being comparatively small.
This can arise via numerical values of coefficients
or under suitable physical circumstances 
such as ultra-high neutrino energies.
A flavor-blind treatment may therefore be appropriate
for time-of-flight studies,
for instance.
The flavor-blind cases are also useful 
as toy models of Lorentz-violating effects
and as a stepping stone 
to the more general models considered below.
Note that experimental sensitivity to oscillation-free effects
is generically reduced because no interferometry is involved.
In this subsection,
we consider two classes of flavor-blind models
distinguished according to whether 
they allow single-flavor neutrino-antineutrino oscillations
or are oscillation free.

\subsubsection{Flavor-blind and single-flavor models}

The effective hamiltonian 
$h^{\rm fb}_{\rm eff}$ 
for the flavor-blind limit
is the restriction of Eq.\ \rf{hresult}
to three copies of a single flavor.
Since $h^{\rm fb}_{\rm eff}$ splits into three identical pieces, 
we can suppress the labeling of the three flavors
or consider only a single flavor. 
The effective hamiltonian then takes the form
\beq
h^{\rm fb}_{\rm eff} 
= |\mbf p| + \fr{|m_l|^2}{2|\mbf p|} - \fr{\ce}{|\mbf p|}
+ \fr{1}{|\mbf p|}
\begin{pmatrix}
\ae & -\ge \\
-\ge^* & -\ae
\end{pmatrix} .
\label{fbham}
\eeq
The coefficients $H_{M+}$, $e_l$, $H_l$
are absent in this limit because 
$e_M$, $f_M$, $H_{M+}$ are antisymmetric in flavor space.
The effective components $\ae$ and $\ce$ are real,
while $\ge$ is complex.

Although neutrino-neutrino oscillations are absent
in flavor-blind models,
the coefficient $\ge$ generates
neutrino-antineutrino mixing for each flavor. 
This is a CPT-odd effect.
Even if the coefficient $\ghat_{M+}^\mn$ vanishes,
the mass-induced coefficient $\ahat_l^\mu$ 
can contribute to $\ge$ 
and induce oscillations between neutrinos and antineutrinos.
Mixing is absent only in the CPT-even limit  
and in the special CPT-violating case 
with both $\ghat_{M+}^\mn$ and $m_l$ vanishing.
Examples of single-flavor models 
with only renormalizable coefficients
are presented in 
Secs.\ IV B 1 and IV B 2 of Ref.\ \cite{km}.

The flavor-blind hamiltonian \rf{fbham}
can be diagonalized in the form 
\beq
h^{\rm fb}_{\rm eff} =
\begin{pmatrix} C & S\\ -S^* & C \end{pmatrix}
\begin{pmatrix} E^{\rm fb}_+ & 0 \\ 0 & E^{\rm fb}_- \end{pmatrix}
\begin{pmatrix} C & -S\\ S^* & C \end{pmatrix} .
\eeq
The eigenvalues are
\beq
E^{\rm fb}_\pm = |\mbf p| + \fr{|m_l|^2}{2|\mbf p|} 
- \fr{\ce}{|\mbf p|} \pm \fr{\la}{|\mbf p|} 
\label{fbenergies}
\eeq
with
\beq
\la = \sqrt{\ae^2 + |\ge|^2} ,
\eeq
and the components of the mixing matrix are 
\bea
C &=& \sqrt{\fr{\la+\ae}{2\la}},
\quad 
S = \fr{\ge}{\sqrt{2\la(\la+\ae)}} .
\label{fbmixing}
\eea

For experimental investigations,
it is useful to have a description of flavor-blind models
accounting for properties under rotation transformations.
As before,
this can be achieved by decomposition into spherical harmonics.
The effective components $\ae$, $\ce$, $\ge$ 
can be expanded in flavor-blind coefficients as
\bea
\ae &=& 
\sum_{djm}  |\mbf p|^{d-2}\, Y_{jm}(\phat) \afb{d}{jm} , 
\nn\\
\ce &=& 
\sum_{djm}  |\mbf p|^{d-2}\, Y_{jm}(\phat) \cfb{d}{jm} ,
\nn\\
\ge &=& 
\sum_{djm}  |\mbf p|^{d-2}\, \syjm{+1}{jm}(\phat) \gfb{d}{jm} ,
\label{fbexp}
\eea
where
\bea
\afb{d}{jm}{}\hskip-10pt ^* &=& (-1)^m \afb{d}{j(-m)} , 
\nn\\
\cfb{d}{jm}{}\hskip-10pt ^* &=& (-1)^m \cfb{d}{j(-m)} .
\eea
Table \ref{fbcoeffs}
lists the allowed ranges of $d$ and $j$
for these coefficients,
along with the number of independent real components
they contain.

\begin{table}
\begin{tabular}{c|c|c|c}
coefficient & $d$ & $j$ & number\\
\hline
$\afb{d}{jm}$ & odd,  $\geq 3$ & $d-1\geq j\geq 0$ & $d^2$ \\
$\cfb{d}{jm}$ & even, $\geq 4$ & $d-2\geq j\geq 0$ & $(d-1)^2$ \\
$\gfb{d}{jm}$ & even, $\geq 2$ & $d-1\geq j\geq 1$ & $2(d^2-1)$ \\
\hline
$\aof{d}{jm}$ & odd,  $\geq 3$ & $d-2\geq j\geq 0$ & $(d-1)^2$ \\
$\cof{d}{jm}$ & even, $\geq 4$ & $d-2\geq j\geq 0$ & $(d-1)^2$ 
\end{tabular}
\caption{\label{fbcoeffs}
Spherical coefficients for flavor-blind models.}
\end{table}

\subsubsection{Oscillation-free models}
\label{Oscillation-free models}

For certain physical applications
and to gain intuition within a simple theoretical framework,
it can be useful to restrict attention to coefficients
that cause no mixing at all.
These oscillation-free models 
are achieved starting from the above flavor-blind models
and imposing vanishing neutrino-antineutrino mixing.
The form of the dispersion relation \rf{fbenergies}
then implies that the general oscillation-free model
can be obtained by setting to zero
the coefficients $\geff{d}{jm}{}$.
Oscillation-free models
therefore amount to flavor-blind models
that conserve lepton number.
The oscillation-free spherical coefficients 
$\aof{d}{jm}$ and $\cof{d}{jm}$
appear in expansions of the form \rf{fbexp}
but have index ranges 
limited to $d\geq 3$ and $4$ with $d-2\geq j\geq 0$,
as shown in Table \ref{fbcoeffs}.
Note that most of these coefficients
describe anisotropic effects,
so a generic oscillation-free model 
predicts direction dependence and sidereal variations
in neutrino and antineutrino properties. 

Denoting the neutrino energy in oscillation-free models
by $E^{\rm of}_\nu$,
we can expand in spherical coefficients to obtain
\bea
E^{\rm of}_{\nu} 
&=& |\mbf p| + \fr{|m_l|^2}{2|\mbf p|} 
\nn\\
&&
+ \sum_{djm} \pmag^{d-3} Y_{jm}(\phat)
\Big[\aof{d}{jm} - \cof{d}{jm}\Big] .
\qquad
\label{ofenergies}
\eea
The antineutrino energy $E^{\rm of}_{\nub}$
is obtained by changing the sign
of the coefficients $\aof{d}{jm}$.

One application of oscillation-free models
is the study of neutrino propagation.
A useful concept in this context is the group velocity
$\vgof = \prt {E^{\rm of}_{\nu}}/{\prt |\mbf p|}$,
which for a neutrino becomes 
\bea
\vgof 
&=& 1 - \fr{|m_l|^2}{2\pvec^2} 
+ \sum_{djm} (d-3) \pmag^{d-4}\, 
Y_{jm}(\phat)\,
\nn\\
&&
\hskip 70pt
\times
\Big[\aof{d}{jm} - \cof{d}{jm}\Big],
\quad
\label{vgof}
\eea
The antineutrino group velocity 
$\ol \vgof = \prt {E^{\rm of}_{\nub}}/{\prt |\mbf p|}$
takes the same form but with a sign change
for the coefficient $\aof{d}{jm}$.

\subsection{Diagonalizable models}
\label{Diagonalizable models}

Another useful simple limit is the class of
diagonalizable models,
for which the effective hamiltonian $h^{\rm d}_{\rm eff}$
is obtained from the general expression \rf{hresult}
by requiring all terms to be simultaneously diagonalizable.
The mass-squared matrix is diagonalized 
using a momentum-independent mixing matrix $U$,
so in the diagonalizable limit
each Lorentz-violating term must take a special form 
that is also diagonalizable with the constant matrix $U$.
In the flavor basis,
this implies every Lorentz-violating operator commutes
with all others and with the mass-squared matrix.

The definition of the diagonalizable limit 
implies that in the mass basis
the neutrino behavior is governed by three copies 
of the single-flavor limit 
discussed in the previous subsection.
The copies are distinct in that they can involve 
different masses and coefficients for Lorentz violation.
For the effective hamiltonian $h^{\rm d}_{\rm eff}$ 
in the flavor basis,
we can therefore write
\bea
\hskip -150pt
h^{\rm d}_{\rm eff} &=&
\begin{pmatrix} U^\dag & 0\\ 0 & U^T \end{pmatrix}
\begin{pmatrix} 
C^{\rm d} & S^{\rm d}\\ -S^{{\rm d}*} & C^{\rm d} 
\end{pmatrix}
\begin{pmatrix} E^{\rm d}_+ & 0 \\ 0 & E^{\rm d}_- \end{pmatrix}
\nn\\
&&
\hskip 60pt
\times
\begin{pmatrix} 
C^{\rm d} & -S^{\rm d}\\ S^{{\rm d}*} & C^{\rm d} 
\end{pmatrix} 
\begin{pmatrix} U & 0\\ 0 & U^* \end{pmatrix} ,
\eea
where $E^{\rm d}_\pm$, $C^{\rm d}$, $S^{\rm d}$
are all $3\times 3$ diagonal matrices 
obtained by combining the three distinct copies of 
the single-flavor results \rf{fbenergies} and \rf{fbmixing}.

Diagonalizable models offer potentially interesting opportunities
to construct realistic models
for neutrino oscillations and propagation
involving perturbative Lorentz-violating effects 
because the conventional mass matrix 
in the three-neutrino massive model (\msm)
can be adopted together with small Lorentz-violating terms. 
It may also be possible to construct 
more ambitious diagonalizable models
in which Lorentz violation plays a key role,
perhaps completely replacing one or more neutrino mass terms
in a vein similar to the puma model
\cite{puma}.
Certain diagonalizable models may be useful as toy models
or as approximations suitable for describing 
a more complete theory in specific physical regimes.
For example,
one simple variation arises if $U$ is taken to be the identity.
Each neutrino in the resulting diagonalizable model 
is then controlled 
by a different set of single-flavor coefficients
and there are no neutrino-neutrino oscillations,
although CPT-odd neutrino-antineutrino mixing can still arise. 

Diagonalizable models with spatial isotropy
represent a special restriction
considered in Sec.\ \ref{Isotropic diagonalizable models} below.
The more general class of diagonalizable models 
having nontrivial rotation behavior 
are conspicuously absent from the literature.
Many interesting signals are predicted by these models,
such as direction dependence of time-of-flight measurements
and of oscillations.

To study the rotation properties of diagonalizable models,
it is again useful to perform a decomposition 
in spherical harmonics.
The treatment proceeds most easily by working 
in the diagonal basis.
Denoting indices in this basis with primes,
the expansion in spherical harmonics becomes
\bea
\ae^{a'b'} &=& 
\de^{a'b'} \sum_{djm}  
|\mbf p|^{d-2}\, Y_{jm}(\phat) \adia{d}{jm}{a'} , 
\nn\\
\ce^{a'b'} &=& 
\de^{a'b'} \sum_{djm}  
|\mbf p|^{d-2}\, Y_{jm}(\phat) \cdia{d}{jm}{a'} , 
\nn\\
\ge^{a'b'} &=& 
\de^{a'b'} \sum_{djm}  
|\mbf p|^{d-2}\, \syjm{+1}{jm}(\phat) \gdia{d}{jm}{a'} .
\eea
Table \ref{diagcoeffs} lists
the spherical coefficients for diagonalizable models
in the diagonal basis.
It also provides the range of $d$ and $j$
and shows the number of independent real components
for each coefficient.

\begin{table}
\begin{tabular}{c|c|c|c}
coefficient & $d$ & $j$ & number\\
\hline
$\adia{d}{jm}{a'}$ & odd,  $\geq 3$ & $d-1\geq j\geq 0$ & $3d^2$ \\
$\cdia{d}{jm}{a'}$ & even, $\geq 4$ & $d-2\geq j\geq 0$ & $3(d-1)^2$ \\
$\gdia{d}{jm}{a'}$ & even, $\geq 2$ & $d-1\geq j\geq 1$ & $6(d^2-1)$
\end{tabular}
\caption{\label{diagcoeffs}
Spherical coefficients for diagonalizable models.}
\end{table}

\subsection{Isotropic models}
\label{Isotropic models}

The class of isotropic models,
sometimes called `fried-chicken' models 
due to their popularity and simplicity,
is generated by restricting attention
to the comparatively few Lorentz-violating operators
that maintain rotation symmetry.
Since observer boosts mix with rotations, 
any isotropic model is well defined
only if its preferred observer inertial frame is specified.
All observers boosted with respect to this frame
see anisotropic effects.
A popular choice for the preferred frame
is the frame of the cosmic microwave background (CMB),
but other choices are possible.
Note that choosing the CMB frame for any isotropic model
implies anisotropies in the canonical Sun-centered inertial frame
\cite{sunframe,labframe}
and in Earth-based experiments.

\subsubsection{Generic isotropic models}
\label{Generic isotropic models}

The expansion \rf{effcoeff} of $\deh$ 
in spherical harmonics
is ideally suited for investigations 
of generic isotropic models. 
Only coefficients with $j=0$ can contribute
in the preferred frame.
Inspection of Table \ref{sphercoeff} and Eq.\ \rf{sphereffcoeff}
reveals that $\ge$ and $\He$ must vanish,
so isotropic models contain no operators
mixing neutrinos with antineutrinos.
The effective hamiltonian therefore
breaks into two $3\times 3$ blocks,
one for neutrinos and one for antineutrinos. 

For neutrinos,
we can write 
\beq
\hri_\nu = |\mbf p| + \fr{m_lm_l^\dag}{2|\mbf p|}
+ \fr{\ae}{|\mbf p|} - \fr{\ce}{|\mbf p|} ,
\label{fcham}
\eeq
where the ring diacritic is used here and below
to denote isotropic quantities
in the preferred frame
\cite{km}.
The effective hamiltonian for antineutrinos 
is obtained by transposing in flavor space 
and changing the sign of the CPT-odd terms,
\beq
\hri_{\bar\nu} = |\mbf p| + \fr{m_l^\dag m_l}{2|\mbf p|} 
- \fr{\ae^T}{|\mbf p|} - \fr{\ce^T}{|\mbf p|} .
\eeq
The expansion of the Lorentz-violating terms 
in spherical harmonics takes the simple form
\bea
\ae^{ab} &=& 
\sum_{d} |\mbf p|^{d-2}\, \afc{d}{ab} ,
\nn\\
\ce^{ab} &=& 
\sum_{d} |\mbf p|^{d-2}\, \cfc{d}{ab} .
\label{icoeff}
\eea
The match to the spherical coefficients
appearing in the expansion \rf{effcoeff} is
\bea
\afc{d}{ab} &=& 
\fr{1}{\sqrt{4\pi}}\aeff{d}{00}{ab} ,
\nn\\
\cfc{d}{ab} &=& 
\fr{1}{\sqrt{4\pi}}\ceff{d}{00}{ab} .
\eea
Both sets of coefficients are hermitian in flavor space,
giving 9 real degrees of freedom for each value of $d$.
Since the CPT-even effects occur only for even $d$
while CPT-odd ones occur only for odd $d$,
only one of the coefficients $\afc{d}{ab}$ and $\cfc{d}{ab}$ 
is present at any fixed $d$.
Table \ref{isocoeffs} summarizes these basic features.

\begin{table}
\renewcommand{\arraystretch}{1.5}
\begin{tabular}{c|c|c|c}
isotropic type &coefficient & $d$ & number per $d$\\
\hline
generic & $\afc{d}{ab}$ & odd, $\geq 3$ & 9 \\
& $\cfc{d}{ab}$ & even, $\geq 4$ & 9 \\
\hline
diagonalizable & $\afc{d}{a'}$ & odd, $\geq 3$ & 3 \\
& $\cfc{d}{a'}$ & even, $\geq 4$ & 3 \\ 
\hline
oscillation-free & $\afc{d}{}$ & odd, $\geq 3$ & 1 \\
& $\cfc{d}{}$ & even, $\geq 4$ & 1 
\end{tabular}
\caption{\label{isocoeffs}
Spherical coefficients for isotropic models.}
\end{table}

Despite their simplicity,
isotropic models retain sufficient complexity
to offer interesting prospects 
as global models for neutrino behavior.
An interesting example is the class of puma models
\cite{puma},
which provide viable alternatives to the \msm\
as a global description of existing neutrino-oscillation data.
For instance,
the \ceafm\ puma model is isotropic in the Sun-centered frame
and is specified by three parameters,
consisting of one mass and two coefficients for Lorentz violation.
The ratio of the two coefficients acts 
like an effective mass at high energies
via the Lorentz-violating seesaw mechanism
\cite{km}.
This ratio and the mass parameter can be chosen
to reproduce all accepted neutrino oscillation results,
while the third degree of freedom
naturally generates the MiniBooNE anomalies
\cite{MiniBooNE1,MiniBooNE2}
that cannot be accommodated in the \msm.
The effective hamiltonian in the \ceafm\ puma model 
takes the form of Eqs.\ \rf{fcham} and \rf{icoeff}
with the explicit choices 
\bea
(m_lm_l^\dag)_{ab} &=& m^2
\begin{pmatrix} 
1 & 1 & 1 \\ 
1 & 1 & 1 \\ 
1 & 1 & 1 
\end{pmatrix} ,
\nn\\
\afc{5}{ab} 
&=& \ari^{(5)}
\begin{pmatrix} 
1 & 1 & 1 \\ 
1 & 0 & 0 \\ 
1 & 0 & 0 
\end{pmatrix} ,
\nn\\
\cfc{8}{ab}
&=& - \cri^{(8)}
\begin{pmatrix} 
1 & 0 & 0 \\ 
0 & 0 & 0 \\ 
0 & 0 & 0 
\end{pmatrix} ,
\label{puma}
\eea 
and all other coefficients zero.
The numerical values giving excellent agreement
with experimental data are 
\bea
m^2 &=& 2.6\times10^{-23} {\rm ~GeV}^2, 
\nn\\
\ari^{(5)} &=& -2.5\times10^{-19} {\rm ~GeV}^{-1}, 
\nn\\
\cri^{(8)} &=& 1.0\times10^{-16} {\rm ~GeV}^{-4},
\label{c8a5m}
\eea
Extensions of this model can also accommodate
the LSND anomaly
\cite{LSND}
and CPT asymmetries of the MINOS type
\cite{MINOSanomaly}. 
Note that the puma models 
lie outside the class of diagonalizable models
described in the previous subsection
because the nontrivial texture \rf{puma} in flavor space
implies the different terms fail to commute.

\subsubsection{Isotropic diagonalizable models}
\label{Isotropic diagonalizable models}

Combining the diagonalizable and isotropic restrictions
described in Secs.\ \ref{Diagonalizable models}
and \ref{Isotropic models}
yields a very simple class of models.
These models must 
both have simultaneously diagonalizable Lorentz-violating operators 
and also exhibit rotational invariance in a preferred frame.

For these models,
it is convenient to work in the diagonal basis
and in the preferred frame,
although care is required in applications 
to Earth-based experiments or observations
because these are boosted relative to any preferred inertial frame
and therefore necessarily exhibit anisotropic effects.
In the diagonal basis and preferred frame,
the energy of a neutrino of species $a'$ 
can be written in the form
\beq
\Eri_{\nu, a'} = 
|\mbf p| + \fr{|m_{l}|_{a'}^2}{2|\mbf p|} 
+ \sum_{d}  |\mbf p|^{d-3}\, 
(\acfc{d}{a'} - \ccfc{d}{a'}) .
\label{disoen}
\eeq
Three coefficients for Lorentz violation appear for each $d$, 
one for each neutrino species.

In isotropic diagonalizable models,
the same three coefficients control 
the behavior of antineutrinos at each $d$,
but the antineutrino dispersion relation 
is CPT conjugated.
This changes the sign
of the $\acfc{d}{a'}$ coefficients,
so that the antineutrino energy is
\beq
\Eri_{\nub, a'} = 
|\mbf p| + \fr{|m_{l}|_{a'}^2}{2|\mbf p|} 
- \sum_{d}  |\mbf p|^{d-3}\, 
(\acfc{d}{a'} + \ccfc{d}{a'}) .
\eeq

Since the Lorentz violation 
in these simple models
leads to a power series in positive powers
of momentum $|\mbf p|$ 
with a single coefficient at each $d$ 
for each species,
the group velocity 
$\vgri_{a'} = \prt {\Eri_{\nu, a'}}/{\prt |\mbf p|}$
for each species $a'$ can be obtained immediately.
For neutrinos,
we find 
\bea
\vgri_{a'} &=& 1 - \fr{|m_l|_{a'}^2}{2|\mbf p|^2} 
+ \sum_{d} (d-3) |\mbf p|^{d-4}\, 
(\acfc{d}{a'} - \ccfc{d}{a'}) .
\qquad
\eea
For antineutrinos,
the group velocity
$\ol \vgri_{a'} = \prt {\Eri_{\nub, a'}}/{\prt |\mbf p|}$
takes the CPT-conjugate form
with an opposite sign for the coefficient $\acfc{d}{a'}$.
We remark that 
provided the coefficients $\acfc{d}{a'}$
dominate the deviations from lightspeed,
this CPT-conjugation property implies
that neutrinos can be superluminal 
while antineutrinos are subluminal
or vice versa.
This may be useful in attempts to model time-of-flight measurements 
from laboratory neutrinos and supernova antineutrinos.
Notice also that although isotropic violations
with operators of mass dimension $d=3$ 
alter the phase velocity,
these have no effect on the group velocity 
because they generate only a constant shift 
in the neutrino energy.

\subsubsection{Isotropic oscillation-free models}
\label{Isotropic oscillation-free models}

The above construction for isotropic diagonalizable models
also incorporates another special class of simple models,
consisting of the isotropic limit of the oscillation-free models
described in Sec.\ \ref{Oscillation-free models}.
In this case,
all neutrino species are assumed 
to have the same isotropic properties
in the preferred frame.

For this heavily restricted limit,
the $a'$ index appearing in isotropic diagonalizable models
can be disregarded,
and only one coefficient for Lorentz violation
appears for each value of $d$.
For example,
the neutrino energy in these isotropic oscillation-free models 
becomes
\beq
\Eri_{\nu} = 
|\mbf p| + \fr{|m_{l}|^2}{2|\mbf p|} 
+ \sum_{d} |\mbf p|^{d-3}\, 
(\acfc{d}{} - \ccfc{d}{}) ,
\label{isoofen}
\eeq
and the corresponding group velocity is 
\beq
\vgri = 1 - \fr{|m_l|^2}{2|\mbf p|^2} 
+ \sum_{d} (d-3) |\mbf p|^{d-4}\, 
(\acfc{d}{} - \ccfc{d}{}) .
\label{isoofv}
\eeq
Note that these isotropic oscillation-free models 
coincide with the isotropic flavor-blind models
and with isotropic single-flavor models
because the isotropic requirement
forces all Lorentz-violating Majorana couplings to vanish. 

Despite the simultaneous conditions
of isotropy, diagonalizability, flavor independence,
and no oscillations,
these models can still exhibit CPT violation
because all their odd-$d$ Lorentz-violating operators are CPT odd.
Indeed,
the even powers of the momentum appearing 
in the dispersion relation for any isotropic flavor-blind model
are associated with CPT violation
and so the corresponding terms for antineutrinos must change sign,
a potentially important feature for phenomenology 
that is often overlooked in the literature.

We remark in passing that 
any of the above simple isotropic deformations 
of the usual dispersion relations for neutrinos
can lead to physical and observable effects
only if some other sector is conventional
or exhibits different Lorentz violation.
Attempts to invoke a common deformation of Lorentz symmetry
across all species merely generate conventional physics
in an unconventional guise
\cite{ck,akgrav,kmnonmin,lnjvjs}.
Deformed Lorentz transformations 
that depend on different species
are discussed in Ref.\ \cite{kmnonmin}
and are naturally described within the SME framework.

\section{Applications to oscillations}
\label{Applications to oscillations}

Neutrino oscillations offer a powerful tool
for investigations of physics beyond the SM
because their interferometric nature
makes them highly sensitive to unconventional couplings. 
Next,
we apply the formalism developed in the previous sections 
to explore oscillation effects 
due to Lorentz-violating operators of arbitrary dimension,
and we obtain explicit constraints 
on a variety of coefficients for Lorentz violation
for $d\leq 10$.

Any diagonal terms in the effective hamiltonian \rf{hresult}
have no effect on oscillations.
We can therefore drop the diagonal momentum term in $(\heff)_0$.
Inspecting Eqs.\ \rf{effliham} and \rf{efflvham}
shows that the effective hamiltonian $\hosc$
controlling oscillations is given by
\beq
\hosc =
\fr 1 {\pmag}\begin{pmatrix}
\half m_l^\dag m_l + \ae - \ce &
\He - \ge \\
\He^\dag - \ge^\dag & 
\half m_l^\dag m_l - \ae^T -\ce^T
\end{pmatrix} .
\eeq
The oscillation amplitudes can be found 
from the time-evolution operator $S(t) \equiv \exp(-i\hosc t)$.
For practical applications,
the time $t$ can be identified with the experimental baseline $L$,
so we can write $S(L) = \exp(-i\hosc L)$.

An exact treatment of oscillations is typically infeasible.
One potential issue is that
possible decay processes including neutrino splitting
can introduce nonlinear effects in certain regimes.
However,
no nonlinear effects from neutrino decay or other processes
have been detected in experiments to date.
A linear treatment using $\hosc$ is therefore
a realistic and feasible approach 
for obtaining robust and conservative constraints.
In what follows,
two regimes of practical interest are considered.
The first is the short-baseline approximation,
which applies when the baseline $L$
is short compared to the effective hamiltonian $\hosc$
and so the transition amplitudes are small.
This approximation is discussed 
in Sec.\ \ref{Short-baseline approximation}.
For short baselines, 
$S(L)$ can be expanded in powers of $\hosc$.
The second regime,
considered in Sec.\ \ref{Perturbative Lorentz violation},
is the limit of perturbative Lorentz violation.
It applies when oscillations are primarily
due to the mass matrix $m_l^\dag m_l$.
In this case,
Lorentz and CPT violation can be treated as a perturbation
on mass-induced oscillations.
The illustrative example of two-flavor maximal mixing
with Lorentz violation,
which also offers intuition about CPT breaking,
is presented in Sec.\ \ref{Two-flavor maximal mixing}.

\subsection{Short-baseline approximation}
\label{Short-baseline approximation}

Expanding $S(L)$ in powers of $\hosc L$ 
yields simple leading-order approximations 
for the transition probabilities 
from flavor $a$ to flavor $b\neq a$,
\bea
P_{\nu_b\to\nu_a} &=& \Big|
\big(\half m_l^\dag m_l + \ae - \ce\big)_{ab}
\Big|^2 \fr{L^2}{\pvec^2} , 
\nn\\
P_{\nub_b\to\nub_a} &=& \Big|
\big(\half m_l^\dag m_l - \ae - \ce\big)_{ab}
\Big|^2 \fr{L^2}{\pvec^2} ,
\nn\\
P_{\nub_b\to\nu_a} &=& \Big|
\big(\He - \ge\big)_{ab}
\Big|^2 \fr{L^2}{\pvec^2} , 
\nn\\
P_{\nu_b\to\nub_a} &=& \Big|
\big(\He + \ge\big)_{ab}
\Big|^2 \fr{L^2}{\pvec^2} .
\eea
The survival probabilities can be found by summing over
possible transitions.
These equations generalize Eq.\ (2) of Ref.\ \cite{km-sb},
where attention was restricted to operators
of renormalizable dimension
and to situations where mass-induced oscillations
are negligible.

The above expressions can be used 
to search for the unconventional energy and direction dependences
associated with Lorentz and CPT violation.
By convention,
the standard inertial frame used to express and compare 
results for the coefficients for Lorentz violation
is a Sun-centered frame 
in which the $Z$ axis is aligned with the Earth's rotation axis
and the $X$ axis points towards the vernal equinox
\cite{tables,sunframe,labframe}.
A beam of neutrinos generated on the Earth
rotates about the $Z$ axis of the Sun-centered frame
once each sidereal day,
so direct analysis of neutrino oscillations in this frame
requires an expression for the beam direction $\phat$ 
as a function of time.

A more convenient approach adopts instead
a standard laboratory frame 
in which the $x$ axis points south, 
$y$ points east, 
and $z$ points vertically upwards 
\cite{labframe}.
Typically,
the source or the detector is chosen as the frame origin.
In this laboratory frame,
the beam direction $\phat\equiv\phat_{\rm lab}$ 
is a constant vector,
while the coefficients for Lorentz violation
vary in time instead.

The spherical-harmonic decomposition 
developed in the previous sections
is well suited to analyze this situation.
Taking $\ae$ as an example 
and momentarily suppressing the flavor indices,
we can write
\beq
\ae = \sum_{djm}  \pmag^{d-2}\, Y_{jm}(\phat_{\rm lab})
\aeff{d}{jm}{{\rm lab}} .
\eeq
Neglecting effects from the Earth's boost,
the coefficients $\aeff{d}{jm}{{\rm lab}}$ in the laboratory frame
are related by a time-dependent rotation
to the constant coefficients $\aeff{d}{jm}{}$
in the Sun-centered frame.
This rotation can be expressed in terms of Wigner matrices
$D^{(j)}_{mm'}(\al,\be,\ga)$,
where $\al$, $\be$, and $\ga$ are
Euler angles relating the two frames.
Denoting the sidereal rotation frequency as $\om_\oplus$
and the local sidereal time as $T_\oplus$,
we obtain
\bea
\aeff{d}{jm}{{\rm lab}}
&=& \sum_{m'} D^{(j)}_{mm'}(0,-\ch,-\om_\oplus T_\oplus) 
\aeff{d}{jm'}{}
\nn\\
&=& \sum_{m'} e^{im'\om_\oplus T_\oplus} 
d^{(j)}_{mm'}(-\ch) \aeff{d}{jm'}{} ,
\eea
where $\ch$ is the angle between
the Sun-frame $Z$ axis and the laboratory-frame $z$ axis,
corresponding in the northern hemisphere
to the colatitude of the laboratory.
The quantities $d^{(j)}_{mm'}$ are the `little' Wigner matrices.
Explicit expressions for the Wigner matrices
in the conventions used here
are given in Eqs.\ (134)-(136) of Ref.\ \cite{kmnonmin}.

Restoring the flavor indices,
we obtain the time-dependent expression  
\bea
\ae^{ab} &=& 
\hskip-6pt
\sum_{djmm'} \pmag^{d-2}
e^{im\om_\oplus T_\oplus} 
Y_{jm'}(\phat_{\rm lab})
d^{(j)}_{m'm}(-\ch) \aeff{d}{jm}{ab}.
\nn\\
\eea
Each term in this equation depends
on the dimension of the Lorentz-violating operator
through the power of $\pmag$,
on the local sidereal time
through a harmonic of the sidereal frequency,
on the direction $\phat_{\rm lab}$
through the spherical harmonics and the Wigner matrices,
and on the coefficients for Lorentz violation 
in the Sun-centered frame.
The equation thereby determines
the linear combinations of coefficients of Lorentz violation
that can be accessed by a given experiment
for each flavor transition.

The dependence on the direction $\phat_{\rm lab}$
of the beam with respect to the Earth
can be compactly encoded by defining the quantities
\beq
\N{s}{jm} 
\equiv \sum_{m'}
\syjm{s}{jm'}(\phat_{\rm lab})\,
d^{(j)}_{m'm}(-\ch) ,
\label{nfactors}
\eeq
which obey the reality conditions
\beq
\N{s}{jm}^* = (-1)^{m+s} \N{-s}{j(-m)}.
\label{nfactorreality}
\eeq
The factors 
$\N{s}{jm} = \N{s}{jm} (\phat_{\rm lab},\ch)
=\N{s}{jm} (\th,\ph,\ch)$
vary with the laboratory polar angles $(\th,\ph)$
and the colatitude angle $\ch$.
Applying the above line of reasoning also 
to the other effective coefficients appearing in $\hosc$
then yields 
\bea
\ae^{ab} &=& 
\sum_{dm} \pmag^{d-2}\, e^{im\om_\oplus T_\oplus} \,
\SBamp{a_\text{eff}}{d}{m}{} , 
\nn\\
\ce^{ab} &=& 
\sum_{dm}  \pmag^{d-2}\, e^{im\om_\oplus T_\oplus} \,
\SBamp{c_\text{eff}}{d}{m}{} , 
\nn\\
\ge^{ab} &=& 
\sum_{dm}  \pmag^{d-2}\, e^{im\om_\oplus T_\oplus} \,
\SBamp{g_\text{eff}}{d}{m}{} , 
\nn\\
\He^{ab} &=& 
\sum_{dm} \pmag^{d-2}\, e^{im\om_\oplus T_\oplus} \,
\SBamp{H_\text{eff}}{d}{m}{} ,
\eea
where
\bea
\SBamp{a_\text{eff}}{d}{m}{} &=& 
\sum_j \N{0}{jm}\, \aeff{d}{jm}{ab} ,
\nn\\
\SBamp{c_\text{eff}}{d}{m}{} &=& 
\sum_j \N{0}{jm}\, \ceff{d}{jm}{ab} ,
\nn\\
\SBamp{g_\text{eff}}{d}{m}{} &=& 
\sum_j \N{+1}{jm}\, \geff{d}{jm}{ab} ,
\nn\\
\SBamp{H_\text{eff}}{d}{m}{} &=& 
\sum_j \N{+1}{jm}\, \Heff{d}{jm}{ab} .
\eea
For completeness,
we also define an amplitude for mass
\beq
\SBamp{m_l}{2}{0}{} = \half (m_l^\dag m_l)^{ab} .
\eeq
The transition probabilities
can then be written compactly as
\bea
P_{\nu_b\to\nu_a} &=& \bigg|\sum_{dm}
L\, \pmag^{d-3} e^{im\om_\oplus T_\oplus}
\SBamp{\nu\nu}{d}{m}{} \bigg|^2 ,
\nn\\
P_{\nub_b\to\nub_a} &=& \bigg|\sum_{dm}
L\, \pmag^{d-3} e^{im\om_\oplus T_\oplus}
\SBamp{\nub\nub}{d}{m}{} \bigg|^2 ,
\nn\\
P_{\nub_b\to\nu_a} &=& \bigg|\sum_{dm}
L\, \pmag^{d-3} e^{im\om_\oplus T_\oplus}
\SBamp{\nub\nu}{d}{m}{} 
\bigg|^2 ,
\nn\\
P_{\nu_b\to\nub_a} &=& \bigg|\sum_{dm}
L\, \pmag^{d-3} e^{im\om_\oplus T_\oplus}
\SBamp{\nu\nub}{d}{m}{} 
\bigg|^2 ,
\label{probs}
\eea
where
\bea
\SBamp{\nu\nu}{d}{m}{} &=&
\SBamp{m_l}{d}{m}{} 
+ \SBamp{a_{\rm eff}}{d}{m}{}
- \SBamp{c_{\rm eff}}{d}{m}{} ,
\nn\\
\SBamp{\nub\nub}{d}{m}{} &=&
\SBamp{m_l}{d}{m}{} 
- \SBamp{a_{\rm eff}}{d}{m}{}
- \SBamp{c_{\rm eff}}{d}{m}{} ,
\nn\\
\SBamp{\nub\nu}{d}{m}{} &=&
\SBamp{H_{\rm eff}}{d}{m}{}
- \SBamp{g_{\rm eff}}{d}{m}{} ,
\nn\\
\SBamp{\nu\nub}{d}{m}{} &=&
\SBamp{H_{\rm eff}}{d}{m}{}
+ \SBamp{g_{\rm eff}}{d}{m}{} .
\eea

\begin{table*}
\renewcommand{\arraystretch}{1.6}
\begin{tabular}{ c @{\hspace{5pt}} |
@{\hspace{2pt}} c @{\hspace{2pt}} |
@{\hspace{2pt}} c @{\hspace{2pt}} |
@{\hspace{2pt}} c @{\hspace{2pt}} ||
@{\hspace{5pt}} c @{\hspace{5pt}} |
@{\hspace{4pt}} c @{\hspace{2pt}} |
@{\hspace{2pt}} c @{\hspace{2pt}} |
@{\hspace{2pt}} c @{\hspace{2pt}} ||
@{\hspace{5pt}} c @{\hspace{5pt}} |
@{\hspace{4pt}} c @{\hspace{2pt}} |
@{\hspace{2pt}} c @{\hspace{2pt}} |
@{\hspace{2pt}} c }
coefficient & LSND & MB & $\ol{\rm MB}$ &
coefficient & LSND & MB & $\ol{\rm MB}$ &
coefficient & LSND & MB & $\ol{\rm MB}$ \\
\hline\hline
$	(m_l^\dag m_l)^{e\mu}	$ & $	2.9	$ & $	3.0	$ & $	3.1	$ & $	\re\aeff{5}{00}{e\mu}	$ & $	8.1\times 10^{4}	$ & $	110	$ & $	26	$ & $	\re\ceff{6}{00}{e\mu}	$ & $	2.0\times 10^{6}	$ & $	320	$ & $	43	$ \\
$	\re\ceff{2}{10}{e\mu}	$ & $	15	$ & $	4.1	$ & $	4.3	$ & $	\re\aeff{5}{10}{e\mu}	$ & $	2.4\times 10^{5}	$ & $	89	$ & $	20	$ & $	\re\ceff{6}{10}{e\mu}	$ & $	6.0\times 10^{6}	$ & $	250	$ & $	33	$ \\
$	\re\ceff{2}{11}{e\mu}	$ & $	3.0	$ & $	3.1	$ & $	4.8	$ & $	\re\aeff{5}{11}{e\mu}	$ & $	4.8\times 10^{4}	$ & $	67	$ & $	22	$ & $	\re\ceff{6}{11}{e\mu}	$ & $	1.2\times 10^{6}	$ & $	190	$ & $	37	$ \\
$	\im\ceff{2}{11}{e\mu}	$ & $	3.0	$ & $	2.6	$ & $	5.1	$ & $	\im\aeff{5}{11}{e\mu}	$ & $	4.8\times 10^{4}	$ & $	56	$ & $	24	$ & $	\im\ceff{6}{11}{e\mu}	$ & $	1.2\times 10^{6}	$ & $	150	$ & $	39	$ \\
$		$ & $		$ & $		$ & $		$ & $	\re\aeff{5}{20}{e\mu}	$ & $	8.1\times 10^{4}	$ & $	150	$ & $	34	$ & $	\re\ceff{6}{20}{e\mu}	$ & $	2.0\times 10^{6}	$ & $	420	$ & $	56	$ \\
$	\re\aeff{3}{00}{e\mu}	$ & $	130	$ & $	15	$ & $	9.2	$ & $	\re\aeff{5}{21}{e\mu}	$ & $	1.1\times 10^{5}	$ & $	40	$ & $	13	$ & $	\re\ceff{6}{21}{e\mu}	$ & $	2.7\times 10^{6}	$ & $	110	$ & $	22	$ \\
$	\re\aeff{3}{10}{e\mu}	$ & $	380	$ & $	11	$ & $	7.1	$ & $	\im\aeff{5}{21}{e\mu}	$ & $	1.1\times 10^{5}	$ & $	33	$ & $	14	$ & $	\im\ceff{6}{21}{e\mu}	$ & $	2.7\times 10^{6}	$ & $	92	$ & $	24	$ \\
$	\re\aeff{3}{11}{e\mu}	$ & $	76	$ & $	8.7	$ & $	8.1	$ & $	\re\aeff{5}{22}{e\mu}	$ & $	4.3\times 10^{4}	$ & $	-	$ & $	-	$ & $	\re\ceff{6}{22}{e\mu}	$ & $	1.1\times 10^{6}	$ & $	-	$ & $	-	$ \\
$	\im\aeff{3}{11}{e\mu}	$ & $	76	$ & $	7.2	$ & $	8.5	$ & $	\im\aeff{5}{22}{e\mu}	$ & $	4.3\times 10^{4}	$ & $	-	$ & $	-	$ & $	\im\ceff{6}{22}{e\mu}	$ & $	1.1\times 10^{6}	$ & $	-	$ & $	-	$ \\
$	\re\aeff{3}{20}{e\mu}	$ & $	130	$ & $	20	$ & $	12	$ & $	\re\aeff{5}{30}{e\mu}	$ & $	1.1\times 10^{5}	$ & $	570	$ & $	130	$ & $	\re\ceff{6}{30}{e\mu}	$ & $	2.8\times 10^{6}	$ & $	1600	$ & $	210	$ \\
$	\re\aeff{3}{21}{e\mu}	$ & $	170	$ & $	5.2	$ & $	4.8	$ & $	\re\aeff{5}{31}{e\mu}	$ & $	6.3\times 10^{4}	$ & $	40	$ & $	13	$ & $	\re\ceff{6}{31}{e\mu}	$ & $	1.6\times 10^{6}	$ & $	110	$ & $	22	$ \\
$	\im\aeff{3}{21}{e\mu}	$ & $	170	$ & $	4.3	$ & $	5.1	$ & $	\im\aeff{5}{31}{e\mu}	$ & $	6.3\times 10^{4}	$ & $	33	$ & $	14	$ & $	\im\ceff{6}{31}{e\mu}	$ & $	1.6\times 10^{6}	$ & $	92	$ & $	23	$ \\
$	\re\aeff{3}{22}{e\mu}	$ & $	69	$ & $	-	$ & $	-	$ & $	\re\aeff{5}{32}{e\mu}	$ & $	8.4\times 10^{4}	$ & $	-	$ & $	-	$ & $	\re\ceff{6}{32}{e\mu}	$ & $	2.1\times 10^{6}	$ & $	-	$ & $	-	$ \\
$	\im\aeff{3}{22}{e\mu}	$ & $	69	$ & $	-	$ & $	-	$ & $	\im\aeff{5}{32}{e\mu}	$ & $	8.4\times 10^{4}	$ & $	-	$ & $	-	$ & $	\im\ceff{6}{32}{e\mu}	$ & $	2.1\times 10^{6}	$ & $	-	$ & $	-	$ \\
$		$ & $		$ & $		$ & $		$ & $	\re\aeff{5}{40}{e\mu}	$ & $	1.1\times 10^{5}	$ & $	110	$ & $	24	$ & $	\re\ceff{6}{40}{e\mu}	$ & $	2.8\times 10^{6}	$ & $	300	$ & $	40	$ \\
$	\re\ceff{4}{00}{e\mu}	$ & $	3200	$ & $	41	$ & $	15	$ & $	\re\aeff{5}{41}{e\mu}	$ & $	6.5\times 10^{4}	$ & $	72	$ & $	24	$ & $	\re\ceff{6}{41}{e\mu}	$ & $	1.6\times 10^{6}	$ & $	200	$ & $	40	$ \\
$	\re\ceff{4}{10}{e\mu}	$ & $	9600	$ & $	32	$ & $	12	$ & $	\im\aeff{5}{41}{e\mu}	$ & $	6.5\times 10^{4}	$ & $	59	$ & $	25	$ & $	\im\ceff{6}{41}{e\mu}	$ & $	1.6\times 10^{6}	$ & $	1600	$ & $	42	$ \\
$	\re\ceff{4}{11}{e\mu}	$ & $	1900	$ & $	24	$ & $	13	$ & $	\re\aeff{5}{42}{e\mu}	$ & $	6.8\times 10^{4}	$ & $	-	$ & $	-	$ & $	\re\ceff{6}{42}{e\mu}	$ & $	1.7\times 10^{6}	$ & $	-	$ & $	-	$ \\
$	\im\ceff{4}{11}{e\mu}	$ & $	1900	$ & $	20	$ & $	14	$ & $	\im\aeff{5}{42}{e\mu}	$ & $	6.8\times 10^{4}	$ & $	-	$ & $	-	$ & $	\im\ceff{6}{42}{e\mu}	$ & $	1.7\times 10^{6}	$ & $	-	$ & $	-	$ \\
$	\re\ceff{4}{20}{e\mu}	$ & $	3300	$ & $	55	$ & $	20	$ & $		$ & $		$ & $		$ & $		$ & $	\re\ceff{6}{50}{e\mu}	$ & $	2.0\times 10^{6}	$ & $	230	$ & $	31	$ \\
$	\re\ceff{4}{21}{e\mu}	$ & $	4400	$ & $	14	$ & $	8.0	$ & $		$ & $		$ & $		$ & $		$ & $	\re\ceff{6}{51}{e\mu}	$ & $	2.6\times 10^{6}	$ & $	770	$ & $	150	$ \\
$	\im\ceff{4}{21}{e\mu}	$ & $	4400	$ & $	12	$ & $	8.5	$ & $		$ & $		$ & $		$ & $		$ & $	\im\ceff{6}{51}{e\mu}	$ & $	2.6\times 10^{6}	$ & $	630	$ & $	160	$ \\
$	\re\ceff{4}{22}{e\mu}	$ & $	1700	$ & $	-	$ & $	-	$ & $		$ & $		$ & $		$ & $		$ & $	\re\ceff{6}{52}{e\mu}	$ & $	1.4\times 10^{6}	$ & $	-	$ & $	-	$ \\
$	\im\ceff{4}{22}{e\mu}	$ & $	1700	$ & $	-	$ & $	-	$ & $		$ & $		$ & $		$ & $		$ & $	\im\ceff{6}{52}{e\mu}	$ & $	1.4\times 10^{6}	$ & $	-	$ & $	-	$ \\
$	\re\ceff{4}{30}{e\mu}	$ & $	4500	$ & $	200	$ & $	76	$ & $		$ & $		$ & $		$ & $		$ & $		$ & $		$ & $		$ & $		$ \\
$	\re\ceff{4}{31}{e\mu}	$ & $	2500	$ & $	14	$ & $	8.0	$ & $		$ & $		$ & $		$ & $		$ & $		$ & $		$ & $		$ & $		$ \\
$	\im\ceff{4}{31}{e\mu}	$ & $	2500	$ & $	12	$ & $	8.4	$ & $		$ & $		$ & $		$ & $		$ & $		$ & $		$ & $		$ & $		$ \\
$	\re\ceff{4}{32}{e\mu}	$ & $	3400	$ & $	-	$ & $	-	$ & $		$ & $		$ & $		$ & $		$ & $		$ & $		$ & $		$ & $		$ \\
$	\im\ceff{4}{32}{e\mu}	$ & $	3400	$ & $	-	$ & $	-	$ & $		$ & $		$ & $		$ & $		$ & $		$ & $		$ & $		$ & $		$ \\
\end{tabular}
\caption{\label{sb}
Maximal attained sensitivities 
on the modulus of coefficients with $d\leq 6$
from LSND antineutrinos at $1\si$
\cite{lsndlv}
and from MiniBooNE neutrinos (MB)
and antineutrinos ($\ol{\rm MB}$) at $2\si$
\cite{MiniBooNElv}.
The units are $10^{-20}$ GeV$^{4-d}$.}
\end{table*}

\begin{table}
\renewcommand{\arraystretch}{1.4}
\begin{tabular}{c|c|c|c}
coefficient & LSND & MB & $\ol{\rm MB}$ \\
\hline\hline
$	(m_l^\dag m_l)^{e\mu}	$ & $	2.9	$ & $	3	$ & $	3.1	$ \\
$	\afc{3}{e\mu}	$ & $	36	$ & $	4.2	$ & $	2.6	$ \\
$	\cfc{4}{e\mu}	$ & $	910	$ & $	12	$ & $	4.3	$ \\
$	\afc{5}{e\mu}	$ & $	2.3\times 10^4	$ & $	32	$ & $	7.2	$ \\
$	\cfc{6}{e\mu}	$ & $	5.7\times 10^5	$ & $	90	$ & $	12	$ \\
$	\afc{7}{e\mu}	$ & $	1.4\times 10^7	$ & $	250	$ & $	20	$ \\
$	\cfc{8}{e\mu}	$ & $	3.6\times 10^8	$ & $	690	$ & $	33	$ \\
$	\afc{9}{e\mu}	$ & $	8.9\times 10^9	$ & $	1900	$ & $	56	$ \\
$	\cfc{10}{e\mu}	$ & $	2.2\times 10^{11}	$ & $	5400	$ & $	93	$ \\
\end{tabular}
\caption{\label{sbiso}
Maximal attained sensitivities 
on the modulus of isotropic coefficients with $d\leq 10$
from LSND antineutrinos at $1\si$
\cite{lsndlv}
and from MiniBooNE neutrinos (MB)
and antineutrinos ($\ol{\rm MB}$) at $2\si$
\cite{MiniBooNElv}.
The units are $10^{-20}$ GeV$^{4-d}$.}
\end{table}

As an explicit example,
consider the study of Lorentz violation 
by the LSND collaboration 
\cite{lsndlv}.
This yielded constraints on amplitudes 
for $\nub_\mu\to\nub_e$ oscillations
with sidereal harmonics $m\leq 2$.
We discuss here the reported quadratic bound 
on the squares of the amplitudes,
which in current terminology and for fixed dimension $d$
can be written as
\bea
&& \hskip -20pt
\pmag^{2(d-3)}
\bigg(
\Big[\SBamp{\nub\nub}{d}{0}{\rm LSND}\Big]^2
+ 2 \SBamp{\nub\nub}{d}{1}{\rm LSND} 
\SBamp{\nub\nub}{d}{-1}{\rm LSND}
\nn\\
&&
\hskip 30pt
+ 2 \SBamp{\nub\nub}{d}{2}{\rm LSND} 
\SBamp{\nub\nub}{d}{-2}{\rm LSND}
\bigg)
\nn\\
&&
\hskip 50pt
= 10.5\pm2.4\pm1.4 \times 10^{-19}\ {\rm GeV} .
\eea
The LSND analysis assumed that the relevant terms 
in the effective hamiltonian are real,
so we must impose the conditions
$\aeff{d}{j(-m)}{e\mu} = (-1)^m {\aeff{d}{jm}{e\mu}}^*$
and
$\ceff{d}{j(-m)}{e\mu} = (-1)^m {\ceff{d}{jm}{e\mu}}^*$.
Combining the $1\si$ errors in quadrature
gives a $1\si$ absolute bound of $13.3\times 10^{-19}$ GeV.
The relevant polar angles are
$\th\simeq 99.0^\circ$,
$\ph\simeq 82.6^\circ$,
and the colatitude is $\ch\simeq 54.1^\circ$,
while the neutrino energy is in the neighborhood of 20-60 MeV.
Taking $\pmag = 40$ MeV as the representative energy,
we estimate the maximal sensitivity achieved 
to individual coefficients for Lorentz violation
by setting all but one to zero
and considering separately any real and imaginary parts.
The results of this calculation
are displayed in Tables \ref{sb} and \ref{sbiso}.

Additional limits can be obtained 
from the recent study of Lorentz violation
by the MiniBooNE collaboration 
\cite{MiniBooNElv}.
This analysis placed bounds on amplitudes 
for both $\nu_\mu\to\nu_e$ and
$\nub_\mu\to\nub_e$ transitions
with sidereal harmonics $m=0$ and $m=1$.
For fixed dimension $d$,
the experiment gives $2\si$ limits
for $\nu\to\nu$ oscillations of 
\bea
\pmag^{d-3} \Big| 
\SBamp{\nu\nu}{d}{0}{\rm MB} \Big|
&<& 4.2\times 10^{-20}\ {\rm GeV} ,
\nn\\
\pmag^{d-3} \Big| 
\SBamp{\nu\nu}{d}{1}{\rm MB}
+\SBamp{\nu\nu}{d}{-1}{\rm MB} \Big|
&<& 4.0\times 10^{-20}\ {\rm GeV} ,
\nn\\
\pmag^{d-3} \Big| i \SBamp{\nu\nu}{d}{1}{\rm MB}
- i \SBamp{\nu\nu}{d}{-1}{\rm MB} \Big|
&<& 3.3\times 10^{-20}\ {\rm GeV} .
\nn\\
\eea
For antineutrinos, 
the $2\si$ limits are 
\bea
\pmag^{d-3} \Big| 
\SBamp{\nub\nub}{d}{0}{\rm MB} \Big|
&<& 2.6\times 10^{-20}\ {\rm GeV} ,
\nn\\
\pmag^{d-3} \Big| 
\SBamp{\nub\nub}{d}{1}{\rm MB}
+\SBamp{\nub\nub}{d}{-1}{\rm MB}\Big|
&<& 3.7\times 10^{-20}\ {\rm GeV} ,
\nn\\
\pmag^{d-3} \Big| 
i \SBamp{\nub\nub}{d}{1}{\rm MB}
- i \SBamp{\nub\nub}{d}{-1}{\rm MB} \Big|
&<& 3.9\times 10^{-20}\ {\rm GeV} .
\nn\\
\eea
The MiniBooNE analysis also assumes 
real Lorentz-violating terms in $\hosc$, 
which implies we must impose conditions
on the coefficients as before.
The polar angles for the beam direction 
are $\th\simeq 89.8^\circ, \ph\simeq 180^\circ$,
and the colatitude is $\ch\simeq 48.2^\circ$.
The average neutrino energy is $0.36$ GeV,
while the average antineutrino energy is $0.60$ GeV.
Proceeding as above,
we extract estimated maximal sensitivities
to individual coefficients for Lorentz violation.
The neutrino and antineutrino results
obtained in this way
are also compiled in Tables \ref{sb} and \ref{sbiso}.

The two tables contain many first limits on neutrino
coefficients for Lorentz violation
with $d=2$ and with $5\leq d\leq 10$.
Several options exist for achieving 
improvements and extensions of these results.
One possibility would be to reanalyze
the LSND and MiniBooNE data for higher harmonics
according to Eq.\ \rf{probs}.
This would generate first constraints
on many additional coefficients.
Another possibility is to use existing data
from other short-baseline experiments.
For example,
data from the MINOS near detector
have already been used to constrain Lorentz violation
within the short-baseline approximation
\cite{minoslv1},
with sensitivities at the level of $10^{-20}$ GeV 
to two coefficients $a_L^{(3)\mu}$ for CPT-odd violation 
and at $10^{-21}$
to seven coefficients $c_L^{(4)\mn}$ for CPT-even violation.
The same data could be analyzed for harmonics
to yield measurements of coefficients 
with other values of $d$. 
Future short-baseline experiments
such as the recent DAE$\de$ALUS proposal 
\cite{DAEdALUS}
would also offer interesting possibilities
for searching for Lorentz violation
along these lines,
as would analyses of data from reactor experiments 
such as Double Chooz
\cite{doublechooz},
Daya Bay
\cite{dayabay},
and RENO
\cite{reno}.

\subsection{Perturbative Lorentz violation}
\label{Perturbative Lorentz violation}

The short-baseline approximation is appropriate
for null experiments 
when little or no neutrino oscillation is detected.
However, 
many experiments observe significant oscillations.
In this case,
a general theoretical analysis is challenging.
One approach is via model building,
which in the context of Lorentz violation
involves designing special Lorentz-violating models 
that can qualitatively reproduce 
the observed global features of oscillations,
perhaps including also one or more of the neutrino anomalies, 
using only a few parameters
\cite{bicycle,tandem,bmw,puma,blmw}.
Another strategy 
assumes oscillations are primarily due to the mass matrix
and treats Lorentz violation as a small perturbation,
seeking to identify or constrain
small deviations from the conventional picture 
that may indicate Lorentz and CPT violation.
For the renormalizable limit of the SME,
this approach is presented 
in Ref.\ \cite{dkm}.

In the present subsection,
we generalize the perturbative treatment 
to include Lorentz-violating operators of arbitrary dimension.
Our primary focus is on beam experiments,
in which the signal dependence on propagation direction 
typically manifests as sidereal time dependence,
but the basic analysis is applicable to other situations
such as the azimuthal signal dependence 
used in the recent search for Lorentz violation
by the IceCube collaboration
\cite{IceCube}.
For example,
future searches with IceCube and Super-Kamiokande
\cite{superk} 
could adopt the methods presented here 
to extract competitive limits on a variety
of SME coefficients for Lorentz and CPT violation.
Other possible applications include
searches for anomalous annual variations 
in solar-neutrino oscillations
beyond those due to the Earth's orbital eccentricity. 

The perturbative approach is based on
time-dependent perturbation theory,
which expands the time-evolution operator
$S(t)$ in powers of the Lorentz-violating part $\de h$
of the effective hamiltonian \rf{hresult}.
This requires first specifying the unperturbed system,
which implies adopting values 
for the conventional mass-squared differences 
and mixing angles.
To enable a straightforward match to previous results,
we adopt here the notation of Ref.\ \cite{dkm}
with upper-case indices $AB$ indicating components 
of the full $6\times 6$ effective hamiltonian,
and lower-case unbarred and barred indices  
$ab$, $a\bar b$, $\bar a b$, $\bar a \bar b$
indicating components of the $3\times 3$ blocks.
As before,
unprimed indices refer to the flavor basis
while primed indices indicate the diagonal energy basis.
In this notation,
we write the conventional energy eigenvalues as $E_{A'}$,
with CPT invariance guaranteeing the condition 
$E_{a'} = E_{\bar a'}$.
The $6\times 6$ Lorentz-invariant effective hamiltonian
$(\heff)_0{}_{AB}$ of Eq.\ \rf{effliham}
is diagonalized by the $6\times 6$ mixing matrix 
\beq
U_{A'B} = \begin{pmatrix}
U_{a'b} & 0 \\ 0 & U_{\bar a'\bar b} \end{pmatrix} ,
\eeq
where $U_{a'b}^{} = U_{\bar a'\bar b}^*$
is the familiar $3\times 3$ neutrino mixing matrix.
Note that the unperturbed system could be defined to include,
for example,
the effects of matter for neutrinos propagating through the Earth.
These would alter the energies $E_{A'}$ 
and the mixing matrix $U_{A'B}$.
Since the interactions of neutrinos with matter
are described by coefficients $(a_L^{(3)t})_{ab}$ 
for CPT-odd Lorentz violation
\cite{km},
the oscillations of neutrinos and antineutrinos in matter
involve different energy spectra and mixing matrices.

The perturbation calculation generates
an expansion of the oscillation probabilities
in powers of the Lorentz-violating perturbation \rf{efflvham}
with components $\de h_{AB}$.
The expansion is
\cite{dkm}
\beq
P_{\nu_B\to\nu_A}
=P^{(0)}_{\nu_B\to\nu_A}
+P^{(1)}_{\nu_B\to\nu_A}
+P^{(2)}_{\nu_B\to\nu_A} + \ldots ,
\eeq
where 
\beq
P^{(0)}_{\nu_B\to\nu_A} = \big|S^{(0)}_{AB}\big|^2,
\quad
S^{(0)}_{AB} = \sum_{A'} U_{A'A}^* U_{A'B}^{} e^{-iE_{A'} t} 
\eeq
are the unperturbed oscillation probability 
and time-evolution operator.
The first- and second-order perturbations take the form 
\bea
P^{(1)}_{\nu_B\to\nu_A} &=&
2t\, \im\Big({S^{(0)}_{AB}}^* \H{1}{AB}\Big) ,
\nn\\
P^{(2)}_{\nu_B\to\nu_A} &=&
-t^2\, \re\Big({S^{(0)}_{AB}}^* \H{2}{AB}\Big)
+t^2\, \big|\H{1}{AB}\big|^2 
\qquad
\eea
with 
\bea
\H{1}{AB} &=& \sum_{CD} \M{1}{AB}{CD} \de h_{CD} , 
\nn\\
\H{2}{AB} &=& \sum_{CDEF} \M{2}{AB}{CDEF} \de h_{CD}\de h_{EF} ,
\eea
where $\M{1}{AB}{CD}$ and $\M{2}{AB}{CDEF}$
are factors depending only on unperturbed quantities
and the experimental setup.
They are given explicitly 
by Eqs.\ (32) and (34) of Ref.\ \cite{dkm}.
Note that the first-order probability
$P^{(1)}_{\nu_B\to\nu_A}$ vanishes 
when the conventional transition is zero, 
$S^{(0)}_{AB}=0$.
Under these conditions,
the second-order probability $P^{(2)}_{\nu_B\to\nu_A}$
governs the dominant Lorentz-violating effects.
Examples of this arise when the only nonzero coefficients
are of Majorana type 
and hence lie in the off-diagonal blocks of $\de h$,
causing mixing between neutrinos and antineutrinos.

\subsubsection{First-order perturbation}

The first-order perturbations are governed by $\H{1}{AB}$.
Since the unperturbed system 
contains no neutrino-antineutrino mixing,
only $\H{1}{ab}$ and its CPT conjugate $\H{1}{\bar a\bar b}$ 
contribute to the first-order probabilities.
Moreover,
only $\de h_{ab}$ enters the expression for $\H{1}{ab}$ 
because $U_{A'A}$ is block diagonal.
Analogous results hold for the other blocks of $\H{1}{AB}$.
Note that all four blocks appear 
in the second-order probabilities. 

We can construct explicit expressions for the four blocks
of $\H{1}{AB}$ in terms of spherical coefficients.
For example,
for the neutrino-neutrino block we obtain
\bea
\H{1}{ab} &=& 
\sum_{ce} \M{1}{ab}{ce} \de h_{ce}
\nn\\
&&
\hskip -30pt
= \sum_{\scriptstyle{ce}\atop\scriptstyle{djm}} \M{1}{ab}{ce}
\pmag^{d-3} Y_{jm}(\phat) 
\big[\aeff{d}{jm}{ce} - \ceff{d}{jm}{ce}\big].
\nn\\
\eea
To analyze a given beam experiment,
it is convenient to adopt as before 
the standard laboratory frame 
with the $x$ axis pointing south, 
$y$ pointing east, 
and $z$ pointing vertically upwards 
\cite{labframe}.
In this frame,
the beam direction is constant
and the coefficients for Lorentz violation
acquire time dependence due to the rotation of the Earth.
To express the result in terms of
coefficients in the canonical Sun-centered frame,
we adopt the strategy employed in the short-baseline case
in the previous subsection.
The factors $\M{1}{ab}{ce}$ and 
the beam direction $\phat_{\rm lab}$ 
are determined in the laboratory frame.
The time dependence is revealed
by using the Wigner matrices 
to rotate the coefficients for Lorentz violation 
to the Sun-centered frame.

Implementing this procedure for all four blocks of $\H{1}{AB}$ 
gives 
\bea
\H{1}{ab} &=&
\sum_{dm} \pmag^{d-3} e^{im\om_\oplus T_\oplus}
\Big( \LBamp{a}{d}{m}{ab}
- \LBamp{c}{d}{m}{ab} \Big) ,
\nn\\
\H{1}{\bar a\bar b} &=&
\sum_{dm} \pmag^{d-3} e^{im\om_\oplus T_\oplus}
\Big( -\LBamp{a}{d}{m}{\bar a\bar b}
- \LBamp{c}{d}{m}{\bar a\bar b} \Big) ,
\nn\\
\H{1}{a\bar b} &=&
\sum_{dm} \pmag^{d-3} e^{im\om_\oplus T_\oplus}
\Big(\LBamp{H}{d}{m}{a\bar b}
- \LBamp{g}{d}{m}{a\bar b} \Big) ,
\nn\\
\H{1}{\bar ab} &=&
\sum_{dm} \pmag^{d-3} e^{im\om_\oplus T_\oplus}
\Big(\LBamp{H}{d}{m}{\bar ab}
+ \LBamp{g}{d}{m}{\bar ab} \Big) .
\qquad 
\label{foprob}
\eea
The various amplitudes appearing in these equations 
take the compact forms 
\bea
\LBamp{a}{d}{m}{ab} &=& 
\sum_{cej} \N{0}{jm}\,
\M{1}{ab}{ce}\, \aeff{d}{jm}{ce} ,
\nn\\
\LBamp{c}{d}{m}{ab} &=& 
\sum_{cej} \N{0}{jm}\,
\M{1}{ab}{ce}\, \ceff{d}{jm}{ce} ,
\nn\\
\LBamp{a}{d}{m}{\bar a\bar b} &=& 
\sum_{cej} \N{0}{jm}\,
\M{1}{\bar a\bar b}{\bar c\bar e}\, \aeff{d}{jm}{ec} ,
\nn\\
\LBamp{c}{d}{m}{\bar a\bar b} &=& 
\sum_{cej} \N{0}{jm}\,
\M{1}{\bar a\bar b}{\bar c\bar e}\, \ceff{d}{jm}{ec} ,
\nn\\
\LBamp{g}{d}{m}{a\bar b} &=& 
\sum_{cej} \N{+1}{jm}\,
\M{1}{a\bar b}{c\bar e}\, \geff{d}{jm}{ce} ,
\nn\\
\LBamp{H}{d}{m}{a\bar b} &=& 
\sum_{cej} \N{+1}{jm}\,
\M{1}{a\bar b}{c\bar e}\, \Heff{d}{jm}{ce} ,
\nn\\
\LBamp{g}{d}{m}{\bar ab} &=& 
\sum_{cej} \N{-1}{jm}\,
\M{1}{\bar ab}{\bar ce}\, (-1)^m \big[\geff{d}{j(-m)}{ce}\big]^* ,
\nn\\
\LBamp{H}{d}{m}{\bar ab} &=& 
\sum_{cej} \N{-1}{jm}\,
\M{1}{\bar ab}{\bar ce}\, (-1)^m \big[\Heff{d}{j(-m)}{ce}\big]^* .
\nn\\
\label{foamps}
\eea
In these expressions,
the factors $\N{s}{jm}$
are defined by Eq.\ \rf{nfactors},
as before.

The above results can be directly applied 
to searches for Lorentz violation with beam experiments.
For example,
consider a study of Lorentz violation  
with $\nu_\mu\to\nu_e$ or $\nub_\mu\to\nub_e$ oscillations
in a long-baseline experiment.
The explicit numerical values 
of the relevant factors 
$\M{1}{e\mu}{ab}$ and 
$\M{1}{\bar e\bar\mu}{\bar a\bar b}$
for a variety of long baseline experiments
are given in Table I of Ref.\ \cite{dkm}.
For any given experiment,
the beam polar angles and the colatitude 
can be used to determine the direction factors $\N{s}{jm}$
via Eq.\ \rf{nfactors}.
Substituting the results into Eq.\ \rf{foamps}
gives the amplitudes 
$\LBamp{a}{d}{m}{e\mu}$,
$\LBamp{c}{d}{m}{e\mu}$,
$\LBamp{a}{d}{m}{\bar e\bar \mu}$, and
$\LBamp{c}{d}{m}{\bar e\bar \mu}$
relevant for the first order probabilities
in terms of the effective spherical coefficients
$\aeff{d}{jm}{ab}$ and $\ceff{d}{jm}{ab}$.
For a chosen mass dimension $d$ and a specific 
sidereal harmonic $m$,
the contribution to the first-order probability 
for $\nu_\mu\to\nu_e$ mixing
is then given by 
\bea
P^{(1)}_{\nu_\mu\to\nu_e} &=&
2L\, \pmag^{d-3}
\nn\\
&&
\hskip -40pt
\times \Big(
\re\big[{S^{(0)}_{e\mu}}^*\LBamp{a}{d}{m}{e\mu}
-{S^{(0)}_{e\mu}}^*\LBamp{c}{d}{m}{e\mu}\big]
\sin m\om_\oplus T_\oplus
\nn\\
&&
\hskip -30pt
+ \im\Big[{S^{(0)}_{e\mu}}^*\LBamp{a}{d}{m}{e\mu}
-{S^{(0)}_{e\mu}}^*\LBamp{c}{d}{m}{e\mu}\Big]
\cos m\om_\oplus T_\oplus 
\Big).
\nn\\
\label{egprob}
\eea
The corresponding result for $\nub_\mu\to\nub_e$ mixing 
is immediately obtained by changing the sign
of the CPT-odd amplitude $\LBamp{a}{d}{m}{e\mu}$
and replacing $e\mu$ with $\bar e\bar\mu$
in the above expression.

The result \rf{egprob} shows that the given long-baseline experiment
can use measurements of sidereal variations 
at various harmonics $m$
to place constraints on linear combinations 
of the real and imaginary parts
of the coefficients
$\aeff{d}{jm}{ab}$ and $\ceff{d}{jm}{ab}$
for Lorentz violation.
If a definite signal is found,
a variety of independent experimental configurations
may be required to identify 
which individual coefficients are nonzero.
In the absence of a signal,
a useful approach is to take one coefficient nonzero at a time.
The experimental bound can then be interpreted
as a maximal attained sensitivity to each coefficient in turn,
yielding tables analogous to Tables \ref{sb} and \ref{sbiso}
obtained above for the short-baseline approximation. 
Reporting the results in this way
facilitates the direct comparison of different experiments
and provides an effective guide to 
constraints on theoretical model building
\cite{tables}.

\subsubsection{Second-order perturbation}

The second-order probability becomes important 
when conventional oscillations are negligible.
One example is mixing between neutrinos and antineutrinos.
Also,
for certain energy and baseline combinations,
$\nu\to\nu$ and $\nub\to\nub$ transitions 
may involve suppressed first-order effects.

For neutrino-antineutrino mixing,
the first-order probabilities vanish.
The second-order probabilities 
take the simple form
\bea
P^{(2)}_{\nu_b\to\nub_a} &=&
t^2\, \big|\H{1}{\bar ab}\big|^2 , 
\quad
P^{(2)}_{\nub_b\to\nu_a} =
t^2\, \big|\H{1}{a\bar b}\big|^2 .
\eea
Expressions for 
$\H{1}{\bar ab}$ and $\H{1}{a\bar b}$
are given in Eq.\ \rf{foprob},
with the corresponding amplitudes
expanded in terms of the effective spherical coefficients
$\geff{d}{jm}{ab}$ and $\Heff{d}{jm}{ab}$
in Eq.\ \rf{foamps}.
These results can be applied to searches for Lorentz violation
in neutrino-antineutrino mixing,
using a procedure 
similar to that outlined in the previous subsection
for long-baseline experiments 
on neutrino-neutrino and antineutrino-antineutrino mixings.
Since neutrino-antineutrino oscillations cannot arise from
the coefficients $\aeff{d}{jm}{ab}$ and $\ceff{d}{jm}{ab}$,
direct studies of this mixing channel
can be expected to yield clean constraints on
the coefficients $\geff{d}{jm}{ab}$ and $\Heff{d}{jm}{ab}$.

In principle,
another possibility for achieving sensitivity
to the coefficients $\geff{d}{jm}{ab}$ and $\Heff{d}{jm}{ab}$
is to consider instead second-order effects in 
neutrino-neutrino and antineutrino-antineutrino oscillations.
The corresponding probabilities 
\bea
P^{(2)}_{\nu_b\to\nu_a} &=&
-t^2\, \re\Big({S^{(0)}_{ab}}^* \H{2}{ab}\Big)
+t^2\, \big|\H{1}{ab}\big|^2 ,
\nn\\
P^{(2)}_{\nub_b\to\nub_a} &=&
-t^2\, \re\Big({S^{(0)}_{\bar a\bar b}}^* \H{2}{\bar a\bar b}\Big)
+t^2\, \big|\H{1}{\bar a\bar b}\big|^2 
\eea
depend on the second-order combinations
\bea
\H{2}{ab} 
\hskip -5pt
&=& 
\hskip -5pt
\sum_{cefg}\Big[
\M{2}{ab}{cefg} \de h_{ce} \de h_{fg}
+ \M{2}{ab}{c\bar e\bar fg}
\de h_{c\bar e} \de h_{\bar fg} \Big] , 
\nn\\ 
\H{2}{\bar a\bar b} 
\hskip -5pt
&=&
\hskip -5pt
\sum_{cefg} \Big[
\M{2}{\bar a\bar b}{\bar c\bar e\bar f\bar g}
\de h_{\bar c\bar e} \de h_{\bar f\bar g}
+ \M{2}{\bar a\bar b}{\bar cef\bar g}
\de h_{\bar ce} \de h_{f\bar g} \Big] .
\nn\\
\label{qcomb}
\eea
This line of attack is more complicated
because the combinations \rf{qcomb}
are quadratic in coefficients for Lorentz violation
and because they contain 
all four $3\times 3$ blocks of $\de h$.
Although the diagonal blocks $\de h_{ab}$
and $\de h_{\bar a\bar b}$ also
contribute to the first-order probabilities,
which implies their appearance in Eq.\ \rf{qcomb}
represents an effect only at subleading order,
the combinations of coefficients appearing
in the probabilities may be distinct.
As a consequence,
the contributions of the diagonal blocks
cannot be omitted in an exact treatment.

Nonetheless,
to gain intuition about the leading-order contributions 
to Eq.\ \rf{qcomb} arising from 
$\geff{d}{jm}{}$ and $\Heff{d}{jm}{}$,
we can choose to focus on the off-diagonal blocks
$\de h_{a\bar b}$, $\de h_{\bar ab}$
while disregarding the contributions 
from $\de h_{ab}$, $\de h_{\bar a \bar b}$.
For $\H{2}{ab}$,
we thereby obtain
\begin{widetext}
\bea
\H{2}{ab} &=& 
\sum_{dd'mm'}
\pmag^{d+d'-6} e^{i(m+m')\om_\oplus T_\oplus}
\Big[
\LBAMP{HH}{dd'}{mm'}{ab}
-\LBAMP{gg}{dd'}{mm'}{ab}
+\LBAMP{Hg}{dd'}{mm'}{ab}
-\LBAMP{gH}{dd'}{mm'}{ab}\Big] ,
\nn\\
\LBAMP{HH}{dd'}{mm'}{ab} &=&
\sum_{cefg}\sum_{jj'} \M{2}{ab}{c\bar e\bar fg}\,
\N{+1}{jm}\, \N{-1}{j'm'} (-1)^{m'}\,
\Heff{d}{jm}{ce} \big[\Heff{d'}{j'(-m')}{fg}\big]^* ,
\nn\\
\LBAMP{gg}{dd'}{mm'}{ab} &=&
\sum_{cefg}\sum_{jj'} \M{2}{ab}{c\bar e\bar fg}\,
\N{+1}{jm}\, \N{-1}{j'm'} (-1)^{m'}\,
\geff{d}{jm}{ce} \big[\geff{d'}{j'(-m')}{fg}\big]^*  ,
\nn\\
\LBAMP{Hg}{dd'}{mm'}{ab} &=&
\sum_{cefg}\sum_{jj'} \M{2}{ab}{c\bar e\bar fg}\,
\N{+1}{jm}\, \N{-1}{j'm'} (-1)^{m'}\,
\Heff{d}{jm}{ce} \big[\geff{d'}{j'(-m')}{fg}\big]^*  ,
\nn\\
\LBAMP{gH}{dd'}{mm'}{ab} &=&
\sum_{cefg}\sum_{jj'} \M{2}{ab}{c\bar e\bar fg}\,
\N{+1}{jm}\, \N{-1}{j'm'} (-1)^{m'}\,
\geff{d}{jm}{ce} \big[\Heff{d'}{j'(-m')}{fg}\big]^*  .
\eea
\newpage
\end{widetext}
These equations explicitly confirm 
that data from neutrino-neutrino mixing
also contain information about the coefficients 
$\geff{d}{jm}{ab}$ and $\Heff{d}{jm}{ab}$.
An exact treatment would generate constraints
on complicated quadratic combinations 
of all four sets of coefficients 
$\aeff{d}{jm}{ab}$, $\ceff{d}{jm}{ab}$,
$\geff{d}{jm}{ab}$, and $\Heff{d}{jm}{ab}$.

In the absence of a definitive signal in the data,
an interesting approach 
is to take only one coefficient for Lorentz violation
to be nonzero at a time.
Sidereal variations can be used 
to determine maximal sensitivities
to individual cartesian coefficients for Lorentz violation
\cite{dkm}.
In the present context,
considering a single effective spherical coefficient
implies working with fixed values of $d$, $j$, $m$.
The only terms that survive in the above expressions
then have a unique value of $m=-m'$,
and so no sidereal variations appear.
Moreover,
the cross terms between
$\geff{d}{jm}{ab}$ and $\Heff{d}{jm}{ab}$ vanish.
The exact form of $\H{2}{ab}$
therefore takes the simple form 
\beq
\H{2}{ab} =
\begin{cases}
\pmag^{2(d-3)} \LBAMP{HH}{dd}{m(-m)}{ab},
& d {\rm ~odd} ,\\
-\pmag^{2(d-3)} \LBAMP{gg}{dd}{m(-m)}{ab}, 
& d {\rm ~even} .
\end{cases}
\eeq
Note that the contributions involving odd $d$ come from
operators that are CPT even,
while those involving even $d$ come from operators that are CPT odd.
Also, 
only even powers of the energy appear.
The unconventional energy dependence offers 
a potential experimental handle 
for studies of $\geff{d}{jm}{ab}$ and $\Heff{d}{jm}{ab}$.
Another option arises for null experiments 
producing an absolute bound on the total oscillation,
which can use the above simple expression
to extract maximal attained sensitivities
to individual components 
$\geff{d}{jm}{ab}$ and $\Heff{d}{jm}{ab}$.

We remark in passing
that other types of experiments may also yield
direct sensitivity to $\geff{d}{jm}{ab}$ and $\Heff{d}{jm}{ab}$.
For example,
neutrinoless double-beta decay
can arise in the presence of Majorana-like couplings 
that mix neutrinos and antineutrinos,
and so studies of this process
could provide competitive measurements 
for Lorentz violation mediated by Majorana operators.
A treatment of these and related possibilities
would be of definite interest 
but lies outside our present scope.

\subsection{Two-flavor maximal mixing}
\label{Two-flavor maximal mixing}

The special case of two-flavor mixing 
with the maximal mixing angle of $45^\circ$ 
offers a simple example with phenomenological relevance. 
In the \msm,
the smaller of the two mass-squared differences 
becomes less important at higher energies,
so oscillations become dominated 
by the larger mass-squared difference.
This situation applies to the energy regime 
relevant for most experiments 
with atmospheric and accelerator neutrinos.
Observations indicate the \msm\ angle $\th_{13}$ is small
and $\th_{23}\simeq 45^\circ$,
which implies oscillations between $\nu_\mu$ and $\nu_\ta$
can be well approximated by the two-flavor maximal-mixing scenario.

In this limit,
the only nonzero first-order transition probabilities are
$P^{(1)}_{\nu_\mu\to\nu_\ta} = P^{(1)}_{\nu_\ta\to\nu_\mu}$,
given by
\bea
\hskip -10pt
P^{(1)}_{\nu_\mu\to\nu_\ta} 
&=& L\, \sin(\De m^2L/2E)\, \re(\de h_{\mu\ta})
\nn\\
&&
\hskip -40pt
= \half L\, \sin(\De m^2L/2E)
\nn\\
&&
\hskip -30pt
\times
\sum_{djm} \pmag^{d-3}\, Y_{jm}(\phat)\,
\Big[\aeff{d}{jm}{(\mu\ta)} - \ceff{d}{jm}{(\mu\ta)}\Big] ,
\quad 
\eea
where the symmetrized combinations are defined by
\bea
\aeff{d}{jm}{(\mu\ta)} &=& 
\aeff{d}{jm}{\mu\ta} + \aeff{d}{jm}{\ta\mu},
\nn\\
\ceff{d}{jm}{(\mu\ta)} &=&
\ceff{d}{jm}{\mu\ta} + \ceff{d}{jm}{\ta\mu}
\eea
and obey the conjugation relations
\bea
\big[\aeff{d}{jm}{(\mu\ta)}\big]^* &=& 
(-1)^m \aeff{d}{j(-m)}{(\mu\ta)},
\nn\\
\big[\ceff{d}{jm}{(\mu\ta)}\big]^* &=& 
(-1)^m \ceff{d}{j(-m)}{(\mu\ta)}.
\eea
The survival probabilities are given by 
$P^{(1)}_{\nu_\mu\to\nu_\mu}
= P^{(1)}_{\nu_\ta\to\nu_\ta} 
= 1-P^{(1)}_{\nu_\mu\to\nu_\ta}$.
For antineutrinos,
the transition probability 
$P^{(1)}_{\nub_\mu\to\nub_\ta}$
and the survival probabilities
$P^{(1)}_{\nub_\mu\to\nub_\mu}
= P^{(1)}_{\nub_\ta\to\nub_\ta}$
are obtained by changing the sign of the
coefficient $\aeff{d}{jm}{(\mu\ta)}$.

In terms of coefficients in the Sun-centered frame,
the transition probability acquires a sidereal time dependence,
taking the form 
\begin{widetext}
\bea
P^{(1)}_{\nu_\mu\to\nu_\ta}
&=& \half L\, \sin(\De m^2L/2E)\, \sum_{dj} \pmag^{d-3}\, 
\Big[ \N{0}{j0}\aeff{d}{j0}{(\mu\ta)} 
- \N{0}{j0}\ceff{d}{j0}{(\mu\ta)}\Big]
\notag \\
&&
+ L\, \sin(\De m^2L/2E)\, \sum_{dj}\sum_{m>0} \pmag^{d-3}\,
\cos(m\om_\oplus T_\oplus) \re \Big[
\N{0}{jm}\aeff{d}{jm}{(\mu\ta)} 
- \N{0}{jm}\ceff{d}{jm}{(\mu\ta)}\Big]
\notag \\
&&
- L\, \sin(\De m^2L/2E)\, \sum_{dj}\sum_{m>0} \pmag^{d-3}\, 
\sin(m\om_\oplus T_\oplus) \im \Big[
\N{0}{jm}\aeff{d}{jm}{(\mu\ta)} 
- \N{0}{jm}\ceff{d}{jm}{(\mu\ta)}\Big] ,
\eea
\newpage
\end{widetext}
where the factors $\N{0}{jm}$ are given 
by Eq.\ \rf{nfactors}, 
as before.
The combinations of coefficients for Lorentz violation 
that can be measured in this scenario 
are therefore the real and imaginary parts of
$\sum_j \N{0}{jm}\aeff{d}{jm}{(\mu\ta)}$ and  
$\sum_j \N{0}{jm}\ceff{d}{jm}{(\mu\ta)}$.

This scenario also provides a simple framework
for studying the effects of CPT violation.
For example, 
consider the CPT asymmetry
\beq
\ACPT{ab} = 
\fr{P_{\nu_a\to\nu_b}-P_{\nub_b\to\nub_a}}
{P_{\nu_a\to\nu_b}+P_{\nub_b\to\nub_a}}.
\eeq
Assuming identical energies, baselines, and beam directions,
this asymmetry vanishes when CPT is conserved.
A nonzero experimental measurement of $\ACPT{ab}$ 
would therefore provide evidence of CPT violation.
We remark in passing that a zero measurement 
would fail to prove CPT symmetry 
because CPT-violating models 
exist for which $\ACPT{ab}$ vanishes
\cite{km}.

For maximal two-flavor mixing,
the asymmetry $\ACPT{ab}$ can be expressed compactly 
in terms of coefficients for Lorentz violation
in the Sun-centered frame. 
Assuming the conventional zeroth-order transition probability
$P^{(0)}_{\nu_\mu\to\nu_\ta} = \sin^2(\De m^2L/4E)$
is large compared to CPT-violating effects, 
appearance experiments are appropriate
and the CPT asymmetry takes the form
\bea
\ACPT{\mu\ta} &\approx &
L\, \cot(\De m^2L/4E)\,
\nn\\
&&
\times
\sum_{djm} \pmag^{d-3}\, 
e^{im\om_\oplus T_\oplus} \N{0}{jm} \aeff{d}{jm}{(\mu\ta)} ,
\qquad
\eea
If instead the survival probability 
$P^{(0)}_{\nu_\mu\to\nu_\mu}$
is large compared to CPT-violating effects,
then disappearance experiments are useful
and the relevant CPT asymmetry is 
\bea
\ACPT{\mu\mu} &\approx &
-L\, \tan(\De m^2L/4E)\,
\nn\\
&&
\times \sum_{djm} \pmag^{d-3}\, 
e^{im\om_\oplus T_\oplus} \N{0}{jm} \aeff{d}{jm}{(\mu\ta)} .
\qquad
\eea
The above two asymmetries contain the same coefficients
for Lorentz violation
but apply in complementary regimes,
so at least one is applicable to any experiment.
They incorporate direction-dependent effects
and variations with sidereal time
as a consequence of the Lorentz violation
that accompanies CPT violation.
Similar effects accompany CPT violation 
in the oscillations of neutral mesons
\cite{akmesons},
and indeed the above expressions are closely related
to the corresponding CPT asymmetries for mesons 
\cite{mesons}.

The CPT asymmetries contain coefficients 
for Lorentz and CPT violation with $m=0$
that produce effects independent of sidereal time.
One simple way to extract these coefficients,
already used in the meson context
\cite{mesons},
is to measure the time-averaged asymmetries
$\overline{\ACPT{\mu\ta}}$ and $\overline{\ACPT{\mu\mu}}$.
For given energy, baseline, and beam direction, 
these asymmetries take the form 
\bea
\overline{\ACPT{\mu\ta}} &\approx &
L\,  \cot(\De m^2L/4E)\,
\sum_{dj} \pmag^{d-3}\, 
\N{0}{j0} \aeff{d}{j0}{(\mu\ta)} ,
\nn\\
\overline{\ACPT{\mu\mu}} &\approx &
-L\,  \tan(\De m^2L/4E)\,
\sum_{dj} \pmag^{d-3}\, 
\N{0}{j0} \aeff{d}{j0}{(\mu\ta)} .
\nn\\
\eea
The presence of the factors $\N{0}{j0}$ 
shows that these expressions depend on the beam direction
despite the time averaging.
As a result,
distinct experiments can be expected to have
different sensitivities to different coefficients 
for Lorentz and CPT violation. 
Over the range of existing and planned long-baseline experiments,
a given factor $\N{0}{j0}$ can change sign 
and can vary by more than an order of magnitude,
so the difference in attained sensitivities can be substantial.

\section{Applications to kinematics}
\label{Applications to kinematics}

Neutrino oscillations can yield observable signals
of Lorentz violation 
because they involve comparing
the propagation of one neutrino flavor against another.
Another possibility for detecting physical effects
of Lorentz violation
is to compare neutrino propagation
to the propagation of a different kind of particle.
Kinematic tests of this kind come in several varieties.
One conceptually straightforward approach is to measure 
the difference in the times of flight of neutrinos 
and photons or other particles.
A more subtle possibility is 
to study decay processes involving neutrinos,
which can be modified when the dispersion relations 
of neutrinos and other species differ in their Lorentz properties.
The effects on decays can be striking,
with certain processes becoming forbidden or allowed
according to the energies of the particles involved.

In this section,
several kinds of kinematic tests are considered.
To focus the discussion,
we assume oscillations are negligible or zero.
This implies working within the context 
of the oscillation-free models
described in Secs.\ \ref{Oscillation-free models}
and \ref{Isotropic oscillation-free models}.
Both generic oscillation-free models
and the isotropic flavor-blind limit
are treated.
We consider time-of-flight measurements,
threshold effects in pion and kaon decays,
and \cv\ radiation,
using existing data to obtain explicit results 
for coefficients for Lorentz violation.

\subsection{Time-of-flight measurements}

\begin{table*}
\begin{tabular}{c|c|c|c|c|c|c}
\multicolumn{7}{c}{MINOS} \\
$m$ & $0$ & $1$ & $2$ & $3$ & $4$ & $\cdots$ \\
\hline
$\N{0}{0m}$ &
$ 0.28$ &  &  &  &  &  \\
$\N{0}{1m}$ &
$0.32$ & $0.22+0.14i$ &  &  &  &  \\
$\N{0}{2m}$ &
$0.09$ & $0.32+0.21i$ & $0.09+0.20i$ &  &  &  \\
$\N{0}{3m}$ &
$-0.21$ & $0.24+0.15i$ & $0.16+0.35i$ & $-0.02+0.18i$ &  &  \\
$\N{0}{4m}$ &
$-0.36$ & $ 0.00+0.00i$ & $0.16+0.35i$ & $-0.04+0.35i$ & $-0.09+0.11i$ &  \\
$\vdots$ & $\vdots$ & $\vdots$ & $\vdots$ & $\vdots$ & $\vdots$ & $\vdots$ \\
\multicolumn{7}{c}{} \\
\multicolumn{7}{c}{OPERA}\\
$m$ & $0$ & $1$ & $2$ & $3$ & $4$ & $\cdots$ \\
\hline
$\N{0}{0m}$ & 
$0.28$ &  &  &  &  &  \\
$\N{0}{1m}$ &
$-0.20$ & $-0.16-0.27i$ &  &  &  &  \\
$\N{0}{2m}$ &
$-0.16$ & $0.14+0.25i$ & $-0.16+0.28i$ & & & \\
$\N{0}{3m}$ &
$0.33$ & $0.02+0.04i$ & $0.18-0.30i$ & $0.32+0.00i$ &  & \\
$\N{0}{4m}$ &
$-0.11$ & $-0.16-0.28i$ & $-0.03+0.04i$ & $-0.39-0.00i$ & $-0.15-0.27i$ &  \\
$\vdots$ & $\vdots$ & $\vdots$ & $\vdots$ & $\vdots$ & $\vdots$ & $\vdots$ \\
\multicolumn{7}{c}{} \\
\multicolumn{7}{c}{T2K} \\
$m$ & $0$ & $1$ & $2$ & $3$ & $4$ & $\cdots$ \\
\hline
$\N{0}{0m}$ &
$ 0.28$ &  &  &  &  &  \\
$\N{0}{1m}$ &
$0.01$ & $-0.01+0.35i$ &  &  &  &  \\
$\N{0}{2m}$ &
$-0.32$ & $0.00+0.01i$ & $-0.39-0.01i$ &  &  &  \\
$\N{0}{3m}$ &
$-0.02$ & $0.01-0.32i$ & $-0.01+0.00i$ & $0.02-0.42i$ &  &  \\
$\N{0}{4m}$ &
$0.32$ & $ 0.00-0.02i$ & $0.33+0.01i$ & $0.00-0.02i$ & $0.44+0.03i$ &  \\
$\vdots$ & $\vdots$ & $\vdots$ & $\vdots$ & $\vdots$ & $\vdots$ & $\vdots$
\end{tabular}
\caption{\label{Nbeams}
Numerical values of direction factors $\N{0}{jm}$ 
for some long-baseline experiments.
Values are given to two decimal places.
An explicit $0.00$ indicates a small nonzero value.}
\end{table*}

\begin{table*}
\renewcommand{\arraystretch}{1.5}
\begin{tabular}{c|c|c|c|c}
coefficient& OPERA & MINOS & Fermilab & Fermilab \\
\hline
$\cof{4}{00}$&$-8.4\pm1.1^{+1.2}_{-0.9}\times 10^{-5}$&$-1.8\pm1.0\times 10^{-4}$&$<1.4\times 10^{-4}$&--\\
$\cof{4}{10}$&$11.8\pm1.6^{+1.7}_{-1.2}\times 10^{-5}$&$-1.6\pm0.9\times 10^{-4}$&$<1.6\times 10^{-4}$&--\\
$\cof{4}{20}$&$15.2\pm2.1^{+2.2}_{-1.5}\times 10^{-5}$&$-5.6\pm3.2\times 10^{-4}$&$<6.2\times 10^{-4}$&--\\
$\aof{5}{00}$&$25.7\pm3.3^{+3.5}_{-2.5}\times 10^{-7}$&$3.0\pm1.7\times 10^{-5}$&$<2.4\times 10^{-6}$&$<2.1\times 10^{-6}$\\
$\aof{5}{10}$&$-34.7\pm4.7^{+5.0}_{-3.5}\times 10^{-7}$&$2.7\pm1.5\times 10^{-5}$&$<2.7\times 10^{-6}$&$<2.3\times 10^{-6}$\\
$\aof{5}{20}$&$-44.8\pm6.1^{+6.4}_{-4.5}\times 10^{-7}$&$9.4\pm5.3\times 10^{-5}$&$<1.0\times 10^{-5}$&$<9.0\times 10^{-6}$\\
$\aof{5}{30}$&$21.1\pm2.9^{+3.0}_{-2.1}\times 10^{-7}$&$-4.1\pm2.3\times 10^{-5}$&$<2.1\times 10^{-6}$&$<1.8\times 10^{-6}$\\
$\cof{6}{00}$&$-9.7\pm1.3^{+1.4}_{-1.0}\times 10^{-8}$&$-6.7\pm3.8\times 10^{-6}$&$<5.3\times 10^{-8}$&--\\
$\cof{6}{10}$&$13.6\pm1.8^{+2.0}_{-1.3}\times 10^{-8}$&$-5.9\pm3.4\times 10^{-6}$&$<5.9\times 10^{-8}$&--\\
$\cof{6}{20}$&$17.6\pm2.4^{+2.5}_{-1.8}\times 10^{-8}$&$2.1\pm1.2\times 10^{-5}$&$<2.3\times 10^{-7}$&--\\
$\cof{6}{30}$&$-8.3\pm1.1^{+1.2}_{-0.8}\times 10^{-8}$&$9.1\pm5.4\times 10^{-6}$&$<4.6\times 10^{-8}$&--\\
$\cof{6}{40}$&$24.2\pm3.3^{+3.5}_{-2.5}\times 10^{-8}$&$5.2\pm3.0\times 10^{-6}$&$<5.6\times 10^{-8}$&--
\end{tabular}
\caption{\label{accelconds}
Single-coefficient measurements and modulus bounds
from accelerator time-of-flight experiments.
Units are GeV$^{4-d}$.}
\end{table*}

Time-of-flight experiments compare the group velocity
of neutrinos with that of photons or other particles.
Here,
we work with four measurements
involving time-of-flight comparisons with photons:
the recent OPERA result
\cite{opera},
the prior MINOS bound 
\cite{minostof},
early constraints from experiments at Fermilab
\cite{kbfaja},
and limits from the supernova SN1987A
\cite{sn1987a}.
Lorentz violation in the photon sector
involving operators 
of both renormalizable and nonrenormalizable dimensions 
is tightly constrained
\cite{tables}.
For definiteness,
we assume below a conventional photon dispersion relation,
with any Lorentz violation confined to the neutrino sector.

\subsubsection{Generic case}

In generic oscillation-free models,
the group velocity $\vgof$ 
for neutrinos propagating in direction $\phat$
is given by Eq.\ \rf{vgof}.
For astrophysical neutrinos,
this result is directly applicable.
However,
for neutrinos in a beam experiment,
sidereal variations in the signal
are induced by the rotation of the Earth.
These can conveniently be handled via the methods discussed
in Sec.\ \ref{Short-baseline approximation}.
Working as before in the standard laboratory frame
\cite{labframe},
the Wigner rotation matrices can be used to
display the sidereal variations explicitly 
and to express the group velocity 
in terms of spherical coefficients for Lorentz violation 
in the canonical Sun-centered frame
\cite{sunframe}.
The effect of this procedure on the group velocity \rf{vgof}
is to perform the substitution $Y_{jm}(\phat) \to
e^{im\om_\oplus T_\oplus} \N{0}{jm}$,
where the factor $\N{0}{jm}$
is defined in Eq.\ \rf{nfactors}.
We thereby find for beam experiments
the neutrino group velocity
\bea
\vgof 
&=& 1 - \fr{|m_l|^2}{2\pvec^2} 
+ \sum_{djm} (d-3) \pmag^{d-4}\, 
e^{im\om_\oplus T_\oplus} \N{0}{jm}
\nn\\
&&
\hskip 70pt
\times
\Big[\aof{d}{jm} - \cof{d}{jm}\Big],
\label{gvofsidereal}
\eea
which displays explicitly the dependence 
on the sidereal rotation frequency $\om_\oplus$
and the local sidereal time $T_\oplus$.
We remind the reader that for antineutrinos
the sign of the coefficients
$\aof{d}{jm}$ 
changes in all expressions for the group velocity.

The OPERA collaboration reported a difference 
between the speed of light and the speed of muon neutrinos of
$\de \vg = 2.37 \pm 0.32^{+0.34}_{-0.24} \times 10^{-5}$
\cite{opera}.
Averaging over sidereal time,
this velocity defect yields the condition
\bea
\sum_{dj} (d-3) \pmag^{d-4}\, \N{0}{j0}
\Big[\aof{d}{j0} - \cof{d}{j0}\Big]
&&
\nn\\
&&
\hskip -100pt
= 2.37 \pm 0.32^{+0.34}_{-0.24} \times 10^{-5} 
\qquad
\label{operacond}
\eea
on the oscillation-free spherical coefficients
for Lorentz violation.
Here,
$\N{0}{j0}$ is the directional factor for the OPERA beam,
which must be computed using the beam angles 
$\th = 86.7^\circ, \ph= 52.4^\circ$
at the colatitude $\ch=47.5^\circ$
of the detector.
Numerical values for relevant values of $\N{0}{j0}$ 
for OPERA are listed in Table \ref{Nbeams}.

To gain intuition and for purposes of comparison 
with other experiments,
we can extract from the condition \rf{operacond}
a set of constraints on individual oscillation-free coefficients
for Lorentz violation
under the assumption that only one is nonzero at a time.
Using for $\pmag$ the average energy of the beam,
$\vev\pmag\simeq 17$ GeV,
the resulting constraints are listed in Table \ref{accelconds}
for dimensions $d\leq 6$.

The shape of the observed neutrino spectrum 
matches the expected form to a high degree,
which implies little or no dispersion in the group velocity
at OPERA energies.
This is reflected in the comparison
of neutrino group velocities
at low ($\vev\pmag\simeq 13.9$ GeV)
and high ($\vev\pmag\simeq 42.9$ GeV) energies.
The arrival-time difference reported by OPERA 
is $14.0\pm 26.2$ ns,
which translates into a velocity difference $\De v$
between the two datasets of approximately
$\De v\simeq 6\pm 11 \times 10^{-6}$.
As an illustration,
we can use this to place 
a comparatively reliable constraint
on dimension $d=5$ operators for Lorentz and CPT violation,
assuming Lorentz-violating operators at other values of $d$ 
are negligible.
We thereby find the condition
\beq
\sum_j \N{0}{j0} \, \aof{5}{j0}
\simeq 10\pm 19 \times 10^{-8}\ \text{GeV}^{-1}
\eeq
on a beam-dependent combination of coefficients with $d=5$.
The lack of dispersion could in principle also be used 
to bound coefficients for $d>5$, 
but obtaining reliable constraints requires access 
to more detailed information about the observed energy spectrum.

A prior time-of-flight experiment by the MINOS collaboration 
also measured the group velocity for the muon neutrino
compared to the speed of light,
with the result 
$\de \vg = 5.1\pm 2.9 \times 10^{-5}$
\cite{minostof}.
For generic oscillation-free coefficients,
we obtain the condition
\bea
\sum_{dj} (d-3) \pmag^{d-4}\, \N{0}{j0}
\Big[\aof{d}{j0} - \cof{d}{j0}\Big]
&&
\nn\\
&&
\hskip -70pt
= 5.1\pm 2.9 \times 10^{-5} ,
\qquad
\eea
where we have again averaged over sidereal time.
The numerical values of the directional factors $\N{0}{j0}$ 
for MINOS are listed in Table \ref{Nbeams}.
They are computed using the polar angles 
$\th=86.7^\circ$, $\ph=203.9^\circ$ of the beam 
and the colatitude $\ch=42.2^\circ$ of the detector.
Adopting the average beam energy as $\vev{\pmag} = 3$ GeV,
we can extract single-coefficient constraints
taken one at a time.
For dimensions $d\leq 6$,
the results are listed in Table \ref{accelconds}.

\begin{table*}
\begin{tabular}{c|c|c||c|c|c}
& time-of-flight & \quad dispersion \quad &
& time-of-flight & \quad dispersion \quad \\
\quad coefficient \quad & bound & bound &
\quad coefficient \quad & bound & bound \\
\hline
$\cof{4}{00}$	&	$7.1\times 10^{-9}$	&		&	$\cof{6}{00}$	&	$2.4\times 10^{-5}$	&	$7.9\times 10^{-9}$	\\
$\cof{4}{10}$	&	$4.4\times 10^{-9}$	&		&	$\cof{6}{10}$	&	$1.5\times 10^{-5}$	&	$4.9\times 10^{-9}$	\\
$\re\cof{4}{11}$	&	$7.7\times 10^{-8}$	&		&	$\re\cof{6}{11}$	&	$2.6\times 10^{-4}$	&	$8.6\times 10^{-8}$	\\
$\im\cof{4}{11}$	&	$8.2\times 10^{-9}$	&		&	$\im\cof{6}{11}$	&	$2.7\times 10^{-5}$	&	$9.1\times 10^{-9}$	\\
$\cof{4}{20}$	&	$3.9\times 10^{-9}$	&		&	$\cof{6}{20}$	&	$1.3\times 10^{-5}$	&	$4.3\times 10^{-9}$	\\
$\re\cof{4}{21}$	&	$3.7\times 10^{-8}$	&		&	$\re\cof{6}{21}$	&	$1.2\times 10^{-4}$	&	$4.1\times 10^{-8}$	\\
$\im\cof{4}{21}$	&	$3.9\times 10^{-9}$	&		&	$\im\cof{6}{21}$	&	$1.3\times 10^{-5}$	&	$4.4\times 10^{-9}$	\\
$\re\cof{4}{22}$	&	$2.1\times 10^{-8}$	&		&	$\re\cof{6}{22}$	&	$7.0\times 10^{-5}$	&	$2.3\times 10^{-8}$	\\
$\im\cof{4}{22}$	&	$9.8\times 10^{-8}$	&		&	$\im\cof{6}{22}$	&	$3.3\times 10^{-4}$	&	$1.1\times 10^{-7}$	\\
	&		&		&	$\cof{6}{30}$	&	$1.4\times 10^{-5}$	&	$4.6\times 10^{-9}$	\\
$\aof{5}{00}$	&	$3.5\times 10^{-7}$	&	$3.5\times10^{-10}$	&	$\re\cof{6}{31}$	&	$8.1\times 10^{-5}$	&	$2.7\times 10^{-8}$	\\
$\aof{5}{10}$	&	$2.2\times 10^{-7}$	&	$2.2\times10^{-10}$	&	$\im\cof{6}{31}$	&	$8.7\times 10^{-6}$	&	$2.9\times 10^{-9}$	\\
$\re\aof{5}{11}$	&	$3.8\times 10^{-6}$	&	$3.8\times10^{-9}$	&	$\re\cof{6}{32}$	&	$2.8\times 10^{-5}$	&	$9.5\times 10^{-9}$	\\
$\im\aof{5}{11}$	&	$4.1\times 10^{-7}$	&	$4.1\times10^{-10}$	&	$\im\cof{6}{32}$	&	$1.3\times 10^{-4}$	&	$4.4\times 10^{-8}$	\\
$\aof{5}{20}$	&	$2.0\times 10^{-7}$	&	$2.0\times10^{-10}$	&	$\re\cof{6}{33}$	&	$5.7\times 10^{-4}$	&	$1.9\times 10^{-7}$	\\
$\re\aof{5}{21}$	&	$1.8\times 10^{-6}$	&	$1.8\times10^{-9}$	&	$\im\cof{6}{33}$	&	$1.9\times 10^{-4}$	&	$6.3\times 10^{-8}$	\\
$\im\aof{5}{21}$	&	$2.0\times 10^{-7}$	&	$2.0\times10^{-10}$	&	$\cof{6}{40}$	&	$1.8\times 10^{-5}$	&	$5.9\times 10^{-9}$	\\
$\re\aof{5}{22}$	&	$1.1\times 10^{-6}$	&	$1.1\times10^{-9}$	&	$\re\cof{6}{41}$	&	$6.4\times 10^{-5}$	&	$2.1\times 10^{-8}$	\\
$\im\aof{5}{22}$	&	$4.9\times 10^{-6}$	&	$4.9\times10^{-9}$	&	$\im\cof{6}{41}$	&	$6.9\times 10^{-6}$	&	$2.3\times 10^{-9}$	\\
$\aof{5}{30}$	&	$2.1\times 10^{-7}$	&	$2.1\times10^{-10}$	&	$\re\cof{6}{42}$	&	$1.6\times 10^{-5}$	&	$5.3\times 10^{-9}$	\\
$\re\aof{5}{31}$	&	$1.2\times 10^{-6}$	&	$1.2\times10^{-9}$	&	$\im\cof{6}{42}$	&	$7.3\times 10^{-5}$	&	$2.4\times 10^{-8}$	\\
$\im\aof{5}{31}$	&	$1.3\times 10^{-7}$	&	$1.3\times10^{-10}$	&	$\re\cof{6}{43}$	&	$2.0\times 10^{-4}$	&	$6.8\times 10^{-8}$	\\
$\re\aof{5}{32}$	&	$4.3\times 10^{-7}$	&	$4.3\times10^{-10}$	&	$\im\cof{6}{43}$	&	$6.8\times 10^{-5}$	&	$2.3\times 10^{-8}$	\\
$\im\aof{5}{32}$	&	$2.0\times 10^{-6}$	&	$2.0\times10^{-9}$	&	$\re\cof{6}{44}$	&	$5.3\times 10^{-4}$	&	$1.8\times 10^{-7}$	\\
$\re\aof{5}{33}$	&	$8.6\times 10^{-6}$	&	$8.6\times10^{-9}$	&	$\im\cof{6}{44}$	&	$1.2\times 10^{-3}$	&	$3.9\times 10^{-7}$	\\
$\im\aof{5}{33}$	&	$2.8\times 10^{-6}$	&	$2.8\times10^{-9}$	&		&		&		
\end{tabular}
\caption{\label{snofcond}
Single-coefficient modulus bounds from time-of-flight and dispersion of SN1987A antineutrinos.
Units are GeV$^{4-d}$.}
\end{table*}

An older experiment at Fermilab 
\cite{kbfaja}
reported bounds of $|\de \vg| < 4\times 10^{-5}$
at 95\% C.L. 
using both muon neutrinos and muon antineutrinos.
Averaging over sidereal time,
this gives the two conditions
\bea
\Big|\sum_{dj} (d-3) \pmag^{d-4}\, \N{0}{j0}
\Big[\pm\aof{d}{j0} - \cof{d}{j0}\Big]\Big|
&&
\nn\\
&&
\hskip -70pt
< 4 \times 10^{-5} .
\qquad
\eea
The experiment also yielded a limit on the difference 
between the neutrino and antineutrino group velocities
of $7\times 10^{-5}$.
In the present context,
this generates the limit 
\beq
\Big|\sum_{dj} (d-3) \pmag^{d-4}\, \N{0}{j0} \aof{d}{j0} \Big|
< 3.5 \times 10^{-5} 
\eeq
on a combination of coefficients for CPT-odd Lorentz violation.
As before,
we can place constraints on individual coefficients
taken one at a time.
We adopt the estimated beam angles 
$\th\simeq 90^\circ$, $\ph\simeq140^\circ$
and the colatitude $\ch=48.2^\circ$ for the detector.
In these experiments
the energies ranged from 30 to 200 GeV,
so for our calculations
we take the conservative value $\pmag =30$ GeV.
The results for dimensions $d\leq 6$
are listed in Table \ref{accelconds}.

Observations of the antineutrino burst from supernova SN1987A 
lead to a conservative bound 
on the difference between the speed of light
and the speed of antineutrinos
of $|\de \vg| < 2\times 10^{-9}$
\cite{sn1987a}.
The large propagation distance
implies the electron antineutrinos produced at the source
oscillated many times during their trip to the Earth.
No sidereal effects occur in this case,
so we obtain the condition
\beq
\Big|\sum_{djm} (d-3) \pmag^{d-4}\, Y_{jm}
\Big[\aof{d}{jm} + \cof{d}{jm}\Big]\Big|
< 2 \times 10^{-9} 
\eeq
involving both isotropic and anisotropic coefficients.
This means we can also extract here individual constraints
on coefficients with nonzero values of $m$.  
In the Sun-centered frame,
the propagation direction is given by the polar angles 
$\th=20.7^\circ$, $\ph=263.9^\circ$.
The observed energies range between about 7.5 and 40 MeV,
so we adopt the conservative value $\pmag = 10$ MeV.
The resulting single-coefficient time-of-flight constraints
for values $d\leq 6$
are given in Table \ref{snofcond}.

The observed antineutrinos from SN1987A
have a spread of energies.
However,
they all arrived within a time interval of about 10 s
after traveling for about $5\times 10^{12}$ s,
implying a maximum difference in speed
$\de \vg < 2 \times 10^{-12}$
across the observed energies.
This restricts the possible antineutrino dispersion
and implies additional constraints
on coefficients for Lorentz violation,
independent of the speed of light
\cite{km}.
For an energy spread ranging from 
$|\pvec_1|$ to $|\pvec_2|$,
the dispersion condition is
\bea
\Big|
\sum_{djm} (d-3) 
\De(\pmag^{d-4}) Y_{jm} \Big[\aof{d}{jm} + \cof{d}{jm}\Big]
\Big|
&&
\nn\\
&&
\hskip -80pt
< 2 \times 10^{-12} ,
\eea
where 
$\De (\mbfp^{d-4}) = |\pvec_2|^{d-4} - |\pvec_1|^{d-4}$.
Using the same values of $\th$ and $\ph$ as before
and adopting the conservative choices 
$|\pvec_1|=10$ MeV and $|\pvec_2|=20$ MeV
yields the single-coefficient dispersion constraints
for values $d\leq 6$
listed in Table \ref{snofcond}.

\begin{table*}
\begin{tabular}{c|c|c|c|c|c|c}
coefficient & OPERA & MINOS & Fermilab & Fermilab & SN1987A & SN1987A \\
\hline
$\ccfc{4}{}$	&$-23.7\pm3.2^{+3.4}_{-2.4}\times 10^{-6}$	&$-5.1\pm2.9\times 10^{-5}$	&$<4.0\times 10^{-5}$	&--	&$<2.0\times 10^{-9}$	&		\\
$\acfc{5}{}$	&$69.7\pm9.4^{+10.0}_{-7.1}\times 10^{-8}$	&$8.5\pm4.8\times 10^{-6}$	&$<6.7\times 10^{-7}$	&$<5.8\times 10^{-7}$	&$<1.0\times 10^{-7}$	&	$<1.0\times 10^{-10}$	\\
$\ccfc{6}{}$	&$-27.3\pm3.7^{+3.9}_{-2.8}\times 10^{-9}$	&$-1.9\pm1.1\times 10^{-6}$	&$<1.5\times 10^{-8}$	&--	&$<6.7\times 10^{-6}$	&	$<2.2\times 10^{-9}$	\\
$\acfc{7}{}$	&$12.1\pm1.6^{+1.7}_{-1.2}\times 10^{-10}$	&$4.7\pm2.7\times 10^{-7}$	&$<3.7\times 10^{-10}$	&$<3.2\times 10^{-10}$	&$<5.0\times 10^{-4}$	&	$<7.1\times 10^{-8}$	\\
$\ccfc{8}{}$	&$-56.8\pm7.7^{+8.1}_{-5.8}\times 10^{-12}$	&$-12.6\pm7.2\times 10^{-8}$	&$<9.9\times 10^{-12}$	&--	&$<4.0\times 10^{-2}$	&	$<2.7\times 10^{-6}$	\\
$\acfc{9}{}$	&$27.8\pm3.8^{+4.0}_{-2.8}\times 10^{-13}$	&$3.5\pm2.0\times 10^{-8}$	&$<2.7\times 10^{-13}$	&$<2.4\times 10^{-13}$	&$<3.3\times 10^{0}$	&	$<1.1\times 10^{-4}$	\\
$\ccfc{10}{}$	&$-14.0\pm1.9^{+2.0}_{-1.4}\times 10^{-14}$	&$-10.0\pm5.7\times 10^{-9}$	&$<7.8\times 10^{-15}$	&--	&$<2.9\times 10^{2}$	&	$<4.5\times 10^{-3}$	
\end{tabular}
\caption{\label{isotofcond}
Time-of-flight and dispersion measurements and modulus bounds
on isotropic oscillation-free coefficients.
Units are GeV$^{4-d}$.}	
\end{table*}

In principle,
direction-dependent constraints
can also be extracted from beam experiments on the Earth
\cite{akmesons,km}.
The sidereal variations due to the Earth's rotation
cause the neutrino group velocity to change with time.
Note that the observed group velocity
can oscillate between superluminal and subluminal values. 
The sidereal variations are encoded
in the factor $\exp(im\om_\oplus T_\oplus)$ 
appearing in the group velocity \rf{gvofsidereal}.
For fixed $d$,
we can introduce complex amplitudes associated
with a particular sidereal harmonic as 
\bea
\TOFamp{a}{d}{m}{} &=& 
\sum_j (d-3)\, \N{0}{jm}\, \aof{d}{jm} ,
\nn\\
\TOFamp{c}{d}{m}{} &=& 
\sum_j (d-3)\, \N{0}{jm}\, \cof{d}{jm},
\label{complampl}
\eea
which obey $\TOFamp{a,c}{d}{m}{*} = \TOFamp{a,c}{d}{-m}{}$.
The neutrino velocity defect $\vgof$ in oscillation-free models
can then be written as the expression 
\beq
\de \vgof = 
\sum_{dm} \pmag^{d-4}\, e^{im\om_\oplus T_\oplus} 
\Big[\TOFamp{a}{d}{m}{} - \TOFamp{c}{d}{m}{}\Big] .
\label{veldefof}
\eeq
The antineutrino velocity defect $\ol{\vgof}$
takes the same form
except for the CPT-induced sign change 
of $\TOFamp{a}{d}{m}{}$.

The above sidereal variations of the group velocity
can be used to extract constraints
from existing and future time-of-flight data
obtained in beam experiments. 
The beam direction relative to the Earth
fixes the $\N{0}{jm}$ factors and hence
determines the relevant linear combinations
of coefficients for Lorentz violation.
To illustrate the variations in these combinations
as a function of beam direction,
we list in Table \ref{Nbeams} 
the $\N{0}{jm}$ factors
for the long-baseline experiments
MINOS, OPERA, and T2K \cite{T2K}.
For the MINOS and OPERA experiments,
we use the polar angles and detector colatitudes
given above.
For the T2K experiment,
we adopt the beam angles 
$\th = 88.7^\circ$, $\ph=270^\circ$
and take the detector colatitude as $\ch=53.6^\circ$.
Note that the reality condition \rf{nfactorreality}
implies $\N{0}{jm}^* = (-1)^m \N{0}{j(-m)}$,
which can be used to compute the factors $\N{0}{jm}$ 
for negative $m$ values.
The values in Table \ref{Nbeams}  
can be inserted in the expressions \rf{complampl}
and thereby into the velocity defect \rf{veldefof}
to extract limits on the oscillation-free coefficients
in several time-of-flight experiments. 
Relevant searches include ones
with MINOS, OPERA, and T2K, 
as well as Borexino
\cite{borexino}
and ICARUS
\cite{icarus-nim},
both of which have the same factors $\N{0}{jm}$ as OPERA. 

\subsubsection{Isotropic case}

The discussion in the preceding subsection
holds for generic oscillation-free models.
In contrast,
for isotropic oscillation-free models,
the physics is independent of the beam direction
and so no sidereal variations arise.
The neutrino group velocity $\vgri$
for isotropic oscillation-free models 
is instead given by Eq.\ \rf{isoofv},
which applies equally to astrophysical neutrinos
and neutrinos in beam experiments. 
As before,
the antineutrino group velocity $\ol{\vgri}$
differs by a change of sign for the coefficients $\acfc{d}{}$.

In the isotropic case, 
the OPERA measurement 
\cite{opera}
yields
\beq
\sum_{d} (d-3) \pmag^{d-4}
(\acfc{d}{} - \ccfc{d}{})
= 2.37 \pm 0.32^{+0.34}_{-0.24} \times 10^{-5} .
\eeq
The lack of dispersion in the neutrino pulse
yields a comparatively reliable constraint
on dimension-5 operators for isotropic Lorentz and CPT violation of
\beq
\acfc{5}{}
\simeq 10\pm 19 \times 10^{-8}\ \text{GeV}^{-1} .
\eeq

Using the MINOS result
\cite{minostof},
we obtain the condition 
\bea
\sum_{d} (d-3) \pmag^{d-4}
(\acfc{d}{} - \ccfc{d}{})
&=& 5.1\pm 2.9 \times 10^{-5} 
\qquad
\eea
on isotropic oscillation-free coefficients.
Analogously,
the older Fermilab results
\cite{kbfaja} 
yield the bound 
\bea
\Big|\sum_{d} (d-3) \pmag^{d-4}
(\pm\acfc{d}{} - \ccfc{d}{})\Big|
< 4 \times 10^{-5} ,
\eea
along with the constraint
\beq
\Big|\sum_{d} (d-3) \pmag^{d-4}\, \acfc{d}{} \Big|
< 3.5 \times 10^{-5} 
\eeq
on CPT-odd effects. 
Finally,
the SN1987A observations 
\cite{sn1987a}
give the time-of-flight bound 
\bea
\Big|\sum_{d} (d-3) \pmag^{d-4}\,
(\acfc{d}{} + \ccfc{d}{})\Big|
< 2 \times 10^{-9} 
\eea
and the dispersion bound
\bea
\Big|\sum_{d} (d-3) \De(\pmag^{d-4})
(\acfc{d}{} + \ccfc{d}{})\Big|
< 2 \times 10^{-12} .
\eea

Under the assumption that only one term is nonzero at a time,
we can use the above conditions to extract limits
on isotropic oscillation-free coefficients for Lorentz violation.
For Lorentz-violating operators of dimension $d\leq 10$,
the results are displayed in Table \ref{isotofcond}.
Future long-baseline experiments
and astrophysical observations
can be expected to improve these constraints.

\subsection{Threshold effects}

The decay processes
$\pi^+ \to \mu^+ + \nu_\mu$ and
$K^+ \to \mu^+ + \nu_\mu$
are the dominant sources of muon neutrinos in most experiments.
In the presence of unconventional dispersion relations
for neutrinos,
these decays can exhibit striking threshold effects 
\cite{ftlnu2,ftlnu3,ftlnu4,lgm,byyy,cns,bapion,btv,mmvv,mr2}.
In this subsection,
we use threshold effects in pion and kaon decays
to obtain additional constraints on the spherical coefficients
under the assumption that 
any Lorentz violation appears only in the neutrino sector.

For pion decay,
let $\kvec$ be the pion momentum 
and $\pvec$ be the neutrino momentum.
Energy-momentum conservation implies
the neutrino energy $E(\pvec)$ obeys 
\bea
E(\pvec) &=&
\sqrt{M_\pi^2 + \kvec^2} - \sqrt{M_\mu^2 + (\kvec-\pvec)^2}
\nn\\
&\leq&
\sqrt{M_\pi^2 + \kvec^2} - \sqrt{M_\mu^2 + (\kmag-\pmag)^2}
\nn\\
&\leq&
\sqrt{\De M^2 + \pvec^2} ,
\eea
where $\De M = M_\pi-M_\mu$.
The first inequality is obtained 
by taking $\pvec$ parallel to $\kvec$
to maximize the allowed energy.
The last relation is a reverse triangle inequality.
Similar relations hold for kaon decay,
with the replacement $M_\pi \to M_K$.

If the neutrino energy $E(\pvec)$ is Lorentz invariant,
the above inequalities are satisfied.
However,
a Lorentz-violating neutrino dispersion relation
can cause the inequalities 
to fail above some threshold energy,
in which case the decay becomes forbidden
by energy-momentum conservation.
Experimental observations of high-energy muon neutrinos
therefore constrain modified dispersion relations.
To identify the conditions 
resulting from the energy threshold,
we can write
\beq
E(\pvec) = E_0(\pvec) + \de E(\pvec) ,
\eeq
where $E_0(\pvec)$ is the conventional neutrino energy 
and $\de E(\pvec)$ is the Lorentz-violating contribution.
The reverse triangle inequality then gives
\beq
\de E(\pvec) 
\leq \fr{\De M^2 - m_\nu^2}{2E_0}
\leq \fr{\De M^2}{2\pmag} .
\eeq
The latter inequality can be applied 
to obtain explicit one-sided constraints
on Lorentz violation.

For the generic oscillation-free coefficients,
the expression \rf{ofenergies}
for the neutrino energy $E^{\rm of}_\nu$ 
and the CPT conjugate for $E^{\rm of}_{\nub}$ 
yield 
\beq
\de E^{\rm of} = 
\sum_{djm} \pmag^{d-3}\, Y_{jm}(\phat)
\Big[\pm\aof{d}{jm} - \cof{d}{jm}\Big].
\eeq
We therefore obtain two one-sided constraints,
\beq
\sum_{djm} \pmag^{d-2}\, Y_{jm}(\phat)
\Big[\pm\aof{d}{jm} - \cof{d}{jm}\Big] 
< \half \De M^2.
\label{ofkin}
\eeq
Notice that these bounds depend on the direction
of the neutrino or antineutrino propagation.

The IceCube collaboration observes atmospheric neutrinos 
at high energies up to about 400 TeV
\cite{IceCube11}.
Using the inequality \rf{ofkin},
data from this experiment could be used 
to search for directional effects
involving a reduced muon-neutrino flux
depending on the polar angles $(\th,\ph)$.
This study would be qualitatively different
from the search for Lorentz violation 
in oscillations of atmospheric neutrinos 
already published by the IceCube collaboration 
\cite{IceCube}.
It would provide sensitivity to 
distinct coefficients for Lorentz violation.

For the isotropic oscillation-free case,
the neutrino energy $\Eri_\nu$ given in Eq.\ \rf{isoofen}
and the corresponding antineutrino energy $\Eri_{\nub}$
yield instead the two conditions
\beq
\de \Eri = \sum_d \pmag^{d-3} (\pm\acfc{d}{} - \ccfc{d}{}),
\eeq
which lead to the two one-sided bounds
\beq
\sum_d \pmag^{d-2} (\pm\acfc{d}{} - \ccfc{d}{}) 
< \half \De M^2.
\label{isokin}
\eeq
For pion decays,
the numerical value of the right-hand side is
$\half\De M^2_\pi = 5.7\times 10^{-4}\ \text{GeV}^2$.
For kaon decays,
the result is weaker by roughly two orders of magnitude,
$\half\De M^2_K = 7.5\times 10^{-2}\ \text{GeV}^2$.
The kaon mode typically dominates at higher energies,
but heavier mesons may contribute as well.

Using the inequality \rf{isokin}
reveals that the IceCube observation 
of neutrinos up to about 400 TeV 
suffices by itself to place fairly robust limits 
on isotropic oscillation-free coefficients for Lorentz violation.
We obtain 
\beq
\sum (400\ \text{TeV})^{d-2}
(\pm\acfc{d}{} - \ccfc{d}{})
< 7.5\times 10^{-2}\ \text{GeV}^2 .
\eeq
Since no significant deviation is seen
in the total neutrino and antineutrino flux,
we can reasonably assume
this constraint applies independently 
to both muon neutrinos and antineutrinos.
Taking only one spherical coefficient to be nonzero
at a time
therefore produces two kinds of constraints.
A lower (negative) bound is obtained
on coefficients for CPT-even Lorentz violation,
while a two-sided bound emerges for the CPT-odd case.
Table \ref{icecube} lists 
the resulting estimated bounds
on single isotropic oscillation-free coefficients
for values $d\leq 10$. 

\begin{table}
\begin{tabular}{c|c||c|c}
coefficient & bound & coefficient & bound \\
\hline
$\big|\acfc{3}{}\big|$ & $<1.9\times 10^{-7}$ 
& $\ccfc{4}{}$   & $>-4.7\times 10^{-13}$ \\
$\big|\acfc{5}{}\big|$ & $<1.2\times 10^{-18}$ 
& $\ccfc{6}{}$  & $>-2.9\times 10^{-24}$ \\
$\big|\acfc{7}{}\big|$ & $<7.3\times 10^{-30}$ 
& $\ccfc{8}{}$  & $>-1.8\times 10^{-35}$ \\
$\big|\acfc{9}{}\big|$ & $<4.6\times 10^{-41}$ 
& $\ccfc{10}{}$ & $>-1.1\times 10^{-46}$
\end{tabular}
\caption{\label{icecube}
Estimated bounds on isotropic oscillation-free coefficients
from a threshold analysis of IceCube data.
Units are GeV$^{4-d}$.}
\end{table}

For the specific isotropic oscillation-free model 
with only the coefficient $\ccfc{4}{}$ nonzero,
two bounds from threshold effects in meson decay
have recently been obtained
using the IceCube observation of neutrinos
with energies up to 400 TeV.
The first translates into the one-sided bound
$\ccfc{4}{} \gsim -4\times 10^{-15}$
\cite{byyy},
and the second into the one-sided bound
$\ccfc{4}{}\gsim -10^{-12}$
\cite{cns}.
The first bound is two orders of magnitude tighter
than our result in Table \ref{icecube}
because it assumes IceCube neutrinos originate from pion decay.
The second result is consistent with our analysis.
All the results are many orders of magnitude
more stringent than the value of $\ccfc{4}{}$
obtained from the OPERA result
and listed in the first row of Table \ref{isotofcond}.

\subsection{\cv\ radiation}

Another kinematic feature of Lorentz violation
is the possibility that a neutrino
undergoes \cv\ radiation,
emitting one or more particles 
\cite{cg99,jlm,mmgls,aem,cg,jcjc,mlm,mr1,mr2,llmwz,hllnq,bg,fbhl}.
This can occur when the maximum attainable velocity
of the neutrino exceeds that of the emitted particles.
In this subsection,
we comment on some implications
of neutrino \cv\ radiation
in the context of the present work,
with focus on threshold effects and spectral distortions.
Except where stated otherwise,
Lorentz violation is assumed 
to be confined to the neutrino sector.

\subsubsection{Threshold effects}

Suppose a neutrino with high energy $E(\pvec)$
and with an unconventional dispersion relation 
experiences \cv-like emission to one or more particles.
Potentially significant \cv\ processes 
for neutrino energies up to some tens of GeV include
neutrino splitting 
$\nu_\mu\to\nu_\mu+\nu_e+\nub_e$,
electron-positron pair production 
$\nu_\mu\to\nu_\mu+e^++e^-$,
and photon decay 
$\nu_\mu\to\nu_\mu+\ga$.
However,
as the neutrino energy increases
other processes become important,
such as muon pair production
$\nu_\mu\to\nu_\mu+\mu^++\mu^-$,
tau pair production
$\nu_\mu\to\nu_\mu+\ta^++\ta^-$,
emission of various hadron combinations,
and ultimately even $Z^0$ emission
$\nu_\mu\to\nu_\mu+Z^0$
and Higgs emission
$\nu_\mu\to\nu_\mu+\ph$.

Energy-momentum conservation for the radiation 
of a single particle of mass $M$ and momentum $\kvec$
implies 
\beq
\De E(\pvec,\kvec)\equiv 
E(\pvec) - E(\pvec-\kvec) = \sqrt{M^2+\kvec^2} .
\eeq
Similarly, 
for emission of two particles 
of masses $M_1$, $M_2$ and momenta $\kvec_1$, $\kvec_2$,
we can write
\bea
\De E(\pvec,\kvec) &=& 
E(\pvec) - E(\pvec-\kvec) 
\nn\\
&=& 
\sqrt{M_1^2+\kvec_1^2} +\sqrt{M^2_2+\kvec_2^2}
\nn\\
&\geq &\sqrt{M^2+\kvec^2} 
\equiv \De E_{\rm th}(\kvec) ,
\eea
where $M=M_1+M_2$ and $\kvec = \kvec_1 + \kvec_2$.
We see that $\De E_{\rm th}(\kvec)$ represents a threshold
on the emitted energy,
obtained by treating the radiated
particles as a single composite particle.
This result generalizes 
to larger numbers of conventional decay products.
Note, 
however,
that it relies on Lorentz violation
being confined to the original particle.
In processes of this type,
natural scenarios exist in which modified dispersion laws
for the emitted species exclude 
the presence of a finite threshold 
and hence forbid \cv\ radiation.

A given \cv\ process is forbidden 
when the change $\De E(\pvec,\kvec)$ in neutrino energy 
remains below 
the emitted-energy threshold $\De E_{\rm th}(\kvec)$ 
for all values of $\kvec$.
This condition can be visualized graphically
by plotting $\De E(\pvec,\kvec)$ and 
$\De E_{\rm th}(\kvec)$ versus $\kmag$.
The threshold curve for emission of particles
with nonzero total mass
takes the form of a conventional mass hyperbola,
while for purely massless emission 
it is a conventional light cone.
The \cv\ decay is allowed if the neutrino curve 
passes above the threshold curve.

Suppose the momentum $\kvec$ transferred is small.
The energy emitted is then
$\De E(\pvec,\kvec) \approx \kvec\cdot{\mbf v}(\pvec)$,
where $\mbf \vg(\pvec)$ is the neutrino group velocity.
This is maximized when $\kvec$ lies along
the direction of $\mbf \vg(\pvec)$.
At the threshold, 
we then find $\kvec^2 \mbf \vg^2 = M^2 + \kvec^2$,
which implies 
\beq
\kvec^2 = \fr{M^2}{\mbf \vg^2 -1}
\qquad {\rm (threshold)}.
\eeq
We emphasize that this result for the onset of \cv\ radiation
is independent of the specific form 
of the neutrino dispersion relation.

The recent OPERA measurement of the muon-neutrino velocity defect 
$\de |\mbf \vg| = 2.37 \pm 0.32^{+0.34}_{-0.24} \times 10^{-5}$
\cite{opera}
yields the threshold value 
\beq
|\kvec| \simeq 145 M 
\label{operath}
\eeq
for neutrino \cv\ radiation into species of total rest mass $M$.
For example,
the threshold for electron-positron emission
is $\kmag \simeq 150$ MeV,
while the threshold for muon pair production
is $\kmag \simeq 31$ GeV.
The OPERA neutrinos have average energies
$\vev\pmag\simeq 17$ GeV,
with a spectrum exceeding 40 GeV.
The result \rf{operath} therefore shows in a model-independent way
that the neutrinos observed by OPERA are above threshold 
for \cv\ emission.
The same argument also implies that 
the atmospheric neutrinos exceeding 100 TeV 
observed by the IceCube collaboration 
\cite{IceCube11}
lie well above the threshold 
of $\kmag \simeq 13$ TeV
for \cv\ $Z^0$ emission.
Note that the occurrence of more complicated processes
such as neutrino splitting can only serve to strengthen
these conclusions.

\subsubsection{Spectral distortion}

The neutrino energy loss to \cv\ radiation  
implies a distortion in the energy spectrum 
of the neutrino beam that depends on the baseline.
Nonobservation of this distortion
therefore offers a potential basis
for placing constraints on Lorentz violation.
One measure of relevance for this analysis 
is the energy loss per distance $dE/dx$,
which can formally be obtained 
for arbitrary coefficients for Lorentz violation
using existing techniques for evaluation 
of scattering and decay processes in the SME.
A complete analysis requires determining 
contributions from all \cv\ processes above threshold.

For each \cv\ process,
care is required to account for two distinct kinds 
of Lorentz-violating modifications to conventional results.
One is kinematical effects in phase space,
associated with the unconventional dispersion relations.
These are comparatively straightforward to treat.
The other is changes to the matrix element,
which arise in several ways
and are more subtle to handle.
In processes with external fermions 
like neutrino \cv\ radiation,
the basic spinor solutions can be unconventional
because Lorentz violation generically implies
that spin no longer commutes with the hamiltonian
\cite{ck,ckscat}.
Also,
gauge invariance implies 
that Lorentz-violating neutrino dispersion relations
come with unconventional interactions,
so vertices such as $\nu\nu Z^0$ acquire modifications.
Moreover,
in realistic models
at least part of each neutrino field
lies in an electroweak doublet with a charged lepton,
so Lorentz-violating neutrino properties
imply modifications to charged leptons. 
As a result,
any \cv\ process involving leptons
such as electron-positron pair production
acquires accompanying Lorentz-violating contributions.
These may qualitatively change the physical behavior,
in some circumstances even eliminating \cv\ emission. 

As an example providing some insight,
consider the case of electron-positron pair emission 
in the presence of neutrino-sector Lorentz violation.
If the neutrino energy is sufficiently above 
the threshold \rf{operath} for this process, 
which is true for OPERA energies and higher, 
the neutrino and electron masses can be neglected.
For simplicity,
suppose the only Lorentz violation in neutrino propagation 
is of Dirac type.
The contributions from neutrino propagation
are then controlled by the term $-\Vhat^\mu_L \bsi_\mu$
in the Weyl hamiltonian \rf{dehw},
or equivalently by the effective coefficients
$\ae^{ab}$ and $\ce^{ab}$ 
in the hamiltonian \rf{efflvham}.
Including the effect of Lorentz violation
in the interaction vertices and in the electron sector
would require further development
of the formalism presented in this work,
which lies beyond our present scope.
However,
these effects are of the same order
in Lorentz violation as the effects on propagation, 
so neglecting them can plausibly be expected
to yield results of the correct order of magnitude
except in special circumstances.
As a consequence,
although the assumption of Lorentz violation 
confined to the neutrino sector is strictly inappropriate 
for a complete and realistic analysis of \cv\ radiation,
we can nonetheless expect to obtain reasonable insight
by evaluating effects from modifications of the neutrino spinors
and of the kinematics.

Under these assumptions,
the relevant contribution to the matrix element 
in unitary gauge is
\bea
\hskip-20pt
i{\mathcal M} &=& \fr{-i\sqrt{2}G_F M_Z^2}
{(k+k')^2 - M_Z^2}
\nub(p') \ga^\mu \nu(p)
\nn\\
&&
\hskip 30pt
\times
\bar u(k) \ga_\mu (2s^2 - P_L) v(k'),
\eea
where the incoming neutrino has momentum $p$,
the outgoing one has momentum $p'$,
the electron and positron momenta are $k$ and $k'$,
and $s\equiv \sin\th_W$.
To ensure validity of the results 
at high neutrino energies above the $Z^0$ pole,
we keep the $Z^0$ propagator
instead of adopting the four-Fermi approximation.

In determining the square of the matrix element,
the sum over electron spins yields the conventional result
by assumption.
However,
the neutrino spin sum is modified.
The neutrino spinors obey the modified Weyl equation
$q\cdot\bsi \ph_L = 0$,
where $q^\mu \equiv p^\mu-\Vhat^\mu_L$ 
is an effective lightlike momentum satisfying 
\bea
q_0 &=& E - \Vhat^0_L 
\ = \ |\vec q| 
\ \approx \
|\vec p| - \fr{\vec p\cdot\vec \Vhat_L}{|\vec p|} ,
\nn\\
q/q_0 &=& (1,\hat q) 
\ \approx \ (1,\phat - \vec \Vhat_L/|\vec p| 
+ \vec p\,\vec p\cdot\vec \Vhat_L /|\vec p|^3).
\qquad
\label{qid}
\eea
Adopting the usual normalization for the spinors implies 
$\ph_L \ph_L^\dag = E(1- \hat q \cdot \vec \si)
=Eq\cdot\si/q_0$.
Using this result,
we obtain 
\beq
\sum_{\rm spin} |{\mathcal M}|^2 = 
\fr{32 G_F^2 M_Z^4(1-4s^2+8s^4) q\!\cdot\!k\ q'\!\cdot\!k'}
{[(k+k')^2 - M_Z^2]^2}
\left(\fr{E_pE_{p'}}{q_0q_0'}\right), 
\eeq
where the factor in parentheses
results from the modified neutrino spinors.

\begin{figure}
\begin{center}
\centerline{\psfig{figure=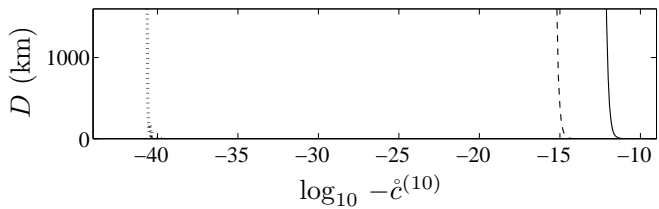,
width=\hsize}}
\caption{ \label{fig}
Effective distortion distance $D$ as a function 
of the isotropic oscillation-free coefficient
$-\ccfc{10}{}$ in units of GeV$^{-6}$
for three neutrino energies:
17 GeV (solid line),
43 GeV (dashed line),
and 100 TeV (dotted line).}
\end{center}
\end{figure}

The energy loss per distance 
is given by the integral
\bea
\fr {dE}{dx} &=& 
-\int
\fr{d^3p'}{(2\pi)^3 2E_{p'}}
\fr{d^3k}{(2\pi)^3 2E_{k}}
\fr{d^3k'}{(2\pi)^3 2E_{k'}}
\nn\\
&&
\hskip 10pt
\times
\fr{(E_k+E_{k'})(2\pi)^4\de^4(p-p'-k'-k) 
\sum |{\mathcal M}|^2}{2E_p}
\nn\\
&=& 
-\fr{C}{8} 
\int 
\sin\th \sin\th' \, 
d\th\, d\ph\, d\th'\, d\ph'\, 
d|\vec\ka| 
\nn\\
&&
\hskip 40pt
\times
\fr{\ka^0 \vec\ka^2 \vec\ka'^2} {(\ka^2-M_Z^2)^2}
\fr{\prt|\vec\ka'|}{\prt\ka^0}
\fr{q\!\cdot\!k\ q'\!\cdot\!k'}{q_0E_{k}q_0'E_{k'}} .
\label{elpd}
\eea
The second integral is constrained by the conditions
$\ka = p - p'$,
$E_p \geq \ka^0 \geq \sqrt{\vec\ka^2+4m^2}$,
and the constant $C$ is given by
$C = {2 G_F^2 M_Z^4(1-4s^2+8s^4)}/{(2\pi)^5}$.
In this second expression
we have used the $\de$-function to perform one integration,
introducing the convenient combinations
$\ka^\mu = k^\mu + k'^\mu$ and 
$\ka'^\mu = k^\mu - k'^\mu$
with spatial spherical polar angles 
$(\th,\ph)$ and $(\th',\ph')$.

The integral \rf{elpd} provides the energy loss per distance
in the presence of Dirac-type Lorentz-violating operators
of arbitrary mass dimension,
including anisotropic effects. 
To gain some feeling for this integral
and to extract estimated bounds
on Lorentz violation from \cv\ spectral distortion,
consider the isotropic oscillation-free model
with neutrino energy $\Eri_\nu$ given by Eq.\ \rf{isoofen}
in the preferred frame.
For this special case,
Eq.\ \rf{qid} shows the factor $q/q_0$ becomes $(1,\phat)$,
so within our assumptions all the Lorentz violation
lies in the kinematics.
Given values of the coefficients
$\acfc{d}{}$, $\ccfc{d}{}$
and a specific neutrino energy $\Eri_\nu$,
the integral \rf{elpd} can be numerically evaluated
and used to calculate the effective distortion distance,
defined by
\beq
D(E) \equiv - \fr{E}{dE/dx}.
\eeq
For fixed energy,
this distance is a rapidly falling function of  
the coefficients $\acfc{d}{}$ and $\ccfc{d}{}$,
as illustrated in Fig.\ \ref{fig} for the case $d=10$.
It therefore offers a useful measure of the spectral distortion
caused by Lorentz violation.

\begin{table}
\begin{tabular}{c|c|c|c}
coefficient & 17 GeV & 43 GeV & 100 TeV \\
\hline\hline
$-\ccfc{4}{}$	&$	3	\times	10^{-5}	$&$	6	\times	10^{-6}	$&$	3	\times	10^{-11}	$\\
$\acfc{5}{}$	&$	1	\times	10^{-6}	$&$	1	\times	10^{-7}	$&$	4	\times	10^{-16}	$\\
$-\ccfc{6}{}$	&$	8	\times	10^{-8}	$&$	3	\times	10^{-9}	$&$	3	\times	10^{-21}	$\\
$\acfc{7}{}$	&$	5	\times	10^{-9}	$&$	6	\times	10^{-11}	$&$	2	\times	10^{-26}	$\\
$-\ccfc{8}{}$	&$	3	\times	10^{-10}	$&$	1	\times	10^{-12}	$&$	2	\times	10^{-31}	$\\
$\acfc{9}{}$	&$	2	\times	10^{-11}	$&$	3	\times	10^{-14}	$&$	2	\times	10^{-36}	$\\
$-\ccfc{10}{}$	&$	9	\times	10^{-13}	$&$	8	\times	10^{-16}	$&$	2	\times	10^{-41}	$
\end{tabular}
\caption{\label{cherenkov}
Estimated values of isotropic oscillation-free coefficients
at which the distortion distance $D(E)$ 
for \cv\ pair emission crosses 1000 km 
for energies 17 GeV, 43 GeV, and 100 TeV.
Coefficient units are GeV$^{4-d}$.}
\end{table}

No substantial spectral distortion is observed 
at the mean OPERA energy of 17 GeV
or ranging up to energies of over 40 GeV
\cite{opera},
and no depletion of atmospheric neutrinos or antineutrinos
is detected in IceCube for energies over 100 TeV
\cite{IceCube11}.
Requiring that the distortion distance $D$ 
at these energies is 1000 km or greater,
we can extract conservative limits 
on isotropic oscillation-free coefficients 
for Lorentz violation.
Table \ref{cherenkov}
shows estimates for the bounds obtained
by taking one coefficient to be nonzero at a time. 
The bounds are one sided
except for the 100 TeV results for
$\acfc{5}{}$, $\acfc{7}{}$, and $\acfc{9}{}$,
which are on the modulus of the coefficients.
The values include ones substantially tighter
than the results for the coefficients
found in time-of-flight experiments,
which are given in Table \ref{isotofcond}.
Note that a complete calculation at 100 TeV  
is likely to produce sharper limits
because at these energies pair emission
is expected to be only a small contribution.
For example,
only about 3\% of the on-shell $Z^0$ emission 
generates electron-positron pairs.

An explicit bound on a coefficient for Lorentz violation 
has recently been obtained
from an analysis of the decay rate and energy loss
for \cv\ pair emission 
\cite{cg}.
This work assumes that Lorentz violation arises 
only from the isotropic oscillation-free coefficient $\ccfc{4}{}$
in the minimal SME,
leading to an isotropic constant shift 
$\de \vg = - \ccfc{4}{}$
of the muon-neutrino group and phase velocities.
The analysis finds the best limit 
on this type of Lorentz violation 
comes from IceCube measurements of 100 TeV neutrinos,
translating to the one-sided bound
$\ccfc{4}{} > -8.5\times 10^{-12}$.
Also within this particular one-coefficient model,
the ICARUS collaboration reports a limit 
based on the nonobservation of \cv-emission products
in the same neutrino beam
\cite{icarus},
which corresponds to the one-sided bound
$\ccfc{4}{} > -2\times 10^{-8}$.
Both these limits are many orders of magnitude
tighter than the nonzero negative value of $\ccfc{4}{}$
implied by the OPERA result for this one-coefficient model
and listed in the first row of Table \ref{isotofcond}.

\section{Summary}

\begin{table}
\renewcommand{\arraystretch}{1.5}
\begin{tabular}{c|c|c|c}
scenario & coefficient & values of $d$ & number per $d$\\
\hline\hline							
effective	&	$\aeff{d}{jm}{ab}$	&	odd, $\geq 3$	&	$9d^2$	\\
	&	$\ceff{d}{jm}{ab}$	&	$\begin{cases} d=2 & \\ \mbox{even, } \geq 4 & \end{cases}$	&	$\begin{array}{c} 27\\[-2pt] 9d^2\end{array}$ 	\\
	&	$\geff{d}{jm}{ab}$	&	even, $\geq 2$	&	$12(d^2-1)$ 	\\
	&	$\Heff{d}{jm}{ab}$	&	odd, $\geq 3$	&	$6(d^2-1)$ 	\\
\hline\hline							
Dirac	&	$\aLcoef{d}{jm}{ab}$	&	odd, $\geq 3$	&	$9(d-1)^2$ 	\\
	&	$\cLcoef{d}{jm}{ab}$	&	even, $\geq 4$	&	$9(d-1)^2$ 	\\
	&	$\mlcoef{d}{jm}{ab}$	&	odd, $\geq 5$	&	$9(d-2)^2$ 	\\
	&	$\elcoef{d}{jm}{ab}$	&	even, $\geq 4$	&	$9(d-2)^2$ 	\\
	&	$\glcoef{d}{jm}{ab}$	&	even, $\geq 4$	&	$9(d-1)^2$ 	\\
	&	$\Hlcoef{d}{jm}{ab}$	&	$\begin{cases} d=3 & \\ \mbox{odd, } \geq 5 \end{cases}$	&	$\begin{array}{c} 27\\[-2pt] 9(d-1)^2\end{array}$ 	\\
\hline							
Majorana	&	$\gMcoef{d}{jm}{ab}$	&	even, $\geq 4$	&	$12d(d-2)$ 	\\
	&	$\HMcoef{d}{jm}{ab}$	&	odd, $\geq 3$	&	$6d(d-2)$ 	\\
	&	$\alcoef{d}{jm}{ab}$	&	odd, $\geq 3$	&	$12d(d-2)$ 	\\
	&	$\clcoef{d}{jm}{ab}$	&	even, $\geq 4$	&	$6d(d-2)$ 	\\
\hline\hline							
renormalizable	&	$\aeff{3}{jm}{}$	&	3	&	$81$	\\
	&	$\ceff{2}{jm}{}$	&	2	&	$27$	\\
	&	$\ceff{4}{jm}{}$	&	4	&	$81$	\\
	&	$\geff{2}{jm}{}$	&	2	&	$36$	\\
	&	$\geff{4}{jm}{}$	&	4	&	$96$	\\
	&	$\Heff{3}{jm}{}$	&	3	&	$48$	\\
\hline\hline							
massless	&	$\aLcoef{d}{jm}{ab}$	&	odd, $\geq 3$	&	$9(d-1)^2$	\\
	&	$\cLcoef{d}{jm}{ab}$	&	even, $\geq 4$	&	$9(d-1)^2$	\\
	&	$\gMcoef{d}{jm}{ab}$	&	even, $\geq 4$	&	$12d(d-2)$	\\
	&	$\HMcoef{d}{jm}{ab}$	&	odd, $\geq 3$	&	$6d(d-2)$ 	\\
\hline\hline							
flavor-blind	&	$\afb{d}{jm}$	&	odd,  $\geq 3$	&	$d^2$	\\
	&	$\cfb{d}{jm}$	&	even, $\geq 4$	&	$(d-1)^2$	\\
	&	$\gfb{d}{jm}$	&	even, $\geq 2$	&	$2(d^2-1)$	\\
\hline							
oscillation-free	&	$\aof{d}{jm}$	&	odd,  $\geq 3$	&	$(d-1)^2$	\\
	&	$\cof{d}{jm}$	&	even, $\geq 4$	&	$(d-1)^2$	\\
\hline\hline							
diagonalizable	&	$\adia{d}{jm}{a'}$	&	odd,  $\geq 3$	&	$3d^2$	\\
	&	$\cdia{d}{jm}{a'}$	&	even, $\geq 4$	&	$3(d-1)^2$	\\
	&	$\gdia{d}{jm}{a'}$	&	even, $\geq 2$	&	$6(d^2-1)$	\\
\hline\hline							
generic isotropic	&	$\afc{d}{ab}$	&	odd, $\geq 3$	&	9	\\
	&	$\cfc{d}{ab}$	&	even, $\geq 4$	&	9	\\
\hline							
isotropic diag.	&	$\afc{d}{a'}$	&	odd, $\geq 3$	&	3	\\
	&	$\cfc{d}{a'}$	&	even, $\geq 4$	&	3	\\
\hline							
isotropic osc.-free	&	$\afc{d}{}$	&	odd, $\geq 3$	&	1	\\
	&	$\cfc{d}{}$	&	even, $\geq 4$	&	1	
\end{tabular}
\caption{\label{summary}
Summary of spherical coefficients. }
\end{table}

In this work,
we study the effects of Lorentz and CPT violation
on the behavior of fermions,
with emphasis on neutrinos.
The starting point in Sec.\ \ref{Fermions}
is the construction 
of the general quadratic Lagrange density \rf{multilag}
for the propagation and mixing of $N$ species of fermions.
This permits Lorentz-violating operators of arbitary mass dimension
to be classified and enumerated.
A procedure to obtain the leading-order hamiltonian 
\rf{leadingham} for fermions is described.

In Sec.\ \ref{Neutrinos},
the general theory is specialized 
to extract the Weyl hamiltonian \rf{rellvham}
describing the propagation and mixing 
of three flavors of left-handed neutrinos
in the presence of Lorentz- and CPT-violation 
involving operators of arbitrary mass dimension.
Block diagonalization of the Weyl hamiltonian
at leading order in mass and Lorentz violation
generates the effective hamiltonian \rf{hresult}
describing dominant modifications
to neutrino propagation and mixing.
This key result depends 
on 10 sets of coefficients for Lorentz violation
that enter the Lorentz-violating piece \rf{efflvham}
of the effective hamiltonian.
These 10 sets and some of their features
are listed in Table \ref{cpttable}.
Six involve Dirac-type operators
for propagation and mixing,
while the other four are Majorana type
and describe neutrino-antineutrino mixing.

Since violations of rotation symmetry
are a central feature of many searches for Lorentz violation,
it is useful to perform 
a decomposition of the effective hamiltonian
and the coefficients using spherical harmonics.
The resulting 10 sets of spherical coefficients for Lorentz violation 
are presented in Sec.\ \ref{Spherical decomposition}.
Six are of Dirac type and four of Majorana type.
This analysis also reveals that 
the fundamental experimental observables
for Lorentz violation in the neutrino sector 
comprise only four sets of effective spherical coefficients,
built from the 10 basic sets 
according to Eq.\ \rf{effcoeff}.
The properties of all 14 coefficient sets are given
in Table \ref{sphercoeff}.

Various special theoretical scenarios
for the spherical coefficients 
can be countenanced,
leading to different experimental predictions.
Section \ref{Special models}
presents several limiting cases of the general formalism. 
We begin in Sec.\ \ref{Renormalizable models}
by matching to the renormalizable sector of the SME,
revealing some qualitatively new effects 
that appear at leading order in both mass and Lorentz violation. 
Some properties of the renormalizable coefficients
are given in Table \ref{rencoeff}.

In Sec.\ \ref{Massless models}
the limit of massless models is described.
This class of models is of interest in part
because it offers the potential for an alternative 
description of neutrino oscillations without invoking mass.
The massless coefficients and some properties
are given in Table \ref{masslesscoeff}.

Another scenario of potential interest,
discussed in Sec.\ \ref{Flavor-blind and oscillation-free models},
is flavor-blind Lorentz violation.
In these models
the different neutrino flavors are assumed 
to experience the same effects,
which is a reasonable approximation 
under some experimental conditions.
For example,
the limit of oscillation-free propagation
is of relevance for certain types of searches
for Lorentz violation,
such as time-of-flight experiments.
Table \ref{fbcoeffs} lists 
the flavor-blind and oscillation-free coefficients
and some of their features.

In Sec.\ \ref{Diagonalizable models}
we consider diagonalizable models,
in which all terms in the effective hamiltonian
are assumed to be simultaneously diagonalizable.
These models offer comparatively simple options
for model building with Lorentz violation,
while avoiding the complications 
that appear for general neutrino mixings.
The diagonalizable coefficients 
are tabulated in Table \ref{diagcoeffs}.

Finally,
various types of isotropic models
are studied in Sec.\ \ref{Isotropic models}.
Isotropy can be enforced only in a preferred inertial frame,
with anisotropies appearing in other boosted frames.
These models are particularly simple 
because they cannot have Majorana mixings
and their Dirac terms must be isotropic.
We discuss generic isotropic models
and their restriction
to isotropic diagonalizable models
and to isotropic oscillation-free models.
The coefficients for each case are 
shown in Table \ref{isocoeffs}.

Table \ref{summary} summarizes 
the various spherical coefficients for Lorentz violation
studied in this work.
The first column of this table lists the theoretical scenario,
the second lists the relevant sets of spherical coefficients,
the third provides the range of operator mass dimension $d$
allowed for each coefficient set,
and the last column indicates
the number of independent coefficients 
for each value of $d$.
The first four rows
concern the effective spherical coefficients
that are the fundamental observables for any experiment
in this framework.
The 10 following rows list 
the 10 basic sets of spherical coefficients 
arising in the general formalism,
which separate into six sets of Dirac type
and four sets of Majorana type.
The rest of the table
concerns the various limiting theoretical scenarios
discussed in the text.

In the penultimate sections of the paper,
we study experimental implications of the theoretical framework
and use existing data on neutrino oscillations and propagation
to extract constraints 
on coefficients for Lorentz and CPT violation.
Section \ref{Applications to oscillations}
treats effects on mixing.
Two experimentally relevant limits are analyzed.
The first is the short-baseline approximation,
for which oscillation effects from all sources are small. 
General expressions 
for short-baseline oscillation probabilities
are given in Eq.\ \rf{probs}.
These expressions are applied to results from 
the short-baseline experiments LSND and MiniBooNE 
to extract maximal attained sensitivities 
to effects from flavor-mixing Lorentz-violating operators 
for $d\leq 10$.
The results for generic spherical coefficients
for Lorentz violation 
are listed in Table \ref{sb},
while those for isotropic coefficients
are given in Table \ref{sbiso}.
The second limit is the perturbative approximation, 
in which the baseline can be arbitrary
but the coefficients for Lorentz and CPT violation 
are assumed small compared to the \msm\ masses.
We derive the mixing probabilities 
up to second order in Lorentz violation,
and we present methods to analyze data 
from long-baseline experiments.
Numerical values of beam-dependent factors
relevant to the MINOS, OPERA, and T2K experiments
are tabulated in Table \ref{Nbeams}.
As a simple example,
we consider the limit of two-flavor mixing
and discuss some asymmetries 
offering sensitivity to CPT violation. 

In Sec.\ \ref{Applications to kinematics},
we discuss several types of kinematic tests 
that are independent of mixing. 
Effects in these tests are controlled by
oscillation-free coefficients for Lorentz violation.
We consider various time-of-flight measurements,
including the OPERA and MINOS experiments,
earlier studies at Fermilab, 
and the supernova SN1987A.
These are used to extract constraints
on generic oscillation-free coefficients 
for $d\leq 6$
and on isotropic oscillation-free coefficients 
for $d\leq 10$.
The results are collected in
Tables \ref{accelconds}, \ref{snofcond}, and \ref{isotofcond}.
We also obtain threshold constraints
from the observation of high-energy neutrinos by IceCube,
presented in Table \ref{icecube},
and we derive the estimated order-of-magnitude bounds 
from neutrino \cv\ radiation
shown in Table \ref{cherenkov}.

The analysis in this paper
provides a general theoretical framework
for the treatment of neutrino propagation and mixing,
along with a guide to its application
in laboratory experiments and astrophysical observations.
We see that searches involving neutrino propagation and mixing
offer excellent sensitivities to numerous 
distinct types of Lorentz and CPT violation.
Many coefficients for Lorentz violation remain unconstrained,
so a substantial region of untouched territory 
is open for future exploration using 
laboratory and astrophysical techniques.

\section*{Acknowledgments}

This work was supported in part
by the Department of Energy
under grant DE-FG02-91ER40661
and by the Indiana University Center for Spacetime Symmetries.

\end{document}